\documentclass[prl,amsmath,amssymb,twocolumn, showpacs, superscriptaddress,10pt]{revtex4-1}

\usepackage[english]{babel}
\usepackage[utf8]{inputenc}
\usepackage[T1]{fontenc}
\usepackage{amsmath}
\usepackage{hyperref}
\usepackage{graphicx}
\usepackage{amsfonts}
\usepackage{amsthm}
\usepackage{cases}
\usepackage{comment}
\usepackage{dsfont}
\usepackage{xcolor}
\usepackage{mwe}

\theoremstyle{remark}

\usepackage{color}
\definecolor{Blue}{rgb}{0.00, 0.00, 1.00}
\definecolor{Red}{rgb}{1.00, 0.00, 0.00}

\newcommand{\nn}{\nonumber}
\newcommand{\be}{\begin{equation}}
\newcommand{\ee}{\end{equation}}
\newcommand{\bea}{\begin{eqnarray}}
\newcommand{\eea}{\end{eqnarray}}

\newcommand{\beq}{\begin{equation}}
\newcommand{\eeq}{\end{equation}}
\newcommand{\beqn}{\begin{eqnarray}}
\newcommand{\eeqn}{\end{eqnarray}}
\begin{document}

\setlength{\abovedisplayskip}{5pt}
\setlength{\belowdisplayskip}{5pt}

\title{Exact short-time height distribution in 1D KPZ equation with Brownian initial condition}

%\author{David S. \surname{Dean}}
%\affiliation{Univ. Bordeaux and CNRS, Laboratoire Ondes et Mati\`ere  d'Aquitaine
%(LOMA), UMR 5798, F-33400 Talence, France}
\author{Alexandre Krajenbrink}
\affiliation{CNRS-Laboratoire de Physique Th\'eorique de l'Ecole Normale Sup\'erieure, 24 rue Lhomond, 75231 Paris Cedex, France}
\author{Pierre Le Doussal}
\affiliation{CNRS-Laboratoire de Physique Th\'eorique de l'Ecole Normale Sup\'erieure, 24 rue Lhomond, 75231 Paris Cedex, France}
%\author{Satya N. \surname{Majumdar}}
%\affiliation{LPTMS, CNRS, Univ. Paris-Sud, Universit\'e Paris-Saclay, 91405 Orsay, France}
%%\author{Alberto \surname{Rosso}}
%%\affiliation{LPTMS, CNRS, Univ. Paris-Sud, Universit\'e Paris-Saclay, 91405 Orsay, France}
%\author{Gr\'egory \surname{Schehr}}
%\affiliation{LPTMS, CNRS, Univ. Paris-Sud, Universit\'e Paris-Saclay, 91405 Orsay, France}

%\title{Large deviations in Johansson's directed polymer in $2$-d}
%\author{Notes prepared by Satya N. Majumdar on large deviations in
%Johansson's model of directed polymer (in interface height language)} 
%\affiliation{LPTMS, CNRS}
\date{\today}

\begin{abstract}
The early time regime of the Kardar-Parisi-Zhang (KPZ) equation in $1+1$ dimension, starting from a Brownian
initial condition with a drift $w$, is studied using the exact Fredholm determinant representation. For large drift 
we recover the exact results for the droplet initial condition, whereas a vanishingly small drift describes the
stationary KPZ case, recently studied by weak noise theory (WNT). 
We show that for short time $t$, the probability distribution $P(H,t)$ of the height $H$ at a given point
takes the large deviation form $P(H,t) \sim \exp{\left(-\Phi(H)/\sqrt{t} \right)}$. We obtain the exact expressions for the 
rate function $\Phi(H)$ for $H<H_{c2}$. Our exact expression for $H_{c2}$ numerically coincides with the value at which
WNT was found to exhibit a spontaneous reflection symmetry breaking. 
We propose two continuations for $H>H_{c2}$,
which apparently correspond to the symmetric and asymmetric WNT solutions.
The rate function $\Phi(H)$
is Gaussian in the center, while it has asymmetric tails, $|H|^{5/2}$ on the negative $H$ side 
and $H^{3/2}$ on the positive $H$ side. 
\end{abstract}


%05.40.-a: Fluctuation phenomena, random processes, noise, and Brownian motion 
%02.10.Yn	Matrix theory
%02.50.-r	Probability theory, stochastic processes, and statistics 

\maketitle

Many works have been devoted to studying 
the 1D continuum KPZ equation \cite{KPZ,directedpoly,reviewCorwin,HairerSolvingKPZ}, 
which describes the stochastic
growth of an interface of height $h(x,t)$ at point $x$ and time $t$ as
\be
\label{eq:KPZ}
\partial_t h = \nu \, \partial_x^2 h + \frac{\lambda_0}{2}\, (\partial_x h)^2 + \sqrt{D} \, \xi(x,t) \;.
\ee
starting from a given initial condition $h(x,t=0)$.
Here
%$\nu > 0$ denotes the strength of the diffusive relaxation, $\lambda_0 > 0$ is the coefficient of the 
%non-linearity and 
$\xi(x,t)$ is a centered Gaussian white noise with 
$\langle \xi(x,t) \xi(x',t')\rangle = \delta(x-x')\delta(t-t')$, and we
use from now on units of space, time and heights such
that $\lambda_0=D=2$ and $\nu=1$ \cite{SuppMat}. A lot of the interest in \eqref{eq:KPZ} is motivated
by the broad universality class of models \cite{johansson,baik2000,spohn2000,corwinsmallreview,reviewSpohn2015}
and experimental systems \cite{HH-TakeuchiReview,myllys,PLDKaz2times},
which share the same asymptotic scaling behavior of their height fluctuations.
It has focused mainly on the limit of large time, for which the universal Tracy-Widom distributions \cite{TWAll} have been
found to emerge \cite{CLR10,DOT10,SS10,ACQ11,PLDflat}.

Recently, the short time behaviour of the KPZ equation has been investigated \cite{CLR10,flatshorttime,Baruch,le2016exact,MeersonParabola,janas2016dynamical}.
It was 
found that the probability distribution function (PDF)  of the %properly shifted 
height $H$ at a given point, see below, takes the following
large deviation form
\be \label{result1}
P(H,t) \sim \exp\left( - \frac{ \Phi(H) }{\sqrt{t}}\right) , \quad H ~ \text{fixed} ~ \text{and} ~ t \ll 1
\ee
where the rate function $\Phi(H)$ depends on the initial condition \cite{footnote1}.
Three types of initial conditions (IC)
have been studied so far, the droplet IC (also called sharp wedge or curved), the flat IC and
the stationary IC. Universal features emerge: (i) the center of the distribution, associated to 
typical fluctuations, is Gaussian, i.e. $\Phi(H) \simeq c (H- H_0)^2$ for $|H-H_0| \ll 1$ 
%{\blue \cite{footnoteGaussian},} 
where here and below $H_0:=\langle H \rangle$, 
and corresponds to the
Edwards-Wilkinson \cite{EW} scaling $H \sim t^{1/4}$ (ii) the tails are asymmetric and exhibit
power law exponents, $\Phi(H) \simeq c_- |H|^{5/2}$ for $H$ large negative, and 
$\Phi(H) \simeq c_+ H^{3/2}$ for large $H$ positive, where the exponents do not depend on the initial condition but the prefactors do.

Two methods have been used to obtain some properties of the rate function. 
The weak noise theory (WNT) uses a saddle point evaluation of the dynamical action associated to
the KPZ equation \eqref{eq:KPZ}, using $1/\sqrt{t}$ as a large parameter
\cite{Baruch,Korshunov,MeersonParabola,janas2016dynamical}. Until now these
saddle point equations have been solved analytically only (i) near the center of the distribution
(ii) in the two tails. This led to predictions for $c, c_{\pm}$, while the complete shape of $\Phi(H)$ could
be obtained only numerically. The second method uses exact formula in terms of a Fredholm determinant
for the moment generating function of $e^H$, valid at any time $t$. These formula however 
are available only for the three IC mentioned above, 
and until now led to the determination of $\Phi(H)$ only for the droplet IC
\cite{le2016exact}, which we refer to as $\Phi_{\rm drop}(H)$,
in agreement with earlier results \cite{CLR10} for the three lowest
cumulants of $H$.
Note that, contrary to the WNT, it yields an exact formula for $\Phi_{\rm drop}(H)$,
recently confirmed in numerical simulations of lattice directed polymer models
\cite{le2016exact,hartmann}. For droplet IC the two methods were found to 
agree, leading to $c_-=\frac{4}{15 \pi}$, $c_+=\frac{4}{3}$ and 
$c=\frac{1}{\sqrt{2 \pi}}$. The flat IC was studied in \cite{Baruch} using the WNT leading to
$c_-=\frac{8}{15 \pi}$, $c_+=\frac{4}{3}$ and 
$c=\frac{\sqrt{\pi}}{2\sqrt{2}}$, showing that the amplitude of the left tail depends on the initial condition.

Interesting connections seem to arise with the (a priori quite different) large deviation
tails observed at {\it large times}. On the positive $H$ side,
the form $P(H,t) \sim \exp( - \frac{4}{3} t (H/t)^{3/2} )$ was shown 
to hold both for droplet and flat IC, implying that the right tail is established at early times
\cite{LargeDevUs}.
On the negative $H$ side a similar feature was recently found in Ref.
\cite{MeersonLatetimes}  for droplet IC.
It would be interesting to understand if these properties hold for a broader class of 
initial conditions. 

Recently, Janas, Kamenev and Meerson \cite{janas2016dynamical} studied stationary initial conditions using the WNT. 
On the negative $H$ side they found $c_-=\frac{4}{15 \pi}$. A surprising feature arises on the positive $H$ side,
where for $H>H_{c2}$ a spontaneous symmetry breaking of reflection invariance occurs, leading
to the coexistence of symmetric and asymmetric solutions. The value $H_{c2} \approx 1.85$
was obtained numerically in \cite{janas2016dynamical}. While the symmetric solution
gives $c_+=4/3$ the asymmetric ones gives $c_+=2/3$, i.e. the same amplitude
as the late time Baik-Rains distribution \cite{baik2000}. An outstanding question is whether
similar results can be obtained using the exact solution.

The aim of this Letter is to use the available exact Fredholm determinant  representation,
valid for all times, to obtain the exact short time rate function $\Phi(H)$ for a broader class of initial conditions,
which interpolate between the droplet and the stationary IC's. We consider the Brownian IC
in presence of a drift
%\vspace{-0.3cm}
\be
%\vspace{-0.3cm}
h(x,0) = B(x) - w |x| \label{ic}
%\vspace{-0.1cm}
\ee
where $B(x)$ is the unit two-sided Brownian motion with $B(0)=0$, and $w$ is the drift. 
The limit $w \to 0^+$ is called the stationary initial condition (the distribution of height differences
at different points being time-independent) while the limit $w \to +\infty$ yields the droplet IC.
We will show that the rate function $\Phi(H)$ depends only on the scaled drift variable
$\tilde w= w t^{1/2}$. For $\tilde w \to +\infty$, we recover the result of \cite{le2016exact} 
as a useful check. The limit $\tilde w \to 0$ continuously leads to the main result of our paper, 
namely the stationary case.

\begin{figure}[!h] 
\includegraphics[width = 1\linewidth]{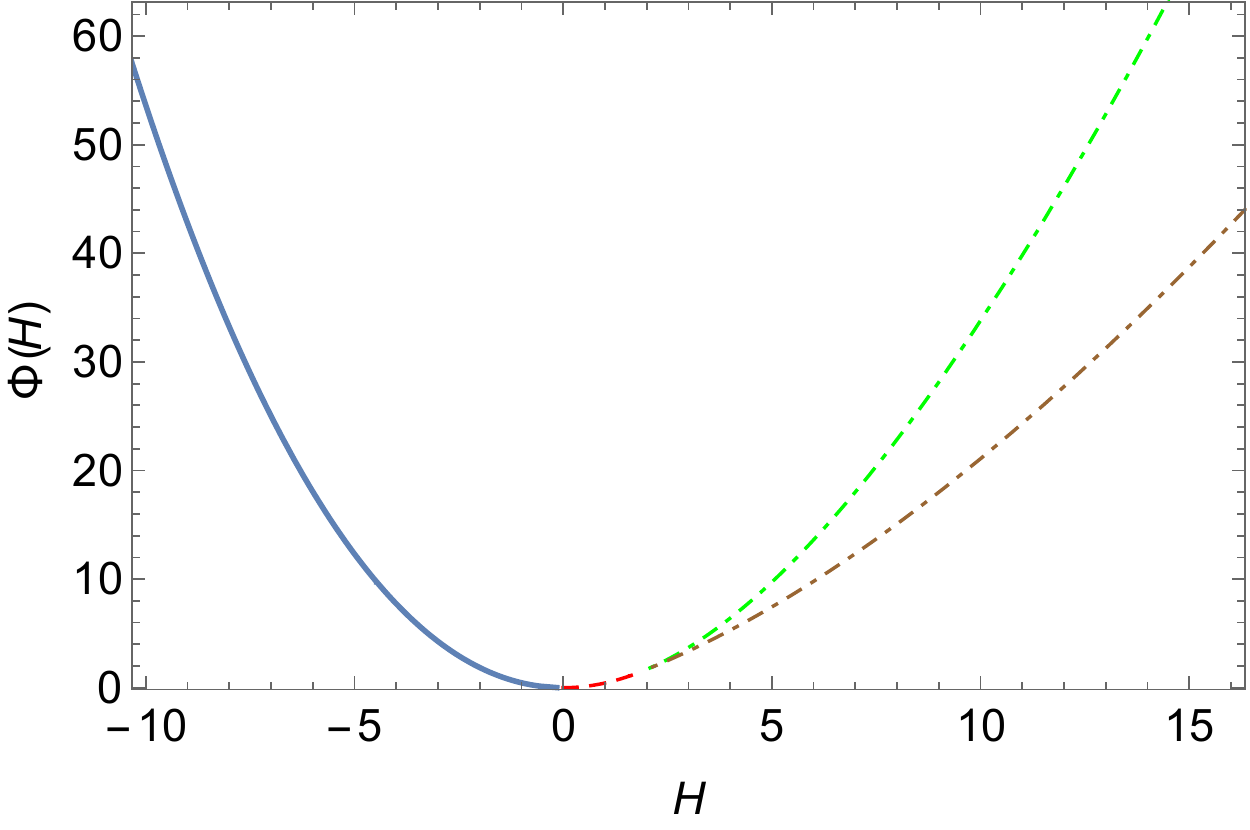}
\centering
\caption{The rate function $\Phi(H)$ defined in \eqref{result1},
which describes the distribution of the KPZ height $H=H(x=0,t)$ at small time for the stationary initial condition ($\tilde w=0$), with $\Phi(0)=0$ and $\langle H \rangle=0$. The blue line corresponds to the exact solution for $H<0$, the dashed red line corresponds to a first analytic continuation for $0<H<H_{c2}$, the dot-dashed green line corresponds to a second symmetric analytic continuation for $H>H_{c2}$ and the dot-dashed brown line corresponds to a second asymmetric non-analytic continuation for $H>H_{c2}$, where $H_{c2}\approx 1.85316$ is discussed in the text. Note the symmetric continuation, with $c_+=4/3$, is not the optimal one in the sense of WNT and the asymmetric continuation with $c_+=2/3$ is regarded as the optimal one.\vspace{-0.2cm}}
\label{fig:PHIdrop}
\end{figure}

As in \cite{le2016exact} we define the shifted height at a point $x$ as
%\cite{footnote2}
%\vspace{-0.3cm}
\bea \label{H}
%\vspace{-0.3cm}
H(x,t) = h(x,t) + \frac{x^2}{4 t} + \frac{t}{12}  \;. 
\eea
and for now we focus on the random variable $H=H(0,t)$. 
We show that its distribution $P(H,t)$ takes the form \eqref{result1}.
From the Fredholm determinant formula we obtain unambiguously the exact form of
$\Phi(H)$ for $H \in ]-\infty, H_c(\tilde w)]$ where $H_c(0)=0$, see Eqs. \eqref{phi1}, \eqref{PhiResult},
which leads to exact formula for the cumulants $\langle H^p\rangle^c$,
see Eq. \eqref{phip2}.
As in the droplet case, a first analytic continuation is required to 
obtain $\Phi(H)$ for $H > H_c(\tilde w)$, and is given in \eqref{PhiResult}.
A new feature arises at the value $H=H_{c2}(\tilde w)$ where
the validity of the first analytic continuations ends.
We obtain $H_{c2}(0) \approx 1.85316$
 consistent with %very close to 
the numerical estimate of \cite{janas2016dynamical},
which %strongly 
suggests that this is the same critical point.
We propose two continuations for $\Phi(H)$ for $H>H_{c2}(\tilde w)$, given in \eqref{resultatphi},
an analytic one which leads to $c_+=4/3$, apparently 
corresponding to the symmetric WNT solution and a non-analytic one which leads to $c_+=2/3$, corresponding to the asymmetric WNT solution.
%At this stage we 
%have no proposal for a continuation which would reproduce the
%asymmetric solutions of \cite{janas2016dynamical} which have $c_+=%\frac{2}{3}$. 
Our result for $\Phi(H)$ is plotted in Fig. \ref{fig:PHIdrop}
and the asymptotic behaviors
of $\Phi(H)$ are for $\tilde w=0$ obtained %explicitly 
{as}
%\bea \label{asympt1} 
%&& \Phi_{{\rm drop}}(H) \simeq \frac{4}{15 \pi} |H|^{5/2} \quad , \quad H \to - \infty \\
%&& \Phi_{{\rm drop}}(H) \simeq \frac{H^2}{\sqrt{2 \pi}} \quad , \quad |H| \ll 1 \label{asympt2}  \\
%&& \Phi_{{\rm drop}}(H) \simeq \frac{4}{3} H^{3/2} \quad , \quad H \to + \infty 
%\label{asympt3} 
%\eea 
\begin{numcases}{\hspace*{-0.5cm}\Phi(H) \simeq}
\frac{4}{15 \pi} |H|^{5/2} \; , \; H \to - \infty  \label{asympt1} \\
\frac{\sqrt{\pi}}{4}( H - H_0)^2 \quad , \quad |H-H_0| \ll 1 \label{asympt2} \\
c_+ H^{3/2} \quad , \quad H \to + \infty \;.
\label{asympt3} 
\end{numcases}
where $c_+=4/3$ for the analytic branch and $c_+=2/3$ for the non-analytic one. Our result for all continuations of $\Phi(H)$ are compared with the numerical determination given in \cite{janas2016dynamical} and we observe \cite{SuppMat} a point to point correspondence between our rate function, the symmetric non-optimal action and asymmetric optimal action of \cite{janas2016dynamical}.
%\begin{figure}[!h] 
%\begin{center}
%\includegraphics[width = 1\linewidth]{Phi(H)2}
%\caption{The rate function $\Phi(H)$ defined in \eqref{result1},
%which describes the distribution of the KPZ height $H=H(x=0,t)$ at small time for the stationary initial condition ($\tilde w=0$), with $\Phi(0)=0$ and $\langle H \rangle=0$. The blue line corresponds to the exact solution for $H<0$, the dashed red line corresponds to a first analytic continuation for $0<H<H_{c2}$, the dot-dashed green line corresponds to a second symmetric analytic continuation for $H>H_{c2}$ and the dot-dashed brown line corresponds to a second asymmetric non-analytic continuation for $H>H_{c2}$, where $H_{c2}\approx 1.85316$ is discussed in the text. Note the symmetric continuation, with $c_+=4/3$, is not the optimal one in the sense of WNT and the asymmetric continuation with $c_+=2/3$ is regarded as the optimal one.\vspace{-0.3cm}}
%\label{fig:PHIdrop}
%\end{center}
%\end{figure}

Let us start by recalling the exact formula obtained  in \cite{SasamotoStationary,SasamotoStationary2,BCFV} for the initial condition \eqref{ic}
with $H=H(x,t)$ and $x=0$ (for details and general $x$ see \cite{SuppMat}).
One needs to introduce $\tilde H = H + \chi$ where
$\chi \in \mathbb{R}$ is a random variable, independent of $H$, with 
a probability distribution 
$p(\chi) d\chi =  e^{-2 w \chi - e^{-\chi}} d\chi/ \Gamma(2 w)$.
Then the moment generating function is given by
\begin{eqnarray}
&&\bigg\langle \exp \left( - e^{\tilde H - s t^{1/3}}  \right) \bigg\rangle = Q_t(s) \label{eq:exact1} \\
&&Q_t(s) := {\rm Det}[ I -  \bar K_{t,s} ] \label{FD1} 
\end{eqnarray} 
where $\langle \ldots \rangle$ denotes an average over the KPZ noise, the random initial condition and the random variable $\chi$. Here $Q_t(s)$ is a Fredholm determinant associated to
the kernel
\be \begin{split} \label{Kt} 
\bar  K_{t,s}(v,v') := &K_{\rm Ai, \Gamma}(v,v')\sigma_{t,s}(v') 
\end{split} \ee
defined in terms of the weight function
\bea \label{sig} 
&& \sigma_{t,s}(u) := \sigma(t^{1/3}(u-s)) \quad , \quad \sigma(v) := \frac{1}{1+e^{- v } } \;.
\eea
and of the deformed Airy kernel
\be \begin{split} \label{KtAi} 
  K_{\rm Ai, \Gamma}(v,v') := &\int_{0}^{+\infty} dr \, \mathrm{Ai}_\Gamma^\Gamma(r+v,t^{-\frac{1}{3}},w,w) \\ &\mathrm{Ai}_\Gamma^\Gamma(r+v',t^{-\frac{1}{3}},w,w) 
\end{split} \ee 
 itself is defined from the deformed Airy function %$\mathrm{Ai}_\Gamma^\Gamma(a,b,c,d)$ 
%\be  \label{aigamma} 
%\mathrm{Ai}_\Gamma^\Gamma(a,b,c,d) := \frac{1}{2i\pi}\int\limits_{-i\infty+\epsilon}^{+i\infty+\epsilon}\mathrm{exp}   (\frac{\eta^3}{3}-a\eta)\frac{\Gamma(-b\eta+d)}{\Gamma(b\eta+c)}  \mathrm{d}\eta \;
%\ee
%where $\epsilon \in [0, \mathrm{Re}(d/b)[$ and $\Gamma$ is the usual Gamma function.

\be  \label{aigamma} 
\mathrm{Ai}_\Gamma^\Gamma(a,b,c,d) := \frac{1}{2 \pi}\int\limits_{-\infty+i \epsilon}^{+\infty+i \epsilon}\mathrm{exp}   (i \frac{\eta^3}{3}+ i a\eta)\frac{\Gamma(i b\eta+d)}{\Gamma(-i b\eta+c)}  \mathrm{d}\eta \;
\ee
where $\epsilon \in [0, \mathrm{Re}(d/b)[$ and $\Gamma$ is the usual Gamma function.

In principle, the formula (\ref{eq:exact1}) allows us to obtain, via a Laplace
inversion, the PDF of $H$ for arbitrary $t$. We now show how to extract the small time behavior directly from 
the generating function (\ref{eq:exact1}). We recall the trace formula for Fredholm determinants
\be
\!  \ln {\rm Det}[ I - \bar K_{t,s} ] = \sum_{p=1}^{+\infty} \frac{-1}{p} {\rm Tr}\,  \bar K_{t,s}^p \label{sump} 
\ee
a convenient form to study the small $t$ limit.
%\begin{comment}
%We now illustrate the small time analysis on the first term $p=1$ 
%of this series, the general term being analyzed in \cite{SuppMat}. 
%One has
%\bea
%&& {\rm Tr}\, \bar K_{t,s}  = \int_{-\infty}^{+\infty} du K_{\rm Ai}(u,u) \sigma(t^{1/3}(u-s)) \nonumber \\
%&& = t^{-1/3} \int_{-\infty}^{+\infty} dv K_{\rm Ai}\left(\frac{v}{t^{1/3}} , \frac{v}{t^{1/3}} \right)  \sigma(v - \tilde s) 
%\label{Tr1} 
%\eea 
%where we have performed the change of variable $u= v/t^{1/3}$ and defined
%$\tilde s=s t^{1/3}$. We see on this equation that the small $t$ limit is controled by
%the large argument behavior of the Airy kernel. Since it is decreasing exponentially
%fast at positive large arguments, we only need its behavior for large negative arguments.
%\end{comment}
 Our strategy throughout will be to consider the small $t$ limit at fixed
$\tilde w= w t^{1/2}$. To calculate the traces in the equation (\ref{sump}) we need the following 
asymptotic estimate, valid for fixed $\hat v<0$, $t\to 0$ and $\kappa,\tilde w$ fixed (see \cite{SuppMat})
\be
K_{\rm Ai,\Gamma}\left(\frac{\tilde{v}}{t^{1/3}} ,\frac{\tilde{v} + t^{1/2} \kappa}{t^{1/3}}\right)  \underset{t \ll 1}{\simeq} 
\frac{ 1}{\pi t^{1/6}} \frac{\sin ( \kappa f_{\tilde w}(\hat v))}{\kappa} \,, \label{estimate} 
\ee
where $\tilde{v}=\hat v- \ln \tilde w^2+ \ln t$, and $f_{\tilde w}(\hat v)=\sqrt{W_0(\tilde w^2 e^{-\hat v+\tilde{w}^2})-\tilde{w}^2}$ and $W_0$ is the first branch of the Lambert $W$ function,
i.e. $y=W_0(x)$ is the solution of $y e^y=x$, with $y$ in $[-1,+\infty[$ and $x \in [-e^{-1},+\infty[$.
For $\tilde w \to +\infty$, $f_{+\infty}(\hat v) = \sqrt{-\hat v}$ and the deformed Airy 
kernel yields the standard one up to a shift, see \cite{SuppMat}, hence both sides of \eqref{estimate} identify 
with Eq. (18) in \cite{le2016exact} (with $\hat v \to v$). Defining $\tilde s=st^{1/3}$, 
the series (\ref{sump}) can be summed up, extending the derivation in \cite{le2016exact} to arbitrary $\tilde w$, leading to \cite{SuppMat}
\bea %\begin{split}\
\label{resQ} 
&&\ln Q_{t}(s) \simeq - \frac{1}{\sqrt{t}} \Psi(t e^{-\tilde s}) \\
&&\Psi(z) :=  \frac{1}{\pi} \int_{0}^{+\infty} dy \left[1+\frac{1}{y+\tilde{w}^2}\right]\sqrt{y} \ln\left(1 + \frac{ze^{-y}}{y+\tilde{w}^2} \right)  \nn
\eea
%\end{split}
%\end{equation}
where the integral $\Psi(z)$ is defined for $z>- \tilde w^2$.
Defining $z=t e^{-\tilde s}$, the exact formula for the generating function (\ref{eq:exact1}) 
takes the following form at small time
\be \label{genfunct} 
\bigg \langle \exp\left( - \frac{z}{t} e^{\tilde H}  \right) \bigg \rangle \sim e^{ -  \frac{1}{\sqrt{ t}} \Psi(z) } .
\ee 
Note that the l.h.s. is finite only for $z>0$ (for
$z<0$ it is infinite). We now want to extract from \eqref{genfunct} information
about the PDF of $H$. To this aim, we now define $\chi'=\chi -\ln (\sqrt{t})$, inserting the assumed 
form (\ref{result1}) into (\ref{genfunct})
for any $z>0$, we obtain $\Phi(H)$ by a saddle point analysis on $z$ and $\chi'$, the latter being exactly solved \cite{SuppMat}. The range of optimization can be enlarged from $z>0$ to $z>-\tilde{w}^2$ as the argument continuously extends on this domain.  The rate function is then given as a generalized Legendre transform of $\Psi$.
\begin{equation}
\begin{split}
\Phi(H)& = \max_{\substack{z \in [-\tilde{w}^2,+\infty[}} [\Psi(z) +2\tilde{w}\ln (\tilde{w}+\sqrt{\tilde{w}^2+  ze^H})\\
&-2\sqrt{\tilde{w}^2+  ze^H} +2\tilde{w}-2\tilde{w}\ln(2\tilde{w})] \label{phi10} 
\end{split}
\end{equation}
This yields a parametric system of equation 
\begin{equation}
\begin{cases}
e^H=z\Psi'(z)^2+2\tilde{w}\Psi'(z)\equiv G(z)\\
\Phi'(H)=-z\Psi'(z)
\end{cases}
\label{phi1}
\end{equation}
to determine $\Phi'(H)$, see Fig. \ref{fig.wz}. Since $G(z)$ is monotonically
decreasing, the solution $z(H)$ is unique.
It is also possible to integrate this system to obtain a parametric equation on $\Phi(H)$
\begin{equation}
\label{phi2}
\Phi(H)= \Psi(z)-2z \Psi'(z)+2\tilde{w}\ln \left|1+\frac{z\Psi'(z)}{2\tilde{w}} \right|
\end{equation}
It is important to note that Eqs. \eqref{phi10}, \eqref{phi1}, \eqref{phi2} are valid 
for $z \in [- \tilde w^2,+\infty[$, hence as for now we have solved the problem
only for $H \in [-\infty, H_c(\tilde w)]$ with 
$H_c(\tilde w)=\ln G(-\tilde w^2)$. The extension is studied below.
Note that from \eqref{phi1}, $1-G(-\tilde w^2)$ is a complete square, hence
$H_c(\tilde w) \leq 0$ for any $\tilde w$. 

We now extract from this solution the cumulants of $H$ and the left tail behavior. 
The most probable value is also the average $H_0=\langle H \rangle$ determined
by $\Phi(H_0)=0$. Noticing that $\Psi(0)=0$ and that $\Psi'(z)$ is bounded, \eqref{phi2} and \eqref{phi1}
imply that $e^{H_0}=2\tilde{w}\Psi'(0)= {\rm Erfc}(\tilde w) e^{\tilde w^2}$ corresponding to $z=0$. Recalling that $G(z)$ is 
decreasing, it implies that $H_0 \leq H_c(\tilde w) \leq 0$. For $\tilde{w} \to 0$, 
$\Psi'(0) \simeq 1/(2 \tilde w)$ \cite{SuppMat} which implies that
the average $H_0=\langle H \rangle=0$ for the stationary case.
Expanding \eqref{phi1} around $H=H_0$ and $z=0$ we obtain iteratively
the derivatives $\Phi^{(q)}(H_0)$, and calculate the leading short time
behavior of the cumulants of the height given by
\be\label{phip1}
\langle H(t)^q \rangle^c \simeq t^{\frac{q-1}{2}} \phi^{(q)}(0) , \quad \phi(p) := \max_H (p H - \Phi(H))  
\ee
where $\phi^{(q)}$ is the $q$-th derivative of the Legendre transform $\phi(p)$ of $\Phi(H)$. 
We display here the first three cumulants \cite{footnote2} for small $\tilde{w}$ (see \cite{SuppMat} for details)
\bea \label{phip2} 
&& \langle H^2 \rangle^c =\big( \frac{2}{\sqrt{\pi}}+(\frac{6}{\pi}-2)\tilde{w} + \mathcal{O}(\tilde w^2) \big) \sqrt{t} \\
&& \langle H^3 \rangle^c =\big( (2-\frac{6}{\pi})-\frac{8(5-3\pi+\sqrt{2}\pi)}{\pi^{3/2}}\tilde{w} + \mathcal{O}(\tilde w^2) \big) t \nn \\
&& \langle H^4 \rangle^c = \frac{8}{\pi^{3/2}}(5+(\sqrt{2}-3)\pi + \mathcal{O}(\tilde w) )t^{3/2} \nn 
\eea
in agreement with the result of \cite{janas2016dynamical} for the second cumulant at $w=0$.
In addition we have checked the predictions for $\langle H^q \rangle^c$, $q=1,2,3$ for arbitrary
$\tilde w$ by a direct small time expansion of the KPZ equation \cite{SuppMat}. 

It is also possible to obtain the left tail of $\Phi(H)$ from \eqref{phi1}. For all $\tilde{w}$, $G(z)$ is decreasing and $\Psi(z)\simeq_{z \to +\infty}\frac{4}{15 \pi}[\ln z]^{5/2}$, which means that as $z$ increases to $+\infty$, $H$ decreases to $-\infty$. Inserting the asymptotics of $\Psi$ into \eqref{phi1}, see \cite{SuppMat}, we obtain the left tail of the rate function $\Phi(H)\simeq_{H\to -\infty} \frac{4}{15\pi}|H|^{5/2}$ i.e. $c_-=\frac{4}{15 \pi}$. This result is valid for all $\tilde w$, and is in
in agreement both with the droplet result \cite{le2016exact,MeersonParabola} (for $\tilde w \to +\infty$)
and the stationary result $w=0$ \cite{janas2016dynamical}.

An important check of our result \eqref{phi10} is that, for $\tilde{w}\to +\infty$,
it recovers the exact formula \cite{le2016exact} for the droplet IC. Indeed the function $\Psi(z)$ in \eqref{resQ} recovers the one of the droplet IC 
\cite{SuppMat}, $\lim_{\tilde w \to +\infty} \Psi(\tilde w^2 z) = - \frac{1}{\sqrt{4 \pi}} {\rm Li}_{5/2}(-z)=\Psi_{\rm drop}(z)$. Performing the transformation $z=\tilde{z}\tilde{w}^2$ we rewrite \eqref{phi1}, \eqref{phi2} to leading order in $\mathcal{O}(1/\tilde{w})$, as 
\begin{equation}
%\begin{cases}
e^H=\frac{2}{\tilde{w}}\Psi'_{\rm drop}(\tilde z)\quad , \quad 
\Phi'(H)=-\tilde z \Psi_{\rm drop}'(\tilde z) \, .
%\Phi(H)= \Psi_{\rm drop}(\tilde z)-\tilde{z}  \Psi'_{\rm drop}(\tilde {z}) 
%\end{cases}
\end{equation}
Defining $\hat H=H+\ln(\tilde w \sqrt{\pi})$, one finds $\lim_{\tilde w \to +\infty} \Phi(H)=\Phi_{\rm drop}(\hat H)$
and the value of $\hat H$ at the branching point $\hat H_c = \ln \zeta(3/2)$ as in \cite{le2016exact}.

We now find an extension of the rate function $\Phi(H)$ for $H> H_c(\tilde w)$. Note that in the stationary limit
$H_c(0^+)=0$.
The trick is to use the analytically continued partner of $\Psi(z)$ which is obtained by adding to $\Psi(z)$ the jump induced by changing the Riemann sheet on which $\Psi(z)$ is defined. We define this jump to be $\Delta_0(z)=\lim_{\epsilon\rightarrow 0}\, [\Psi(z-i\epsilon)-\Psi(z+i\epsilon)]$, its expression is given by \cite{SuppMat}
\bea
\Delta_{0}(z)&=&\frac{4}{3}[\tilde{w}^2-W_{0}(-ze^{\tilde{w}^2})]^{\frac{3}{2}}-4[\tilde{w}^2-W_{0}(-ze^{\tilde{w}^2})]^{\frac{1}{2}} \nn\\
&&+2\tilde{w}\, \ln(\frac{\tilde{w}+[\tilde{w}^2-W_{0}(-ze^{\tilde{w}^2})]^{\frac{1}{2}}}{|\tilde{w}-[\tilde{w}^2-W_{0}(-ze^{\tilde{w}^2})]^{\frac{1}{2}}|})  \label{Delta0} 
%\end{split}
\eea
%\end{equation}
where $W_0$ is the first branch of the Lambert W function. Its derivative is given by 
\begin{equation} \label{De0prime} 
\Delta'_0(z)=-2\frac{[\tilde{w}^2-W_0(-ze^{\tilde{w}^2})]^{1/2}}{z}
\end{equation}
We extend the parametric system \eqref{phi1} by imposing the minimal replacement 
$\Psi(z) \to \Psi(z) +\Delta_0(z)$ in both equations. This produces the natural
extension of $\Phi(H)$. Despite the addition of the jump, %the branching point is analytic,
$\Phi(H)$ is an analytic function at $H=H_c(\tilde w)$, see \cite{SuppMat}. 
As $\Delta_0'(-\tilde{w}^2)=0$, $H$ viewed as a function of $z$ is also continuous. 
Note that \cite{SuppMat} in the limit $\tilde{w}\to +\infty$, $\Delta_0(z)$ converges towards the analytic jump 
obtained in \cite{le2016exact} for the droplet IC $\Delta_0(z) \simeq_{\tilde{w}\to +\infty}\frac{4}{3} [-\ln(-z)]^{3/2}$,
a consistency check on our method. 

It turns out that this analytic partner of $\Psi(z)$ is defined only on a finite interval, $ z \in [-\tilde{w}^2,e^{-1-\tilde{w}^2}]$ as $W_0$ is defined on $[-e^{-1},+\infty[$. Here this implies the existence of a second branching point 
$H_{c2}(\tilde w)<+\infty$, as we now show. Defining $G_0(z)$ the analytic partner of $G(z)$ obtained by doing the minimal replacement $\Psi(z) 
\to \Psi(z) + \Delta_0(z)$ in \eqref{phi1}, we see \cite{SuppMat} that $G_0(z)$ is increasing. Hence as $z$ increases from $-\tilde{w}^2$ to $e^{-1-\tilde{w}^2}$, $H$ increases from $H_c(\tilde{w})$ to $H_{c2}(\tilde{w})=\ln G_0(e^{-1-\tilde{w}^2})$. In
the stationary case, using \eqref{phi1}, \eqref{De0prime}, $H_{c2}(0)$ is given by
\begin{equation}
\begin{cases}
H_{c2}(0)=2\ln[2e-\Psi_0'(e^{-1})]-1\approx 1.85316\\
\Psi_0'(e^{-1})=\frac{1}{\pi} \int_{0}^{+\infty} dy [1+\frac{1}{y}] \frac{\sqrt{y}}{e^{-1}+ye^y}
\end{cases}
\end{equation} 
hence $2 H_{c2} \approx 3.70632$ to be compared with the value $3.7$ in \cite{janas2016dynamical}.
We also find that $\lim_{\tilde{w}\to +\infty} H_{c2}(\tilde{w})=+\infty$,
which means that for the droplet IC, only one continuation is needed, i.e. $\hat H_{c2}=+\infty$,
as found in \cite{le2016exact}.

However, for finite $\tilde{w}$ a {\it second extension} is needed to obtain $\Phi(H)$ for
$H > H_{c2}(\tilde w)$. We now investigate the fundamental reason for this point to be special.
This leads us to identify two extensions, by defining two other real partners of $\Psi(z)$. We now
study their properties, and compare below with the work of \cite{janas2016dynamical}. 
When $z=e^{-1-\tilde{w}^2}$ the Lambert function inside $\Delta_0(z)$ in Eq. \eqref{Delta0}
equals $W_0(-e^{-1})$ which is the point where it exhibits a second-order branch point separating three branches $W_0$, $W_{-1}$ and $W_1$, only the first two being real valued
(see Fig. 4 in \cite{corless1996lambertw}). For this reason, a natural continuation for $\Delta_0(z)$ is the function $\Delta_{-1}(z)$ defined by replacing the first branch of the Lambert function $W_0$ by the second real valued one $W_{-1}$
in \eqref{Delta0}, leading to
\bea
\!  \Delta_{-1}(z)&=&\frac{4}{3}[\tilde{w}^2-W_{-1}(-ze^{\tilde{w}^2})]^{\frac{3}{2}}-4[\tilde{w}^2-W_{-1}(-ze^{\tilde{w}^2})]^{\frac{1}{2}}
\nn \\
&& \label{Delta1} 
+2\tilde{w}\, \ln(\frac{\tilde{w}
+[\tilde{w}^2-W_{-1}(-ze^{\tilde{w}^2})]^{\frac{1}{2}}}{|\tilde{w}-[\tilde{w}^2-W_{-1}(-ze^{\tilde{w}^2})]^{\frac{1}{2}}|})  
\eea
As shown in \cite{SuppMat}, $\Psi(z)$ is then be continued by either of the following minimal replacements $\Psi(z) \to \Psi(z)+\Delta_{-1}(z)$ and $\Psi(z) \to \Psi(z)+\frac{\Delta_0(z)+\Delta_{-1}(z)}{2}$. We call the first replacement the symmetric continuation of $\Psi(z)$ and the second one the asymmetric continuation.
They are defined on the interval $z \in ]0,e^{-1-\tilde{w}^2}]$ as $W_{-1}$ is real valued on the interval $[-e^{-1},0[$. Similarly, we define $G_{-1}(z)$ and $G_{-1/2}(z)$ as the continuations of $G_0(z)$ replacing $\Delta_0(z)$ by $\Delta_{-1}(z)$ and $\frac{\Delta_0(z)+\Delta_{-1}(z)}{2}$.  $G_{-1}(z)$ and $G_{-1/2}(z)$ are now decreasing functions, as $z$ decreases from $e^{-1-\tilde{w}^2}$ to $0^+$, $H$ increases from $H_{c2}(\tilde{w})$ to $+\infty$ for using both symmetric and asymmetric continuations, therefore completing the range of $\Phi(H)$ by two extensions above $H_{c2}(\tilde{w})$. Note that this construction yields a function $\Phi(H)$ with a symmetric continuation
analytic at $H_{c2}(\tilde w)$ and an asymmetric continuation non-analytic at $H_{c2}(\tilde w)$ inducing a discontinuity in the second derivative $\Phi''(H_{c2}(\tilde{w}))$ for any finite $\tilde{w}$, see \cite{SuppMat}. 

We now determine the large positive $H$ tail associated to the symmetric and asymmetric extensions of $\Phi(H)$. 
As $z$ approaches $0^+$, $\Delta_{-1}(z)$ behaves as $\frac{4}{3}[-\ln(z)]^{3/2}$. Inserting this asymptotics in \eqref{phi1} 
we obtain the right tail $\Phi(H)\sim_{H\to +\infty} c_+H^{3/2}$, with $c_+=\frac{4}{3}$ for the symmetric extension and $c_+=\frac{2}{3}$ for the asymmetric one. Both tails hold for any finite $\tilde{w}$.

We now compare with the results of Ref. \cite{janas2016dynamical}. The fact that the value of $H_{c2}(0)$ obtained there coincides, up to their numerical precision, with our exact result strongly suggests that this is the same point. 
In \cite{janas2016dynamical} the authors found that at this point $\Phi(H)$ exhibits a second order 
phase transition, i.e. the second derivative $\Phi''(H)$ has a jump. They observe that this is due to 
a spontaneous breaking of the spatial reflection symmetry $x \to - x$ in the saddle point solutions of the dynamical action
of the WNT. For $H > H_{c2}(0)$ they find three solutions: (i) a symmetric solution, which leads to
a positive $H$ tail with $c_+=\frac{4}{3}$ (ii) a pair of asymmetric
solutions with $c_+=\frac{2}{3}$, and they claim that the asymmetric solutions dominate the dynamical action. The two continuations that we have identified very likely correspond to the two solutions found 
numerically in Ref. \cite{janas2016dynamical}. Indeed, overlapping the plot of the exact expression of $\Phi(H)$ with the numerical estimates of \cite{janas2016dynamical} provided by Janas, Kamenev and Meerson, %and upon a change of units 
we observe \cite{SuppMat} that the non-analytic continuation of $\Phi(H)$ coincides point to point \cite{footnote1} with the value of the action obtained there from the asymmetric solution, and the analytic continuation of $\Phi(H)$ coincides with the symmetric one.
%The outstanding remaining question within our method, is to identify the extensions of $\Phi(H)$
%which would correspond to the asymmetric solutions of Ref. \cite{janas2016dynamical}. 
%Hints of {\it some} symmetry breaking also emerge in our calculation, although it is at this
%point unclear how it could be related to the spatial symmetry breaking in \cite{janas2016dynamical}.
%Let us describe it here for completeness.
%First note that for all $y\in \mathbb{C}\setminus\mathbb{R}^-$, $W_1( y^*)=W_{-1}(y)^*$, 
%where $*$ designates the complex conjugate \cite{corless1996lambertw}. This symmetry is
%spontaneously broken when $y \in \mathbb{R}^-$, where $W_{-1}(y)$ is real valued 
%and $W_1(y)$ is conjugated to $W_{-2}(y)$ instead of $W_{-1}(y)$, i.e. 
%$W_1(y)=W_{-2}(y)^*$. In particular in our calculation, when $z \in ]0,e^{-1-\tilde w^2}]$ the
%argument of the Lambert function is 
%$y=-z e^{\tilde w^2}$ and belongs to the negative real axis, where this symmetry breaking occurs.
%This leads to choose $W_{-1}$ to define a real continuation for $\Psi(z)$,
%which eventually leads to $c_+=\frac{4}{3}$. However it is quite possible that other continuations 
%exist, corresponding to some asymmetric solution with $c_+=\frac{2}{3}$.} {\blue Que fait on de tout ca ? cette discussion sur les branches de W n a plus trop de sens, si ?}

To summarize for the stationary limit, $\tilde{w}=0^+$, we find the following parametric representation for $\Phi(H)$,
made of three branches, the last one being composed of an analytic one and a non-analytic one. We recall the intervals 
\begin{eqnarray}
&&I_1=[0,+\infty] ,\; I_2=[0,e^{-1}],\; I_3=]0,e^{-1}]\nn \\
&&J_1=[-\infty,H_c(0)] ,\; J_2=[H_c(0),H_{c2}(0)] ,\; J_3=[H_{c2}(0),+\infty] \nn
\end{eqnarray}
and the relation between $H$ and $z$ in these intervals
 
\begin{flalign} 
&e^H= z\Psi'(z)^2\quad {\rm 
for}\quad 
z\in I_1\; {\rm and}\; H\in J_1&\\
&e^H= z [\Psi'(z)+\Delta'_0(z)]^2 \quad {\rm for}\;
z\in I_2\; {\rm and}\; H\in J_2\, . \nn&
\end{flalign}
For $z\in I_3$ and $H\in J_3$ there are two distinct relations

\begin{flalign}
&e^H= z [\Psi'(z)+\Delta'_{-1}(z)]^2 \qquad \qquad {\rm \it {( analytic)}}&
\\
&e^H= z [\Psi'(z)+\frac{\Delta'_{-1}(z)+\Delta'_0(z)}{2}]^2 \; {\rm \it {(non\;  analytic)}}. &\nn
\end{flalign}
We then recall the relation between $\Phi(H)$ and $z$ 
\begin{flalign} 
\label{resultatphi}
& \Phi(H)= \Psi(z)-2z \Psi'(z) \quad {\rm 
for}\; z\in I_1 &\\
&\Phi(H)=\Psi(z)-2z \Psi'(z)+\frac{4}{3}[-W_0(-z)]^{\frac{3}{2}} \; {\rm for}\;
z\in I_2 \, . \nn&
\end{flalign}
For $z\in I_3$ there exist two branches for $\Phi(H)$, an analytic one and a non-analytic one with different asymptotics
\begin{flalign} 
&\Phi(H)= \Psi(z)-2z \Psi'(z)+\frac{4}{3}[-W_{-1}(-z)]^{\frac{3}{2}}\quad {\rm \it {(  analytic)}} &
\\
&\Phi(H)= \Psi(z)-2z \Psi'(z)+\frac{2}{3}[-W_{0}(-z)]^{\frac{3}{2}}+\frac{2}{3}[-W_{-1}(-z)]^{\frac{3}{2}}&\nn \\& {\rm \it {(non\;  analytic)}}&\nn 
\end{flalign}
\begin{comment}
\begin{multline} 
\label{PhiResult}
\Phi(H) = ( \;  2\tilde{w}-2\tilde{w}\ln(2\tilde{w})+\\ 
\begin{cases}
 \displaystyle \max_{\substack{z \in [-\tilde{w}^2,+\infty[}} [\Psi(z) +2\tilde{w}\ln (\tilde{w}+\sqrt{\tilde{w}^2+  ze^H})\\
 ~~~~~~~~~~~~~~~  -2\sqrt{\tilde{w}^2+ ze^H} ]), \; H \leq H_{c}(\tilde{w})\\
\\
 \displaystyle \max_{z \in [-\tilde{w}^2,e^{-1-\tilde{w}^2}]} [\Psi(z) +\Delta_0(z)	+2\tilde{w}\ln |\tilde{w}+\epsilon_H \sqrt{\tilde{w}^2+ ze^H}| \\
 ~~~~~~~~~~~~~~~  -2 \epsilon_H \sqrt{\tilde{w}^2+  ze^H} ]), \; H_{c}(\tilde{w})\leq H \leq H_{c2}(\tilde{w})\\
 \\
 \displaystyle \max_{z \in ]0,e^{-1-\tilde{w}^2}]} [\Psi(z) +\Delta_{-1}(z) +2\tilde{w}\ln |\tilde{w}-\sqrt{\tilde{w}^2+  ze^H}| \\
 ~~~~~~~~~~~~~~~  +2\sqrt{\tilde{w}^2+  ze^H}]), \; H \geq H_{c2}(\tilde{w}) 
\end{cases}
%\hspace{-1cm}\text{with} \quad \Delta_{0,-1}(z)=\frac{4}{3}[\tilde{w}^2-W_{0,-1}(-ze^{\tilde{w}^2})]^{3/2}\\
%\hspace{-3.2cm}-4[\tilde{w}^2-W_{0,-1}(-ze^{\tilde{w}^2})]^{1/2}\\
%+2\tilde{w}\, \ln(\frac{\tilde{w}+[\tilde{w}^2-W_{0,-1}(-ze^{\tilde{w}^2})]^{1/2}}{\tilde{w}-[\tilde{w}^2-W_{0,-1}(-ze^{\tilde{w}^2})]^{1/2}})  
\end{multline}
\end{comment}
where $\Delta_{0}(z)$ is given in \eqref{Delta0} and $\Delta_{-1}(z)$ in \eqref{Delta1}
(setting $\tilde w=0^+$ which cancels the logarithmic terms).
%\begin{comment}
%Here
%$\epsilon_H={\rm sgn}(H^*-H)$ where $H^*$ is the unique field at which $\Phi'(H^*)=\tilde w$.
%Since $\Phi'(H)=\tilde w - \epsilon_H \sqrt{\tilde w^2 + z e^H}$ one has $z e^H \geq - \tilde w^2$
%with a unique field such that $z e^{H^*}=- \tilde w^2$.
%\end{comment}
In addition, 
$H_{c}(0)= 0$ and $H_{c2}(0)=2\ln(2e-\Psi_0'(e^{-1}))-1\simeq 1.85316$, where $\Psi_0$ is the function $\Psi$ in the limit $\tilde w \rightarrow 0$. %
%and $W_{0,-1}$ are the two real branches of the Lambert $W$ function. 
%\begin{comment}
%Note that when investigating the small $\tilde{w}$ limit, the limit and the maxima do not commute in this problem due to a lack of convergence of certain integrals. 
%
%\end{comment}
%{\blue
%Note that despite the apparent branches, the function
%$\Phi(H)$ is analytic at $H_{c}(0)$ and has one analytic continuation at $H_{c2}(0)$ and a non-analytic one. The non-analyticity appears in the second derivative $\Phi''(H_{c2})$ where it is continuous for the analytic branch and discontinuous for the non-analytic one. More generally, the existence of a non-analytical branch above $H_{c2}(\tilde{w})$ holds for any finite $\tilde{w}$.}
%\begin{comment}
%It is possible to obtain 
%a parametric representation of $\Phi(H)$ that we give in \cite{SuppMat} in Eq. \eqref{parametric}.
%\end{comment}
From the parametric representation of $\Phi(H)$ one obtains the asymptotic behaviors 
given in Eqs. (\ref{asympt1}-\ref{asympt3})~\cite{SuppMat}.

In conclusion we studied the statistics of the height
fluctuations for the continuum KPZ equation at short time
with the Brownian initial condition with a drift.  We obtained an exact determination of the rate function $\Phi(H)$, which describes the stationary IC
at zero drift, and recovers the droplet IC at large drift. 
It extends, through an exact solution, recent approaches
using weak noise theory for the stationary geometry.
We have obtained exactly the value $H_{c2}$ at which a
spontaneous symmetry breaking was found in WNT, showed that this phase transition should happen for any finite drift, and
identified the symmetric and asymmetric
solutions beyond that point. We hope it provides a further bridge
between quite different methods to address large deviations 
in growth and particle transport problems.

%\newpage

\acknowledgments We thank G. Schehr and S. Majumdar for very helpful discussions, and M. Janas, A. Kamenev and B. Meerson for providing their data from \cite{janas2016dynamical} and for their comments. 

%\newpage

{}

\newpage

.

\begin{widetext} 

\bigskip

\bigskip

\begin{large}
\begin{center}

SUPPLEMENTARY MATERIAL

\end{center}
\end{large}

\bigskip

We give the principal details of the calculations described in the manuscript of the Letter. 

\section{0. Solution of continuum KPZ equation with Brownian initial condition}

In this paper we study the KPZ equation \eqref{eq:KPZ} using everywhere the following units 
of space, time and height
\bea \label{units} 
x^*=(2 \nu)^3/(D \lambda_0^2) \quad , \quad t^*=2(2 \nu)^5/(D^2 \lambda_0^4) \quad , \quad 
h^*=\frac{2 \nu}{\lambda_0}
\eea
which amounts to set $\lambda_0=D=2$ and $\nu=1$ in \eqref{eq:KPZ}. 
Let us recall the solution obtained in \cite{SasamotoStationary,SasamotoStationary2,BCFV} for the Brownian initial condition in its most general form, i.e. with two unequal drifts at an arbitrary point $x$. The initial condition is
\bea
h(x,t=0)=B(x)  + w_- x \, \theta(-x) - w_+ x \, \theta(x)
\eea 
where $\theta(x)$ is the Heaviside unit step function and $B(x)$ a double-sided Brownian motion, with $B(0)=0$. Defining now $H=H(x,t)$
as in \eqref{H}, and $\tilde H = H+\chi$  where
$\chi \in \mathbb{R}$ is a random variable, independent of $H$, with 
a probability distribution 
$p(\chi) d\chi =  e^{-(w_++w_-) \chi - e^{-\chi}} d\chi/ \Gamma(w_++w_-)$,
it was shown in \cite{SasamotoStationary,SasamotoStationary2} that (in our units)
\begin{eqnarray}
&&\bigg\langle \exp \left( - e^{\tilde H - s t^{1/3}}  \right) \bigg\rangle = Q_t(s) \quad , \quad Q_t(s) := {\rm Det}[ I -  P_0 K_{t,s} P_0 ] \label{FD0} 
\end{eqnarray}
where, as in the text, $\langle \ldots \rangle$ denotes an average over the KPZ noise, the random initial condition and the random variable $\chi$. 
Here $Q_t(s)$ is a Fredholm determinant associated to
the kernel
\be \label{Kt0} 
K_{t,s}(v,v') := \int_{-\infty}^{+\infty} dr \, \mathrm{Ai}^\Gamma_\Gamma(r+v,t^{-1/3},w_+ - \frac{x}{2 t}, w_- +  \frac{x}{2 t}) 
 \mathrm{Ai}^\Gamma_\Gamma(r+v',t^{-1/3},w_- +  \frac{x}{2 t},w_+-  \frac{x}{2 t}) \sigma_{t,s}(r) 
\ee
where $\sigma_{t,s}(u)$ is defined in \eqref{sig}, and the deformed Airy functions are defined in \eqref{aigamma}.
Now one can rewrite
\be
 {\rm Det}[ I -  P_0 K_{t,s}  ] = {\rm Det}[ I -  \bar K_{t,s}] 
\ee
in terms of the kernel $\bar K_{t,s}(v,v')=K_{\rm Ai, \Gamma}(v,v')\sigma_{t,s}(v')$ with
\be \begin{split} \label{Ktapp} 
 K_{\rm Ai, \Gamma}(v,v') := \int_{0}^{+\infty} dr \, \mathrm{Ai}_\Gamma^\Gamma(r+v,t^{-\frac{1}{3}},w_+ -  \frac{x}{2 t},w_- +  \frac{x}{2 t}) \mathrm{Ai}_\Gamma^\Gamma(r+v',t^{-\frac{1}{3}},w_- +  \frac{x}{2 t},w_+-  \frac{x}{2 t}) 
\end{split} \ee
This can be seen e.g. by
expanding in powers of ${\rm Tr} (P_0 K)^p$ and exchanging the order of integrations. Specializing to $w_{\pm}=w$ and $x=0$ one 
obtains \eqref{eq:exact1}, \eqref{FD1}, \eqref{Kt} and \eqref{KtAi} in the text. 

\section{1. The Lambert function $W$}
We introduce the Lambert $W$ function \cite{corless1996lambertw} which we use extensively throughout the Letter. Consider the function defined on $\mathbb{C}$ by $f(z)=ze^z$, the $W$ function is composed of all inverse branches of $f$ so that $W(z e^z)=z$. It does have two real branches, $W_0$ and $W_{-1}$ defined respectively on $[-e^{-1},+\infty[$ and $[-e^{-1},0[$. On their respective domains, $W_0$ is strictly increasing and $W_{-1}$ is strictly decreasing. By differentiation
 of $W(z) e^{W(z)}=z$, one obtains a differential equation valid for all branches of $W(z)$
\begin{equation} \label{derW} 
\frac{dW}{dz}(z)=\frac{W(z)}{z(1+W(z))}
\end{equation}
Concerning their asymptotics, $W_0$ behaves logarithmically for large argument $W_0(z)\simeq_{z\to +\infty} \ln(z)-\ln \ln (z)$ and is linear for small argument  $W_0(z) \simeq_{z \to 0} z-z^2+\mathcal{O}(z^3)$. $W_{-1}$ behaves logarithmically for small argument $W_{-1}(z)\simeq_{z \to 0^-} \ln(-z)-\ln(-\ln(-z))$. Both branches join smoothly at the point $z=-e^{-1}$ and have the value $W(-e^{-1})=-1$. These remarks are summarized on Fig. \ref{fig:Lambert}. More details on the
other branches, $W_k$ for integer $k$, can be found in \cite{corless1996lambertw}.

\begin{figure}[h!] 
\begin{center}
\includegraphics[width = 0.5\linewidth]{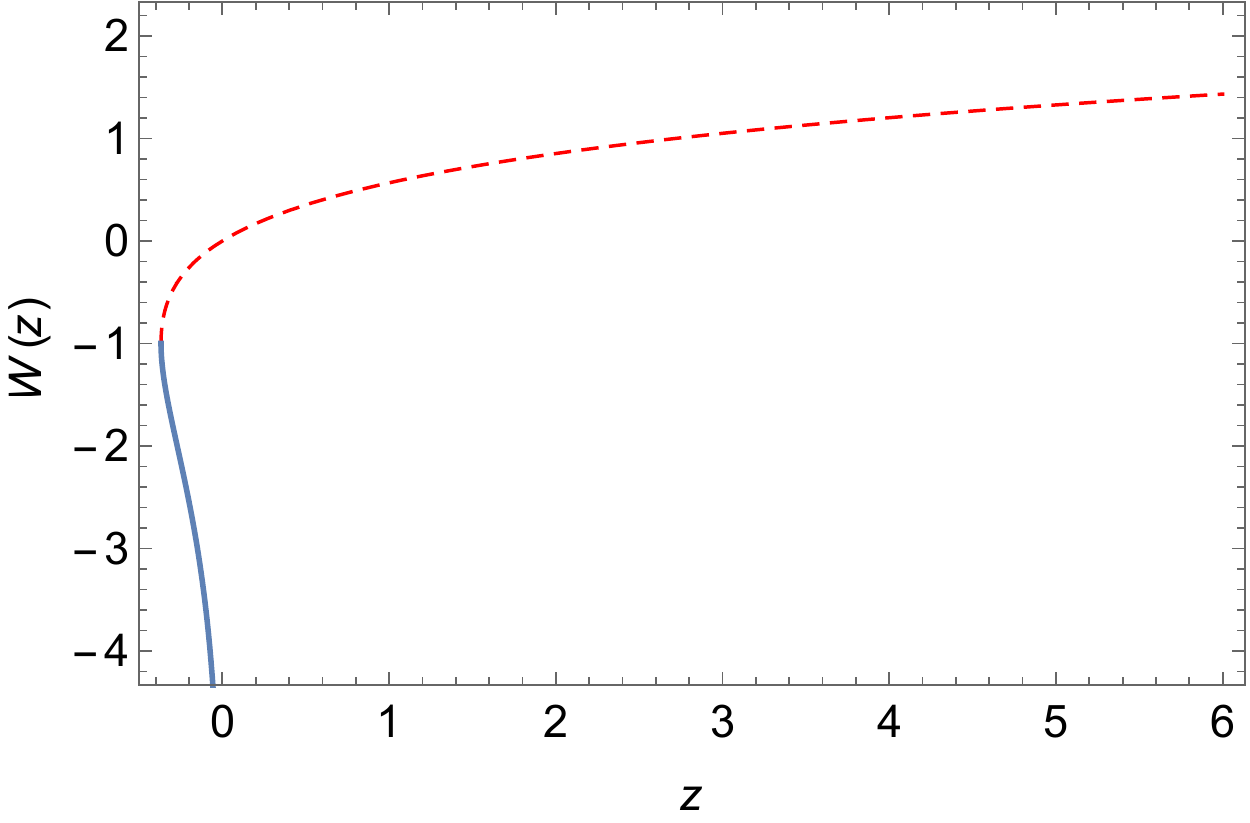}
\caption{The Lambert function $W$. The dashed red line corresponds to the branch $W_0$ whereas the blue line corresponds to the branch $W_{-1}$. }
\label{fig:Lambert}
\end{center}
\end{figure}
\vspace{-1cm}
\section{2. Definition and asymptotics of the deformed Airy function and kernel}
\subsection{2.1 Asymptotics of the Gamma function}
We first recall the asymptotics of the Gamma function to will be used to study the asymptotics of the deformed Airy function and kernel. We define $z=x+iy=\rho e^{i\theta}$. As $|z|\rightarrow \infty$ and $|\mathrm{arg}(z)|<\pi$, we have 
\begin{equation}
\Gamma(z)\sim \sqrt{2\pi} \rho^{x-1/2} e^{-\theta y}e^{-x}e^{-iy+i \theta (x-\frac{1}{2}) +iy \ln(\rho)} \label{asympt}
\end{equation}
with $ \theta=2\arctan(\frac{y}{x+\sqrt{x^2+y^2}})$ which is the natural extension of $\arctan(y/x)$ in case $\theta$ exits $[-\pi /2, \pi /2]$. We notice that $\Gamma(\bar{z})=\overline{\Gamma(z)}$ which yields $\frac{\Gamma(z)}{\Gamma(\bar{z})}=e^{2i \mathrm{Arg}(\Gamma(z))}$.

\subsection{2.2 Asymptotics of the deformed Airy function}

We are interested in the asymptotics of the deformed Airy function \eqref{aigamma} with the arguments of \eqref{KtAi}
which correspond to the case $w_+=w_-=w$ and $x=0$. We %choose to 
scale the arguments 
%of the deformed Airy functon 
so that all terms share the same scaling in time, allowing to apply the steepest descent method. Since the scale of the
first argument $t^{-1/3}$ is imposed so that the weight function in Eq. \eqref{sig} has an argument of order $\mathcal{O}(1)$, that leads to the rescaling of the drift $w=\tilde w t^{-1/2}$, as mentioned in the text, and to a rescaling of
the integration variable, i.e. we define $\tilde{\eta}=\eta t^{1/6}$. We then obtain, using the asymptotics \eqref{asympt}
\bea
\mathrm{Ai}_{\Gamma}^{\Gamma}(\tilde{a}t^{-1/3},t^{-1/3},\tilde{w}t^{-1/2},\tilde{w}t^{-1/2})
&=& \frac{1}{2 \pi t^{1/6}}\int\limits_{-\infty+i \epsilon}^{+\infty+i \epsilon}\mathrm{exp} \left(t^{-1/2}  (i \frac{\tilde \eta^3}{3}
+ i \tilde a \tilde \eta) \right)
\frac{\Gamma(t^{-1/2} (i \tilde \eta+\tilde w))}{\Gamma(t^{-1/2}( -i \tilde \eta+\tilde w))}  \mathrm{d} \tilde \eta \\
&\underset{t \ll 1}{\simeq}&\frac{1}{2\pi t^{1/6}}\int\limits_{-\infty+i\epsilon}^{+\infty+i\epsilon}\mathrm{exp}\left(2it^{-\frac{1}{2}}\phi(\tilde{\eta},\tilde a)-i\arctan \frac{\tilde \eta}{\tilde w}\right) d\tilde{\eta} 
\eea 
where $\epsilon< \tilde w$ and where we have defined (dropping the tilde on $\eta$ from now on for notational simplicity)
\be
\phi(\eta,\tilde a):=\frac{\eta^3}{6}+\frac{\eta}{2}\left(\tilde{a}-\ln t +\ln\tilde{w}^2 \right)+\tilde{w}\left(\arctan (\frac{\eta}{\tilde{w}})-\frac{\eta}{\tilde{w}}\right)+\frac{\eta}{2}\ln\left(1+\frac{\eta^2}{\tilde{w}^2}\right) \label{defphi}
\ee

Note that the explicit time dependent factor is harmless, as it 
can be absorbed by the redefinition $\tilde a  := \hat a  + \ln t -\ln \tilde w^2$, 
and $\hat a$ fixed as $t \to 0$, see below.
%\bea
%&=\frac{1}{2\pi}\int\limits_{-\infty+i\epsilon}^{+\infty+i\epsilon}\mathrm{exp}(\frac{i\eta^3}{3}+i\tilde{a}t^{-1/3}\eta+2i(-\eta t^{-1/3}+\arctan (\frac{t^{1/6}\eta}{\tilde{w}})(\tilde{w}t^{-1/2}-\frac{1}{2})+\frac{\eta t^{-1/3}}{2}\ln(\tilde{w}^2t^{-1}+\eta^2t^{-2/3})) d\eta 
%\eea
%We now define $\tilde{\eta}=\eta t^{1/6}$ and with this scaling all terms except one have the right time scaling, allowing us to apply a steepest descend method.
%\begin{equation}
%\begin{split}
%&\mathrm{Ai}_{\Gamma}^{\Gamma}(\tilde{a}t^{-1/3},t^{-1/3},\tilde{w}t^{-1/2},\tilde{w}t^{-1/2})=\frac{t^{-1/6}}{2\pi}\int\limits_{-\infty+i\epsilon}^{+\infty+i\epsilon}\mathrm{exp}(2it^{-\frac{1}{2}}\phi(\tilde{\eta})-i\arctan \frac{\tilde \eta}{\tilde w}) d\tilde{\eta} \\
%\end{split}
%\end{equation}
%with $
%\phi(\eta)=\frac{\eta^3}{6}+\frac{\tilde{a}\eta}{2}+(\tilde{w}\arctan (\frac{\eta}{\tilde{w}})-\eta)+\frac{\eta}{2}\ln((\tilde{w}^2+\eta^2)t^{-1})$. 
To apply the steepest descend method, we look for the zeros of the derivative of the phase, which are given by
\begin{equation}
\begin{split} \label{speq} 
&2\phi'(\eta_{\rm sp},\tilde a)=\eta_{\rm sp}^2+\tilde{a}-\ln t +\ln\tilde{w}^2+\ln(1+\frac{\eta_{\rm sp}^2}{\tilde{w}^2})=0 \quad \Rightarrow\quad  
\eta_{\rm sp} = \pm \eta_0 \quad , \quad \eta_0 = \sqrt{W(\tilde{w}^2e^{-\hat a+\tilde{w}^2})-\tilde{w}^2}
\end{split}
\end{equation}

where here and below primes denote derivatives w.r.t. the first argument,
and W is the Lambert W function (see Section 1.). For the case of a real $\tilde a$ studied here, 
the argument
of $W$ is positive hence one chooses the branch $W_0$. This leads to a pair of zeroes that are
real for $\hat a <0$, vanish at $\hat a=0$ and become imaginary for $\hat a>0$. The latter case
corresponds to fast decaying behavior which, as in \cite{le2016exact} we claim contributes
subdominantly in the calculation of the traces. Hence we focus on the case $\hat a<0$ 
which leads to oscillating behavior.\\

At the stationary points the phase and its second derivative w.r.t. $\eta$ are given by 
\begin{equation} 
\phi(\eta_{\rm sp},\tilde a)=-\frac{\eta_{\rm sp}^3}{3}-\eta_{\rm sp}+\tilde{w}\arctan(\frac{\eta_{\rm sp}}{\tilde{w}}) 
\qquad \phi''(\eta_{\rm sp},\tilde a)=\eta_{\rm sp}+\frac{\eta_{\rm sp}}{\tilde{w}^2+\eta_{\rm sp}^2}
\label{phase}
\end{equation}

We now expand the integral around the two saddle points and sum their contribution.
\begin{equation}
\label{saddlep1}
\begin{split}
2\pi t^{\frac{1}{6}} \mathrm{Ai}_{\Gamma}^{\Gamma}(\tilde{a}t^{-1/3},t^{-1/3},\tilde{w}t^{-1/2},\tilde{w}t^{-1/2})& \simeq \sum\limits_{\pm}\int\limits_{-\infty}^{+\infty}\exp \left( 2it^{-\frac{1}{2}} [\pm \phi(\eta_0,\tilde a)\pm \frac{1}{2}( \eta\mp \eta_0)^2 \phi''(\eta_0,\tilde a)]\mp i\arctan \frac{\eta_0}{\tilde w}\right)d  \eta  \\
&\simeq 2\sqrt{\frac{\pi  t^{\frac{1}{2}}}{{ \phi''(\eta_0,\tilde a)}}} \cos (2t^{-\frac{1}{2}} \phi(\eta_0,\tilde a)- \arctan \frac{ \eta_0}{\tilde w}+ \frac{\pi}{4})
\end{split}
\end{equation}
%Note that there is an implicit convergence factor $e^{- t^{-1/2} \epsilon \eta^2}$ with $\epsilon=0^+$ {\red tu es d'accord? tu peux le dire ou le faire plus correctement?}
Finally, combining \eqref{phase} and \eqref{saddlep1}, we obtain  
\begin{equation}
\mathrm{Ai}_{\Gamma}^{\Gamma}(\tilde{a}t^{-1/3},t^{-1/3},\tilde{w}t^{-1/2},\tilde{w}t^{-1/2})
\underset{t \ll 1}{\simeq} \frac{t^{\frac{1}{12}}}{\sqrt{\pi}}\sqrt{\frac{\tilde{w}^2+\eta_0^2}{\eta_0(1+\tilde{w}^2+\eta_0^2)}}\cos(\frac{\pi}{4}-\arctan \frac{ \eta_0}{\tilde w}-2t^{-\frac{1}{2}}(\frac{\eta_0^3}{3}+\eta_0-\tilde{w}\arctan(\frac{\eta_0}{\tilde{w}})))
\end{equation}
which, strictly speaking, is valid for $\tilde a = \hat a + \ln t -\ln \tilde w^2$ at fixed $\hat a<0$ and $\tilde w>0$,
where $\eta_0 = \sqrt{W(\tilde w^2 e^{- \hat a+\tilde{w}^2})-\tilde{w}^2}$. We have also tested 
this estimate numerically.

\subsection{2.3 Asymptotics of the deformed Airy kernel}

To calculate the deformed Airy kernel, we first rescale the arguments in exactly the same way as in the
previous calculation for the deformed Airy function. We obtain
\begin{equation}
\begin{split}
&K_{\rm Ai,\Gamma}(v=\frac{\tilde{v}}{t^{1/3}},v'=\frac{\tilde{v}'}{t^{1/3}})=\\
%&\frac{-1}{4\pi^2}\int\limits_{-i\infty+\epsilon}^{+i\infty+\epsilon} \int\limits_{-i\infty}^{+i\infty}\int\limits_0^{+\infty} \mathrm{exp}(\frac{\eta^3}{3}+\frac{x^3}{3}-v\eta-v'x-r(\eta+x))\frac{\Gamma(-t^{-\frac{1}{3}}\eta+\tilde{w}t^{-1/2})}{\Gamma(t^{-\frac{1}{3}}\eta+\tilde{w}t^{-1/2})}\frac{\Gamma(-t^{-\frac{1}{3}}x+\tilde{w}t^{-1/2})}{\Gamma(t^{-\frac{1}{3}}x+\tilde{w}t^{-1/2})} 
%d\eta dx dr \\
				   &=\frac{i}{4\pi^2}\int\limits_{-\infty+i\epsilon}^{+\infty+i\epsilon} \int\limits_{-\infty}^{+\infty}\frac{\mathrm{exp}(\frac{i\eta^3}{3}+i\frac{\eta'^3}{3}+iv\eta+iv'\eta')}{\eta+\eta'} \frac{\Gamma(it^{-\frac{1}{3}}\eta+\tilde{w}t^{-1/2})}{\Gamma(-it^{-\frac{1}{3}}\eta+\tilde{w}t^{-1/2})}\frac{\Gamma(it^{-\frac{1}{3}}\eta'+\tilde{w}t^{-1/2})}{\Gamma(-it^{-\frac{1}{3}}\eta'+\tilde{w}t^{-1/2})}d\eta d\eta' \\
				   &=\frac{it^{-\frac{1}{6}}}{4\pi^2}\int\limits_{-\infty+i\epsilon}^{+\infty+i\epsilon} \int\limits_{-\infty}^{+\infty}\frac{\mathrm{exp}(2it^{-\frac{1}{2}}(\phi(\eta,\tilde{v})+\phi( \eta',\tilde{v}'))-i\arctan \frac{ \eta}{\tilde w}-i\arctan \frac{ \eta'}{\tilde w})}{\eta+ \eta'}d{\eta} d\eta'\\
\end{split}
\end{equation}
where in the last line we have redefined $\eta \to \eta t^{-1/6}$, $\eta' \to \eta' t^{-1/6}$, and
used again the asymptotics \eqref{asympt} of the Gamma function. The function 
$\phi(\eta,\tilde{v})$ is defined in \eqref{defphi}. 
% this leads to the following expression for the kernel\\
%\begin{equation}
%\begin{split}
%&K_{\rm Ai,\Gamma}(u=\frac{\tilde{u}}{t^{1/3}},u'=\frac{\tilde{u}'}{t^{1/3}})=\frac{it^{-\frac{1}{6}}}{4\pi^2}\int\limits_{-\infty+i\epsilon}^{+\infty+i\epsilon} \int\limits_{-\infty}^{+\infty}\frac{\mathrm{exp}(2it^{-\frac{1}{2}}(\phi(\tilde{x},\tilde{u}')+\phi(\tilde{\eta},\tilde{u}))-i\arctan \frac{\tilde \eta}{\tilde w}-i\arctan \frac{\tilde x}{\tilde w})}{\tilde{\eta}+\tilde{x}}d\tilde{z} d\tilde{x}\\
%\end{split}
%\end{equation}
%with $
%\phi(\eta, \tilde{u})=\frac{\eta^3}{6}+\frac{\tilde{a}\eta}{2}+(\tilde{w}\arctan (\frac{\eta}{\tilde{w}})-\eta)+\frac{\eta}{2}\ln((\tilde{w}^2+\eta^2)t^{-1})
%$. 
Applying the steepest descent on $\eta$ and $\eta'$, as in \eqref{speq} in the previous section, we
obtain the saddle points 
%by finding the zeros of the derivatives of the phases as follows
\begin{equation}
\eta_{\rm sp}= \pm \eta_0 \quad , \quad \eta_0=\eta_0(\tilde v)=\sqrt{W(\tilde w^2 e^{-\hat v +\tilde{w}^2})-\tilde{w}^2} \quad , \quad
\eta'_{\rm sp}= \pm \eta'_0 \quad , \quad \eta'_0=\eta_0(\tilde v')=\sqrt{W(\tilde{w}^2e^{-\hat v'+\tilde{w}^2})-\tilde{w}^2} 
\end{equation}
where W is the Lambert W function. Here we choose the branch $W_0$ of the Lambert function which
is the only one leading to a real saddle point. We again defined $\tilde v = \hat v + \ln t -\ln \tilde w^2$
and $\tilde v' = \hat v' + \ln t -\ln \tilde w^2$ and study the case $\hat v, \hat v'<0$ where the above saddle points
are real. We now expand around the four saddle points and sum their contribution.\\
\begin{equation}
\begin{split}
K_{\rm Ai,\Gamma}(v,v') &\simeq \sum\limits_{\pm_1}\sum\limits_{\pm_2} \frac{it^{\frac{1}{3}}}{4\pi}\sqrt{\frac{ 1}{{ \phi''(\eta'_0,\tilde{v}') \phi''(\eta_0,\tilde{v})}}}\\
								&\frac{\exp(\pm_2 2i t^{-\frac{1}{2}} \phi(\eta'_0,\tilde{v}') \mp_2i \arctan \frac{\eta'_0}{\tilde w} \pm_2 i \pi /4\pm_1 2it^{-\frac{1}{2}} \phi(\eta_0,\tilde{v}) \mp_1 i\arctan \frac{\eta  _0}{\tilde w}\pm_1 i \pi /4)} {\pm_2 \eta'_0\pm_1 \eta_0}
\end{split}
\end{equation}
We are left with four terms and we drop the terms with same sign as they decay too quickly and are therefore subdominant, leading to

\begin{equation}
\begin{split} \label{KK1} 
K_{\rm Ai,\Gamma}(v,v')	&\simeq \frac{-t^{\frac{1}{3}}}{2\pi}\sqrt{\frac{ 1}{ \phi''(\eta'_0,\tilde{v}')\phi''(\eta_0,\tilde{v})}}\frac{\sin( 2t^{-\frac{1}{2}} [\phi(\eta'_0,\tilde{v}')-\phi(\eta_0,\tilde{v})]+\arctan \frac{ \eta_0}{\tilde w}-\arctan \frac{\eta'_0}{\tilde w})} { \eta'_0-\eta_0}\\
\end{split}
\end{equation}
where from \eqref{phase}, $\phi(\eta_0,\tilde{v})=-\frac{\eta^3_0}{3}-\eta_0+\tilde{w}\arctan(\frac{\eta_0}{\tilde{w}}) $ and $\phi''(\eta_0,\tilde v)=\eta_0+\frac{\eta_0}{\tilde{w}^2+\eta_0^2}$ and similarly for $\eta'_0$ and $\tilde u'$.
We now introduce $\tilde \kappa$ such that $\tilde{v}'=\tilde{v}+\tilde \kappa$ and study the limit $\tilde \kappa \rightarrow 0$. Taking the derivative w.r.t. $\tilde v$ of $\phi'(\eta_0(\tilde v),\tilde v)=0$ gives
$\frac{d\eta_0}{d\tilde v}=- \eta_0/(2 \phi''(\eta_0,\tilde v))$: approximating $\eta'_0-\eta_0=\tilde \kappa \frac{d\eta_0}{d\tilde v}+ \mathcal{O}(\tilde \kappa^2)$
we see that to leading order in $\tilde \kappa$ in \eqref{KK1} the denominator cancels the second derivatives in the 
square root. Next, since $\frac{d}{d\tilde v} \phi(\eta_0(\tilde v),\tilde v) =\partial_{\tilde v} \phi(\eta_0,\tilde v)$ from the saddle point condition, one has
$\phi(\eta'_0,\tilde{v}')=\phi(\eta_0,\tilde{v})+\frac{\tilde \kappa\eta_0}{2}+\mathcal{O}(\tilde \kappa^2)$. Finally we use
$\arctan \frac{\eta'_0}{\tilde{w}}=\arctan\frac{\eta_0}{\tilde{w}}-\frac{\tilde \kappa \tilde{w}}{2\eta_0(1+\eta_0^2+\tilde{w}^2)}+\mathcal{O}(\tilde \kappa^2)$ and obtain

%\begin{equation}
%\begin{cases}
%\eta'_0= \sqrt{W(te^{-(\tilde{u}'+\kappa)+\tilde{w}^2})-\tilde{w}^2}=\eta_0-\frac{\kappa(\eta_0^2+\tilde{w}^2)}{2\eta_0(1+\eta_0^2+\tilde{w}^2)}+\mathcal{O}(\kappa^2)
%\\
%\phi(\eta'_0,\tilde{u}')=\phi(\eta_0,\tilde{u})+\frac{\kappa\eta_0}{2}+\mathcal{O}(\kappa^2)
%\\
%\arctan	\frac{\eta'_0}{\tilde{w}}=\arctan\frac{\eta_0}{\tilde{w}}-\frac{\kappa \tilde{w}}{2\eta_0(1+\eta_0^2+\tilde{w}^2)}+\mathcal{O}(\kappa^2)
%\end{cases}
%\end{equation}
\begin{equation}
\begin{split}
K_{\rm Ai,\Gamma}(v,v')				&\simeq \frac{t^{\frac{1}{3}}}{\pi}\frac{\sin( t^{-\frac{1}{2}} \tilde \kappa \eta_0+\frac{\tilde \kappa \tilde{w}}{	2\eta_0(1+\eta_0^2+\tilde{w}^2)})} { \tilde \kappa}\\
\end{split}
\end{equation}
We finally define $\kappa=\tilde \kappa t^{-\frac{1}{2}}$ and drop the second term in the sine which is subdominant.
We obtain
\begin{equation}
\begin{split} \label{asympt4} 
K_{\rm Ai,\Gamma}(v=\frac{\tilde{v}}{t^{1/3}},v'=\frac{\tilde{v}+\kappa t^{\frac{1}{2}}}{t^{1/3}})&\simeq \frac{t^{-\frac{1}{6}}}{\pi}\frac{\sin\left( \kappa \sqrt{W(\tilde{w}^2e^{-\hat v+\tilde{w}^2})-\tilde{w}^2}\right)} { \kappa}\\
\end{split}
\end{equation}
which is valid in the limit $t \ll 1$, provided we define $\tilde v = \hat v + \ln t -\ln \tilde w^2$,
and keep $\hat v<0$ and $\tilde w>0$ fixed in the limit. This leads to \eqref{estimate} 
in the text, with the branch $W_0$. Note that the asymptotics \eqref{asympt4} involves only the
value of the saddle point $\eta_0$, suggesting a more general asymptotic formula for kernels of a
similar type.

\section{3. Short time estimate of the Fredholm determinant $Q_t(s)$}
 
\subsection{3.1. Derivation of the function $\Psi(z)$} 

We start by deriving the formula for $Q_t(s)$ given in Eq. (\ref{resQ}) in the Letter. The derivation
follows very closely the one of Ref. \cite{le2016exact}. 
From Eqs. (\ref{FD1}) and (\ref{sump}) given in the Letter, one has
\begin{eqnarray}\label{Q_start_supp}
\ln Q_t(s) = - \sum_{p=1}^\infty \frac{1}{p} {\rm Tr}\, \bar{K}^p_{t,s} \;, \; \bar K_{t,s}(v,v') = K_{\rm Ai,\Gamma}(v,v') \sigma_{t,s}(v')
\end{eqnarray}
where $K_{\rm Ai,\Gamma}(v,v')$, the Airy kernel, and $\sigma_{t,s}$ are given in Eqs. \eqref{KtAi} and \eqref{sig} of the Letter (respectively). 
Hence one has
\bea
&& {\rm Tr} \; {\bar K}_{t,s}^p = \int_{-\infty}^\infty dv_1 \int_{-\infty}^\infty dv_2 \ldots \int_{-\infty}^\infty dv_p K_{\rm Ai,\Gamma}(v_1,v_2) .. K_{\rm Ai,\Gamma}(v_p,v_1)  \sigma_{t,s}(v_1) \ldots  \sigma_{t,s}(v_p)
\eea
The expression of $\sigma_{t,s}(v) = \sigma(t^{1/3}(v-s))$ suggests to perform the change of variable $v_i \to v_i / t^{1/3}$, which yields (setting $\tilde s = s t^{1/3}$):
 \bea\label{trace_Kp}
{\rm Tr} \; {\bar K}_{t,s}^p &=& t^{-p/3} \int_{-\infty}^\infty dv_1 \int_{-\infty}^\infty dv_2 \ldots \int_{-\infty}^\infty dv_p \, K_{\rm Ai,\Gamma}\left(\frac{v_1}{t^{1/3}} ,\frac{v_2}{t^{1/3}} \right) \ldots K_{\rm Ai,\Gamma}\left(\frac{v_p}{t^{1/3}} ,\frac{v_1}{t^{1/3}} \right)  
\sigma(v_1-\tilde s) \ldots \sigma(v_p-\tilde s) \\
\sigma(v) &=& \frac{1}{e^{- v } + 1} \;. \nonumber
\eea 
Let us now recall the representation of the deformed Airy kernel for the case $x=0$ and $w_+=w_-=w$
\be 
  K_{\rm Ai, \Gamma}(v,v') := \int_{0}^{+\infty} dr \, \mathrm{Ai}_\Gamma^\Gamma(r+v,t^{-\frac{1}{3}},w,w)\mathrm{Ai}_\Gamma^\Gamma(r+v',t^{-\frac{1}{3}},w,w) 
 \ee
Recalling the short time asymptotics \eqref{asympt4} of $K_{\rm Ai, \Gamma}$ we get
\bea
K_{\rm Ai,\Gamma}\left(\frac{v}{t^{1/3}} ,\frac{v + t^{1/2} \kappa}{t^{1/3}}\right)  \underset{t \ll 1}{\simeq} 
\frac{ 1}{\pi t^{1/6}} \frac{\sin ( \kappa f(\hat v))}{\kappa} \,, 
\eea 
where $f(\hat v)=\sqrt{W_0(\tilde{w}^2e^{-\hat v+\tilde{w}^2})-\tilde{w}^2}$, $v=\hat v +\ln t-\ln \tilde{w}^2$ and $W_0$ is the first real branch of the Lambert $W$ function (here we drop the subscript $\tilde w$ on $f$ as compared to the text). In particular, we define $v_j=\hat v_j +\ln t-\ln \tilde{w}^2$. We may now use the asymptotics of the deformed Airy kernel for $t\rightarrow 0$ and $\hat v_j<0$ such that $W_0(\tilde{w}^2 e^{-\hat v_j+\tilde{w}^2})-\tilde{w}^2>0$, otherwise the Kernel vanishes exponentially. Hence for $p \geq 2$, 
separating the center of mass coordinate (which we take as $v_1$) 
and the $p-1$ relative coordinates $v_j=v_{j-1}+ t^{1/2} \kappa_j$ 
we obtain
\begin{equation}
\begin{split}
&{\rm Tr} \bar{K}^p \simeq  t^{-p/3}   \int_{-\infty}^{-\ln(\tilde{w}^2t^{-1})} dv_1 \left(\frac{1}{\pi t^{1/6}}\right)^p [\sigma(v_1-\tilde s)]^p 
t^{(p-1)/2}\\
& \times \int_{-\infty}^\infty d\kappa_1  \ldots \int_{-\infty}^\infty d\kappa_p  \frac{\sin ( f(\hat v_1) \kappa_1)}{\kappa_1} \frac{\sin (f(\hat v_1)\kappa_2)}{\kappa_2} \ldots \frac{\sin ( f(\hat v_1)\kappa_p)}{\kappa_p}
\delta(\kappa_1+\kappa_2+\dots+\kappa_p) \nonumber
\\
& = \frac{1}{\pi^p \sqrt{t} }  \int_{-\infty}^{-\ln(\tilde{w}^2t^{-1})} dv_1 f(\hat v_1)[\sigma(v_1-\tilde s)]^p I_p \;, \; I_p = 
\int_{-\infty}^\infty d\kappa_1  \ldots \int_{-\infty}^\infty d\kappa_p  \frac{\sin \kappa_1}{\kappa_1} \frac{\sin \kappa_2}{\kappa_2} .. \frac{\sin \kappa_p}{\kappa_p} \delta(\kappa_1+..+\kappa_p) = \pi^{p-1} \\ 
\end{split}
\end{equation}

%The multiple integral defining $I_p$ can be computed explicitly, using $\sin{x}/x = (1/2) \int_{-1}^1 e^{i k x} dk$ and an integral representation of the delta function to obtain
%\begin{equation}
%I_p= \frac{1}{2^p} \int_{-1}^1 dx_1 \ldots   \int_{-1}^1 dx_p \int_{-\infty}^\infty \frac{dk}{2 \pi} 
%\int_{-\infty}^\infty d\kappa_1 \ldots \int_{-\infty}^\infty d\kappa_p e^{i \sum_{j=1}^p (x_j + k) \kappa_j } = \frac{1}{2^p} (2 \pi)^p \int_{-1}^1  \frac{dk}{2 \pi}  =
%\pi^{p-1}
%\end{equation}
Combining the different results, and recaling that $v=\hat v +\ln t-\ln \tilde{w}^2$,we obtain 
\begin{equation}
\begin{split}
& \ln Q_t(s) \approx -\frac{1}{\sqrt{t}} \Psi(t e^{-\tilde s}) \\ 
& \Psi(z) = \frac{1}{\pi} \sum_{p=1}^\infty \frac{1}{p}  \int_{-\infty}^{0} d\hat v f(\hat v) \frac{1}{(e^{- \hat v }  \tilde w^2 z^{-1}+ 1)^p} \\
\end{split}
\end{equation}
 It is then straightforward to perform the sum over $p$ for $z> - \tilde w^2$, and upon the change $\hat v \to - \hat v$ we obtain
 \begin{equation}
\begin{split}
\Psi(z) &=  \frac{1}{\pi} \int_{0}^{+\infty} d \hat v \sqrt{W_0(\tilde w^2 e^{\hat v+\tilde{w}^2})-\tilde{w}^2} \, \ln\left(1 + \frac{z}{\tilde w^2} e^{-\hat v}\right) \\
\end{split}
\end{equation}
Performing the change of variable $y=W_0(\tilde{w}^2 e^{\hat v+\tilde{w}^2})-\tilde{w}^2$, and using the definition and properties of the Lambert function and its derivative \eqref{derW} we obtain an equivalent
formula
\begin{equation}
\begin{split}
&\Psi(z) =  \frac{1}{\pi} \int_{0}^{+\infty} dy \left[1+\frac{1}{y+\tilde{w}^2}\right]\sqrt{y} \ln\left(1 + \frac{ze^{-y}}{y+\tilde{w}^2} \right)
\end{split} \label{defPsi} 
\end{equation}
leading to \eqref{resQ} in the main text. Note the expression for the derivatives: for $q \geq 1$
\begin{equation}
\begin{split} \label{Psiq} 
\Psi^{(q)}(z) &= (q-1)! (-1)^{q+1} \frac{1}{\pi} \int_{0}^{+\infty} dy (1+\frac{1}{y+\tilde{w}^2})\sqrt{y}\frac{1}{((y+\tilde{w}^2)e^{y}+z)^q} \\
\end{split}
\end{equation}

\subsection{3.2. The function $\Psi(z)$ for the stationary case $\tilde w=0^+$} 

It is useful to study in details the function $\Psi(z)$ for $\tilde w=0$. We now show
that it is non analytic in $z$, but for $z>0$ it can be expanded in a power series in $u=\sqrt{z}$ as
follows
\bea \label{psitaylor}
\Psi(z) = \psi(u= \sqrt{z}) \quad , \quad \psi(u) = \sum_{n \geq 1} \frac{u^n}{n!} \psi^{(n)}(0) \quad , \quad 
\psi^{(n)}(0) = (-1)^{n-1}  \frac{2^{n-1}}{\sqrt{\pi}} \Gamma\left(\frac{n}{2}\right) \left(\frac{n}{2}\right)^{\frac{n-3}{2}}
\eea 
i.e. $\psi(u)$ can be Taylor expanded for $u>0$. To calculate these
derivatives, one can start from the expression \eqref{Psiq} for $q=1$ setting $y=x^2$ 
\bea
&& \psi'(u) = 2 u \Psi'(z=u^2) = \frac{2}{\pi} \int_{-\infty}^{+\infty} dx\; (1+ x^2) \frac{u}{u^2 + x^2 e^{x^2}} \\
&& = \frac{2}{\pi} \int_{-\infty}^{+\infty} dy \;h(y) \frac{u}{u^2 + y^2}  \quad , \quad h(y)=\frac{\sqrt{W(y^2)}}{|y|} 
\eea 
Using that $\frac{u}{\pi(u^2 + y^2)} \underset{u \to 0^+}{\simeq} \delta(y) + .. $ we obtain $\psi'(0)=2$. 
To obtain the higher derivatives we note the formula, for any integer $q \geq 1$
%\be
%\partial_u \frac{u}{u^2 + y^2} = - \partial_y \frac{y}{u^2 + y^2} \quad , \quad 
%\partial_u \frac{y}{u^2 + y^2} =  \partial_y \frac{u}{u^2 + y^2}
%\ee
%hence
\be
 \partial_u^{2 q} \frac{u}{u^2 + y^2} = (-1)^q \; \partial_y^{2 q} \frac{u}{u^2 + y^2} \quad , \quad \partial_u^{2 q-1} \frac{u}{u^2 + y^2} = (-1)^q \;  \partial_y^{2 q-1} \frac{y}{u^2 + y^2} 
\ee
The odd derivatives are obtained using integration by parts
\bea
\psi^{(2 q+1)}(0) =  (-1)^q \frac{2}{\pi} \lim_{u \to 0^+} \int_{-\infty}^{+\infty} dy\;  h(y) \; \partial_y^{2 q} \frac{u}{u^2 + y^2}
= 2 (-1)^q h^{(2 q)}(0) = \frac{(2 q)!}{q!} (q + \frac{1}{2})^{q-1} 
\eea 
where we have used that $h(y)=\sqrt{W(y^2)/y^2}= 
\frac{1}{2} \sum_{n=0}^{+\infty} \frac{(n+ \frac{1}{2})^{n-1}}{n!} (-y^2)^n$. The even derivatives, after
integration by part
are given by
\bea
\psi^{(2 q)}(0) =  
(-1)^{q+1} \frac{2}{\pi} \int_{-\infty}^{+\infty} \frac{dy}{y} h^{(2 q-1)}(y)
\eea
One can further perform integrations by parts, noting that 
\bea
 \psi^{(2 q)}(0) \; &&= (-1)^{q+1} \frac{4 (2q-2)!}{\pi} \int_{0}^{+\infty} dy  \; h_{\rm reg}'(y) \frac{1}{y^{2q-1}} =  (-1)^{q+1} \frac{4 (2q-2)!}{\pi} \left[ \int_0^1 dx \frac{x^{2q-1}}{\left[-\ln(x^2)]^{q-\frac{1}{2}} \right]}\right]_{\rm reg} \\
&& = \frac{2 (-1)^q q^{q-\frac{3}{2}} (2 q-2)! \Gamma\left(\frac{3}{2}-q\right)}{\pi }
\eea
where for $q=1$ we use the change of variable $x=h(y)$ and note that $y=h^{-1}(x)=\sqrt{- \ln(x^2)/x^2}$ with $h(0)=1$ and
$h(+\infty)=0$. This leads to a convergent integral. For $q \geq 2$ we define $h_{\rm reg}(y)=h(y)-
\sum_{n=1}^{2q-2} h^{(n)}(0) y^n/n!$ which leads to the "regularized version" of the (divergent) integral, given by
analytic continuation.
We have checked the correctness of the final formula. Putting all together we obtain the result given in \eqref{psitaylor}.
%\bea
%\psi^{(n)}(0) = (-1)^{n-1}  \frac{2^{n-1}}{\sqrt{\pi}} \Gamma(\frac{n}{2}) (\frac{n}{2})^{\frac{n-3}{2}}
%\eea 

\section{4. Calculation of $\Phi(H)$ for $H \in ]-\infty, H_c(\tilde w)]$}

\subsection{4.1. Saddle point equations} 

Defining $z=t e^{-\tilde s}$, we start from Eq. (\ref{genfunct}) of the text
which takes the following form at small time
\bea  \label{gener2} 
\bigg \langle \exp\left( - \frac{z}{t} e^{\tilde H}  \right) \bigg \rangle \sim e^{ -  \frac{1}{\sqrt{ t}} \Psi(z) } .
\eea 
Recalling that $\tilde H= H+ \chi$, where $\chi$ is a random variable independent from $H$, 
the difficulty is now to extract the leading small time behavior of the cumulants of $H$, equivalently 
the function $\Phi(H)$. One route is to observe that from \eqref{gener2} one easily obtains the
{\it cumulants} of $\tilde Z=e^{\tilde H}$ from the derivatives of the known function $\Psi(z)$ as
$\langle \tilde Z^q \rangle^c = (-1)^{q+1} \Psi^{(q)}(0) t^{q- \frac{1}{2}} + o(t^{q- \frac{1}{2}})$.
In principle, to obtain the cumulants of $Z$ we can now use relations between the {\it moments} of $Z=e^H$ and of $\tilde Z$, i.e. 
$\langle Z^q \rangle = \langle \tilde Z^q \rangle/ \langle e^{q \chi} \rangle =
(2 w-q)_q \langle \tilde Z^q \rangle$, where $(x)_q=x (x+1) \cdots (x+q-1)=\Gamma(x+q)/\Gamma(x)$ is the
Pochhammer symbol. We have performed that exercise up to $q=3$. We checked that indeed the
leading small time behavior of $\langle Z^q \rangle$ and then of $\langle H^q \rangle$,
could be extracted in this manner, and that it agrees the small time expansion of
the KPZ equation (see Section 10.). We have then verified that the limit $\tilde w \to 0$ produces
the correct cumulants for $w=0$ (which is far from a priori obvious in the intermediate steps of the calculation). \\
~\\
A more powerful method, which as we checked reproduces these results and allows
to obtain directly the function $\Phi(H)$ is as follows. We consider the leading behavior for fixed $\tilde w$, which implies that $w=\tilde w/\sqrt{t}$ 
is large. We define $\chi'=\chi -\ln (\sqrt{t})$, use Stirling's formula for the $\Gamma(2w)$ factor in the PDF of $\chi$ 
given in the text and Section 0, and write
\bea \label{rep1} 
e^{-\frac{1}{\sqrt{t}}\Psi(z)} \sim
\bigg\langle \exp \left( - \frac{z e^H e^\chi}{t} \right) \bigg\rangle \simeq
\int_{-\infty}^{+\infty} d\chi' \bigg\langle\exp \left( - \frac{2 \tilde w \chi' + e^{-\chi'} + z e^H e^{\chi'}-2\tilde{w}+2\tilde{w}\ln(2\tilde{w})}{\sqrt{t}} \right)\bigg\rangle
\eea 
where here the second bracket denotes average only on the KPZ noise and initial condition. 

We now define $R(z)$ to be the cumulant generating function of $e^H$, 
\bea \label{defR} 
\bigg\langle \exp \left( - \frac{X e^H}{\sqrt{t}} \right) \bigg\rangle \sim e^{-\frac{1}{\sqrt{t}} R(X)} 
\eea 
In \eqref{rep1} using $1/\sqrt{t}$ as a large parameter we perform a saddle point and obtain the
following relation between the functions $R(X)$ and $\Psi(z)$
\bea
\Psi(z) = \min_{\chi'} ( 2 \tilde w \chi' + e^{-\chi'} + R( z e^{\chi'}) ) - 2 \tilde w + 2 \tilde w \ln(2 \tilde w)
\eea 

~\\ Thus we have, with $X=z e^{\chi'}$,
\be
\Psi(z) = \min_{X} ( 2 \tilde w \ln(X/z) + \frac{z}{X} + R(X) ) - 2 \tilde w + 2 \tilde w \ln(2 \tilde w) 
\ee
which we invert as
\be
R(X) = \max_z ( \Psi(z) - 2 \tilde w \ln(X/z) -  \frac{z}{X} ) + 2 \tilde w - 2 \tilde w \ln(2 \tilde w) 
\ee
On the other hand, by substituting the anticipated form, $P(H,t) \sim 
e^{-\frac{1}{\sqrt{t}}\, \Phi(H)}$ as $t\to 0$ we have
\be \label{LegendreR} 
R(X) = \min_H ( \Phi(H) + X e^H)   \quad , \quad \Phi(H) = \max_X (R(X) - X e^H) 
\ee
hence 
\be
\Phi(H) = \max_{z,X} ( \Psi(z) - 2 \tilde w \ln(X/z) -  \frac{z}{X} - X e^H ) + 2 \tilde w - 2 \tilde w \ln(2 \tilde w) 
\ee 
we can perform the saddle point on the variable $X$ 
\be
X = e^{-H} ( \pm \sqrt{ \tilde w^2 + z e^H} - \tilde w )
\label{saddlechi}
\ee 
By consistency with the droplet case we must take the positive root. Indeed for $\tilde w \to +\infty$
\eqref{saddlechi} gives $X \simeq z/(2 \tilde w)$ and since $X=ze^{\chi'}= z e^{\chi}/\sqrt{t}=z/(2 \tilde w)$ this is consistent with the fact that for large $w$, $\chi$ becomes a deterministic variable
equal to $- \ln(2 w)$ (see Section 9.2). Taking the positive root we obtain the expression of the rate function in terms of the solution of a maximization problem
\bea
\Phi(H) = \max_{z \in [-\tilde w^2,+\infty[} \left( \Psi(z) - 2 \sqrt{ \tilde w^2 + z e^H}  + 2 \tilde w - 2 \tilde w \ln(2 \tilde w) 
+ 2 \tilde w \ln(\tilde w + \sqrt{ \tilde w^2 + z e^H}) \right) \label{Phi200} 
\eea 
From the definition of $z$ the maximization was to be done for $z\geq 0$, yet observing the domain of definition of $\Psi(z)$ and the square root, we actually have weaker constraints
\begin{equation}
\begin{cases}
z\geq-\tilde{w}^2\\
ze^H\geq -\tilde{w}^2 \label{condition2} 
\end{cases}
\end{equation}
As we will show the second constraint is always verified, and we thus have defined in 
\eqref{Phi200}
the the range of optimization by the first constraint.\\

The maximization problem \label{Phi20} is equivalent to the parametric system of equations given in the text
\begin{equation}
\label{parametric}
\begin{cases}
                 e^{H}= z\Psi'(z)^2+2\tilde{w}\Psi'(z)\equiv G(z)\\
                \Phi'(H)=-z\Psi'(z)\\
            \end{cases}
\end{equation}
For completeness, we also have the following parametric relation :  $\Phi'(H)=\tilde{w}-\sqrt{\tilde{w}^2+z e^H}$
(see below however for a modification of this relation in some range of values of $H$).

\subsection{4.2 Analysis of the saddle point equations} 

Now that we solved the optimization problem exactly, we wish to know if it allows us to obtain all values of $H \in \mathbb{R}$. We know that the optimization has to be done in the interval $z\in [-\tilde{w}^2,+\infty[$ so we first investigate the behavior of $\Psi(z)$ and of $G(z)$ on these boundaries, and then use the monotonicity of $G(z)$ to extrapolate the range of $H$.
\vspace{-0.5cm}
\subsubsection{4.2.1 behavior of $\Psi(z)$ for $z\rightarrow +\infty$}

We recall the definition \eqref{defPsi} of $\Psi$ for $z\geq -\tilde{w}^2$ and look for its asymptotics for large positive $z$
and fixed $\tilde w$. After an integration by part we obtain
\begin{equation}
\begin{split} \label{Psinew} 
\Psi(z) &= % \frac{1}{\pi} \int_{0}^{+\infty} dy (1+\frac{1}{y+\tilde{w}^2})\sqrt{y} \ln\left(1 + \frac{ze^{-y}}{y+\tilde{w}^2} \right)\\
%&=
\frac{1}{\pi}\int_{0}^{+\infty} dy(\frac{2}{3}y^{\frac{3}{2}}+2y^{\frac{1}{2}}-2\tilde{w}\arctan{\frac{\sqrt{y}}{\tilde{w}}})(1+\frac{1}{y+\tilde{w}^2})\frac{z}{(y+\tilde{w}^2)e^y+z} \\
%&= \frac{1}{\pi}\int_{\tilde{w}^2}^{+\infty} \mathrm{d}b\left(\frac{2}{3}[W_0(be^{\tilde{w}^2})-\tilde{w}^2]^{\frac{3}{2}}+2[W_0(be^{\tilde{w}^2})-\tilde{w}^2]^{\frac{1}{2}}-2\tilde{w}\arctan({\frac{[W_0(be^{\tilde{w}^2})-\tilde{w}^2]^{\frac{1}{2}}}{\tilde{w}}})\right)\frac{z}{b}\frac{1}{b+z} \\
\end{split}
\end{equation}
In the limit of large $z$ one can show that the fraction $\frac{z}{(y+\tilde{w}^2)e^y+z}$ can be replaced
by either one or zero depending which term in the denominator is larger, the change occurring
for $(y+\tilde{w}^2)e^y=z$ which is equivalent to $y=W_0(z e^{\tilde w^2} ) - \tilde w^2$
(similarly to the computation of the asymptotics of the polylogarithm function \cite{wood1992}), leading to
\bea
&& \Psi(z) \simeq % \frac{1}{\pi} \int_{0}^{+\infty} dy (1+\frac{1}{y+\tilde{w}^2})\sqrt{y} \ln\left(1 + \frac{ze^{-y}}{y+\tilde{w}^2} \right)\\
%&=
\frac{1}{\pi}\int_{0}^{W_0(z e^{\tilde w^2} ) - \tilde w^2} dy(\frac{2}{3}y^{\frac{3}{2}}+2y^{\frac{1}{2}}-2\tilde{w}\arctan{\frac{\sqrt{y}}{\tilde{w}}})(1+\frac{1}{y+\tilde{w}^2}) 
\eea
For a fixed $\tilde w$ one can further neglect the $\arctan$ term in the integrand, which leads to
%where from \eqref{defPsi} to the first line we did an integration by part, and we did the change of variable $y=W_0(be^{\tilde{w}^2})-\tilde{w}^2$ between the first and the second line. Similarly to the computation of the asymptotics of the polylogarithm function \cite{wood1992}, we find the leading term of the expansion for large positive $z$
%\begin{equation}
%\Psi(z)\underset{z\rightarrow +\infty}{\sim}  \frac{1}{\pi}\int_{\tilde{w}^2}^{z} \frac{\mathrm{d}b}{b}\left(\frac{2}{3}[W_0(be^{\tilde{w}^2})-\tilde{w}^2]^{\frac{3}{2}}+2[W_0(be^{\tilde{w}^2})-\tilde{w}^2]^{\frac{1}{2}}-2\tilde{w}\arctan({\frac{[W_0(be^{\tilde{w}^2})-\tilde{w}^2]^{\frac{1}{2}}}{\tilde{w}}})\right)
%\end{equation}
%By integration and using the logarithmic asymptotic of $W_0$, we obtain the asymptotics of $\Psi$ which is valid for all finite $\tilde{w}$.
\begin{equation}
\Psi(z)\underset{z\rightarrow +\infty}{\simeq}  \frac{ 1}{\pi}\left( \frac{4}{15}[W_0(z e^{\tilde w^2} ) - \tilde w^2]^{5/2}+\frac{16}{9}[W_0(z e^{\tilde w^2} ) - \tilde w^2]^{3/2} \right)
\end{equation}
Recalling that $W_0(z)\simeq_{z\to +\infty} \ln(z)-\ln \ln (z)$, and expanding at large $z$ and fixed $\tilde w$ we finally find 
\begin{equation} \label{asymptPsi} 
\Psi(z)\underset{z\rightarrow +\infty}{\simeq} \frac{4}{15 \pi} [\ln (z)]^{5/2} - \frac{2}{3 \pi} [\ln (z)]^{3/2} \ln \ln (z) + \frac{16}{9 \pi} [\ln (z)]^{3/2}
\end{equation}
\vspace{-0.5cm}
\subsubsection{4.2.2 Behavior of $\Psi'(z)$ at $z\rightarrow -\tilde{w}^2$}

From \eqref{defPsi}, the expression of $\Psi'(z)$ at $z=-\tilde{w}^2$ is given by
\begin{equation}
\begin{split}
\Psi'(-\tilde{w}^2) &=  \frac{1}{\pi} \int_{0}^{+\infty} dy (1+\frac{1}{y+\tilde{w}^2})\sqrt{y}\frac{1}{(y+\tilde{w}^2)e^{y}-\tilde{w}^2} \\
\end{split}
\end{equation}
In the small $\tilde{w}$ limit, this integral can be computed and behaves as 
\begin{equation}
\Psi'(-\tilde{w}^2) \underset{\tilde{w}\rightarrow 0}{\sim} \frac{1}{\tilde{w}}+\mathcal{O}(1)
\end{equation}
\vspace{-0.5cm}
\subsubsection{4.2.3 Behavior of $G(z)$}
$G(z)$ is defined on $z \in [-\tilde{w}^2,+\infty[$ as $G(z)=z\Psi'(z)^2+2\tilde{w}\Psi'(z)$
and it can be seen that in that interval it is monotonically decreasing.
One notes that $G(-\tilde{w}^2) = 1 - (1 - \tilde w \Psi'(-\tilde w^2))^2 \leq 1$
and that, using the previous estimates
\begin{equation}
 G(-\tilde{w}^2) \underset{\tilde{w}\rightarrow 0}{\sim}1 \quad \text{and} \quad
 G(z) \underset{z\rightarrow +\infty}{\sim} \frac{4}{9\pi^2}\frac{ [\ln z]^{3}}{ z}
\end{equation}
Hence as one decreases $z$ from $+\infty$ to $-\tilde{w}^2$, $G(z)$ increases monotonically from $0$ to $0<G(- \tilde w^2)\leq 1$. Recalling that $ e^H=G(z)$, we find that for any given $H \in [-\infty,H_c(\tilde w)]$, there is a unique solution $z(H)$, and that $H_c(\tilde w) \leq 0$ (which justifies our neglect of the condition \eqref{condition2}). 
In the small $\tilde{w}$ limit, we find that $H_c(0)=0$. 
\subsubsection{4.2.4 Derivatives of $\Phi(H)$ at $H=H_0$} 

Here we show how to identify the center of the distribution, $H_0=\langle H \rangle$, 
and to calculate iteratively the derivatives of $\Phi(H)$ at $H=H_0$, in order to
obtain the cumulants. From the equations \eqref{parametric} we obtain by integration
\begin{equation}
\label{parametric2}
\begin{cases}
\Phi(H)= \Psi(z)-2z \Psi'(z)+2\tilde{w}\ln \left|1+\frac{z\Psi'(z)}{2\tilde{w}} \right| \\
~\\
				\Phi ( H)- 2\Phi '( H)=\Psi(e^{- H} \Phi '( H)(\Phi'(H)-2\tilde{w}))+2\tilde{w}\ln \left|1-\frac{\Phi'(H)}{2\tilde{w}} \right|\\
            \end{cases}
\end{equation}
where by definition of $H_0$, $\Phi(H_0)=0$, and corresponds to the value $z=0$, i.e. $z(H_0)=0$, 
which implies $\Phi'(H_0)=0$ since $z \Psi'(z) \to 0$ as $z \to 0$.
Expanding the last equation into a series the first non-zero derivatives, we first recover
$e^{H_0}=2\tilde{w}\Psi'(0)$, as given in the text, as well as
\begin{equation}
 \Phi^{(2)}(H_0)=-\frac{2\tilde{w}\Psi'(0)^2}{\Psi'(0)^2+2\tilde{w}\Psi''(0)}, \qquad \Phi^{(3)}(H_0)=-\frac{2(\tilde{w}\Psi'(0)^6+12\tilde{w}^3\Psi'(0)^2\Psi''(0)^2-4\tilde{w}^3\Psi'(0)^3\Psi^{(3)}(0))}{(\Psi'(0)^2+2\tilde{w}\Psi''(0))^3}
\end{equation}

We can now calculate explicitly the derivatives $\Psi^{(q)}(0)$ from \eqref{Psiq} as
\bea
&& \Psi^{(q)}(0) = (q-1)! (-1)^{q+1} \frac{1}{\pi} \int_{0}^{+\infty} dy (1+\frac{1}{y+\tilde{w}^2})\sqrt{y}
e^{-q y} \frac{1}{(y+\tilde{w}^2)^q} \\
&& = \frac{1}{2 \sqrt{\pi} q} (-1)^{q+1} \left( \frac{q^{q-\frac{1}{2}} \Gamma \left(\frac{1}{2}-q\right) \Gamma
   (q) \, _1F_1\left(q;q+\frac{1}{2};q w^2\right)}{\sqrt{\pi
   }}+w^{1-2 q} \Gamma \left(q-\frac{1}{2}\right) \,
   _1F_1\left(\frac{1}{2};\frac{3}{2}-q;q w^2\right) \right) \label{beautiful} 
\eea
This leads to 
\begin{equation}
\begin{split}
&e^{H_0}=\mathrm{Erfc}(\tilde{w})e^{\tilde{w}^2}, \qquad \Phi^{(2)}(H_0)=\frac{- 2\pi\tilde{w}\mathrm{Erfc}(\tilde{w})^2 e^{2\tilde{w}^2}}{-2\sqrt{2\pi}\tilde{w}+e^{2\tilde{w}^2}\pi\mathrm{Erfc}(\tilde{w})^2+e^{2\tilde{w}^2}\pi(4\tilde{w}^2-1)\mathrm{Erfc}(\sqrt{2}\tilde{w})},\\
\end{split}
\end{equation}
while the third derivative is given by
\begin{equation}
\begin{split}
& \Phi^{(3)}(H_0)=\frac{2e^{2\tilde{w}^2}\pi^2\tilde{w} \mathrm{Erfc}(\tilde{w})^2}{(-2\sqrt{2\pi}\tilde{w}+e^{2\tilde{w}^2}\pi\mathrm{Erfc}(\tilde{w})^2+e^{2\tilde{w}^2}\pi(4\tilde{w}^2-1)\mathrm{Erfc}(\sqrt{2}\tilde{w}))^3}\\
& [-4e^{4\tilde{w}^2}\pi\mathrm{Erfc}(\tilde{w})^4-3\left(8\tilde{w}^2-4e^{2\tilde{w}^2}\sqrt{2\pi}\tilde{w}(4\tilde{w}^2-1)\mathrm{Erfc}(\sqrt{2}\tilde{w})+e^{4\tilde{w}^2}\pi (4\tilde{w}^2-1)^2\mathrm{Erfc}(\sqrt{2}\tilde{w})^2\right) +\\
& 4e^{\tilde{w}^2}\mathrm{Erfc}(\tilde{w})\left(2\sqrt{3\pi}\tilde{w}(1-2\tilde{w}^2)+e^{3\tilde{w}^2}\pi (1-4\tilde{w}^2+12\tilde{w}^4)\mathrm{Erfc}(\sqrt{3}\tilde{w}) \right)]\\
\end{split}
\end{equation}

Expanding around $\tilde w=+\infty$ we find
\bea
&& e^{H_0}= \frac{1}{\sqrt{\pi } \tilde w}-\frac{1}{2 \sqrt{\pi }
   \tilde w^3}+\mathcal{O}(\frac{1}{\tilde w^4}), \qquad \Phi^{(2)}(H_0)=\sqrt{\frac{2}{\pi }}+\frac{1}{\pi  \tilde w}-\frac{\pi -1}{\sqrt{2}
   \pi ^{3/2} \tilde w^2}+\mathcal{O}(\frac{1}{\tilde w^3}) \\
&& \Phi^{(3)}(H_0)= \frac{1}{9} \left(27-16 \sqrt{3}\right) \sqrt{\frac{2}{\pi
   }}+\frac{27-16 \sqrt{3}}{3 \pi 
   \tilde w}+\mathcal{O}(\frac{1}{\tilde w^2}) \\
&& \Phi^{(4)}(H_0)=   \frac{1}{9} (319-108 \sqrt{2} - 96 \sqrt{3}) \sqrt{\frac{2}{\pi }} 
+ \frac{1775-432 \sqrt{2} - 672 \sqrt{3}}{9 \pi \tilde w} +\mathcal{O}(\frac{1}{\tilde w^2})
\eea 
Expanding around $\tilde{w}=0$ we obtain up to first order
\bea \label{derPhiH0} 
&& e^{H_0}=1-\frac{2\tilde{w}}{\sqrt{\pi}} +\tilde w^2+\mathcal{O}(\tilde w^3)
, \qquad \Phi^{(2)}(H_0)=\frac{\sqrt{\pi}}{2}+\frac{(\pi-3)}{2}\tilde{w} +\frac{\left(3+\pi  \left(-5-4 \sqrt{2}+3 \pi
   \right)\right) \tilde w^2}{6 \sqrt{\pi }}+\mathcal{O}(\tilde w^3) \\
&&  \Phi^{(3)}(H_0)=\frac{\sqrt{\pi}}{4}(3-\pi)+\frac{-7+6\pi+4\sqrt{2}\pi-3\pi^2}{4}\tilde{w} +\mathcal{O}(\tilde w^2) \\
&& \Phi^{(4)}(H_0)=\frac{\sqrt{\pi}}{8}(7-6\pi -4\sqrt{2}\pi +3\pi^2)+\frac{5}{8}(-3-5\pi+8\sqrt{2}\pi-3\pi^2-8\sqrt{2}\pi^2+3\pi^3)\tilde{w} +\mathcal{O}(\tilde w^2)
\eea
with $H_0=-\frac{2 w}{\sqrt{\pi }}+\left(1-\frac{2}{\pi }\right)
   w^2+\mathcal{O}\left(w^3\right)$. We see that all derivatives of $\Phi(H)$ have a finite limit as $\tilde w=0$
   which coincides with the stationary IC.

\subsubsection{4.2.5 Cumulants of the height}
   
To compute the cumulants of the height $H$ at short times, we first
define the cumulant generating function
\begin{equation}
G(p,t)= \left\langle e^{\frac{p}{\sqrt{t}}\, H}\right\rangle= \int 
e^{\frac{p}{\sqrt{t}}\,
H}\, 
P(H,t)\, dH\, ,
\label{cum1.1}
\end{equation}
where $P(H,t)$ is the height PDF. Substituting the short time form,
$P(H,t)\sim e^{-\frac{1}{\sqrt{t}}\, \Phi(H)}$,
and performing the integral by the saddle point method as $t\to 0$ gives
\begin{equation}
G(p,t)\simeq e^{\frac{1}{\sqrt{t}}\, \phi(p)},\quad {\rm where}\quad 
\phi(p)= \max_{H}\left[p H - \Phi(H)\right]\, ,
\label{cum1.2}
\end{equation}
By definition, the logarithm of $G(p,t)$ generates the height 
cumulants as
\begin{equation}
\ln G(p,t)= \sum_{q=1}^{\infty} \langle H(t)^q \rangle^c\, 
\left[\frac{p}{\sqrt{t}}\right]^q \, .
\label{cum1.3}
\end{equation}
Hence, taking logarithm on both sides of Eq. (\ref{cum1.2}), using
(\ref{cum1.3}) and matching powers of $p$ gives
\begin{equation}
\langle H^q \rangle^c= t^{(q-1)/2}\, \phi^{(q)}(0)\, ,
\label{cum1.4}
\end{equation}
for all $q\ge 1$,
where $\phi^{(q)}(0)$ is the $q$-th derivative of $\phi(p)$ evaluated at $p=0$. The optimization problem \eqref{cum1.2} can be solved exactly and yields the implicit equation
\begin{equation}
\phi(p)=p\phi'(p)-\Phi(\phi'(p))
\label{cum2.0}
\end{equation}
Expanding Eq. \eqref{cum2.0} into a series and using the explicit values of the derivatives of $\Phi$ at $H=0$, one
obtains $\phi^{(q)}(0)$ explicitly. For example, the first three non-trivial
cumulants are given by
\begin{eqnarray}
\phi^{(2)}(0) = \frac{1}{\Phi^{(2)}(H_0)} \quad , \quad \phi^{(3)}(0) = -\frac{\Phi^{(3)}(H_0)}{\Phi^{(2)}(H_0)^3} 
\quad , \quad \phi^{(4)}(0) = \frac{3\Phi^{(3)}(H_0)^2-\Phi^{(2)}(H_0)\Phi^{(4)}(H_0)}{\Phi^{(2)}(H_0)^5}
%\phi^{(5)}(0) & = & 
%120\,\left(-\frac{317}{432}-\frac{1}{\sqrt{2}}+\frac{2}{\sqrt{3}}+
%\frac{16}{25\sqrt{5}}\right) \pi^2 
\label{cum1.5}
\end{eqnarray} 
This leads to the following cumulants, for any $\tilde w$
\bea
&& \langle H \rangle = \ln\left( \mathrm{Erfc}(\tilde{w})e^{\tilde{w}^2} \right) + \mathcal{O}(t^{1/2}) \\
&& \langle H^2 \rangle^c = \frac{2\sqrt{2\pi}\tilde{w}-e^{2\tilde{w}^2}\pi\mathrm{Erfc}(\tilde{w})^2-e^{2\tilde{w}^2}\pi(4\tilde{w}^2-1)\mathrm{Erfc}(\sqrt{2}\tilde{w})}{2\pi\tilde{w}\mathrm{Erfc}(\tilde{w})^2 e^{2\tilde{w}^2}} t^{1/2} + \mathcal{O}(t) 
\eea
and, for small $\tilde w$
\begin{eqnarray}
&& \langle H^2 \rangle^c =  \left( \frac{2}{\sqrt{\pi }}+\left(\frac{6}{\pi }-2\right) \tilde{w}+\frac{2
   \left(24+\left(4 \sqrt{2}-13\right) \pi \right) \tilde{w}^2}{3 \pi^{3/2}} \right) t^{1/2} +\mathcal{O}\left(\tilde{w}^3 t^{1/2},t\right) \\
&&  \langle H^3 \rangle^c = 
\left( \left(2-\frac{6}{\pi }\right)-\frac{8
   \left(5+\left(\sqrt{2}-3\right) \pi \right) \tilde{w}}{\pi
   ^{3/2}}-\frac{4 \left(45+\left(16 \sqrt{2}-37\right) \pi
   \right) \tilde{w}^2}{\pi ^2} \right) t +\mathcal{O}\left(\tilde{w}^3 t , t^{3/2} \right) \\
&&  \langle H^4 \rangle^c = 
\bigg( \frac{8 \left(5+\left(\sqrt{2}-3\right) \pi \right)}{\pi
   ^{3/2}}+\frac{20 \left(21+2 \left(4 \sqrt{2}-9\right) \pi
   \right) \tilde{w}}{\pi ^2}\\
   && +\frac{8 \left(1680+25 \left(36
   \sqrt{2}-73\right) \pi +\left(95-122 \sqrt{2}+48
   \sqrt{3}\right) \pi ^2\right) \tilde{w}^2}{5 \pi
   ^{5/2}} \bigg) t^{3/2} +\mathcal{O}\left(\tilde{w}^3 t^{3/2} , t^2\right)
   \label{cum2.5}
\end{eqnarray} 
For completeness we give a few higher cumulants at $\tilde w=0$ obtained
as described in Section 7.1.

\bea
&& \langle H^5 \rangle^c = -\frac{20 \left(21+2 \left(4 \sqrt{2}-9\right) \pi
   \right) t^2}{\pi ^2} \\
&&  \langle H^6 \rangle^c =  \frac{24 \left(252+140 \left(\sqrt{2}-2\right) \pi
   +\left(15-20 \sqrt{2}+8 \sqrt{3}\right) \pi ^2\right)
   t^{5/2}}{\pi ^{5/2}}
\eea

Finally for large $\tilde w$ we find
\begin{eqnarray}
&& \langle H^2 \rangle^c = \left( \sqrt{\frac{\pi }{2}}-\frac{1}{2 \tilde{w}}+\frac{\sqrt{\frac{\pi
   }{2}}}{2 \tilde{w}^2} \right) t^{1/2} +\mathcal{O}(\frac{1}{\tilde{w^3}} t^{1/2},t) \\
&&  \langle H^3 \rangle^c = 
\left(\frac{8 \pi }{3 \sqrt{3}}-\frac{3 \pi
   }{2}\right) t+\frac{-\frac{1}{4}-\frac{3 \pi }{2}+\frac{8
   \pi }{3
   \sqrt{3}}}{\tilde{w}^2} t +\mathcal{O}(\frac{1}{\tilde{w}^3} t, t^{3/2}) \\
   &&  \langle H^4 \rangle^c = \left( \frac{18 + 15 \sqrt{2} - 16 \sqrt{6}}{3} \pi^{3/2} 
   + \frac{18 + 15 \sqrt{2} - 16 \sqrt{6}}{2 \tilde w^2} \pi^{3/2} \right) t^{3/2} 
   + \mathcal{O}(\frac{1}{\tilde w^3} t^{3/2}, t^2) 
\end{eqnarray} 
and we recover the cumulants obtained in \cite{le2016exact}, and in addition obtain 
their leading corrections at large $\tilde w$.

%\bea
%\Psi^{(q)}(0) \approx \frac{1}{2 \sqrt{\pi}} (-1)^{q+1} \tilde w^{1-2 q} \Gamma \left(q-\frac{1}{2}\right)
%\eea
%Note that
%\bea
%\Psi(z) = \sum_{q=1}^{+\infty} \frac{1}{2 \sqrt{\pi} q!} (-1)^{q+1} \tilde w^{1-2 q} \Gamma \left(q-\frac{1}{2}\right)
%= \sqrt{ \tilde w^2 + z} - \tilde w
%\eea 
%which presumably describes correctly $\Psi(z)$ for small $z$ and $\tilde w$ in the region $z \sim \tilde w$.
%
\subsubsection{4.2.6 Variational problem at $\tilde{w}=0$}
Let us note that the function $\Psi(z)$ is well defined directly for $\tilde w=0$, but is not analytic
at $z=0$, with the behavior at small $z$ obtained in Eq. \eqref{psitaylor}
\bea
\Psi(z) = 2 \sqrt{z} - \frac{1}{ \sqrt{\pi}} z + \mathcal{O}(z^{3/2})
\eea
Nevertheless the variational problem \eqref{Phi200} is well defined  and becomes, for $H \leq H_0=0$
\bea
\Phi(H) = \max_{z \in [0,+\infty[} \left( \Psi(z) - 2 \sqrt{z e^H}   \right) \label{Phi2000} 
\eea 
which corresponds to the parametric system \eqref{parametric} setting $\tilde w=0$.
However since $H_0=H_c(0)=0$ one cannot take strictly the derivatives at $H=H_0$
(only the left derivatives are determined). However the limit $\tilde w=0^+$ allows
to recover the correct derivatives and cumulants, as we have checked these
cumulants $\langle H^q \rangle$ for arbitrary $\tilde w$ for $q=1,2$ and
for $w=0$ from a direct small time expansion on the KPZ equation 
(see Section 10.).\\

Naturally, the question now arises: how do we find a solution for $H>0$ ? The trick is to use the analytically continued partner of $\Psi$ that we now investigate.
\section{5. Analytic continuation of $\Psi$}
Let us obtain an analytic continuation of $\Psi$, we start with the following form of $\Psi(z)$ obtained
from \eqref{Psinew} upon the change of variable $b=(y + \tilde w^2) e^y$, i.e. $y=W_0(be^{\tilde{w}^2})-\tilde{w}^2$
\begin{equation}
\begin{split}
\Psi(z) 
&=  \frac{1}{\pi}\int_{\tilde{w}^2}^{+\infty} \mathrm{d}b\left(\frac{2}{3}[W_0(be^{\tilde{w}^2})-\tilde{w}^2]^{\frac{1}{2}}[W_0(be^{\tilde{w}^2})-\tilde{w}^2+3]-2\tilde{w}\arctan({\frac{[W_0(be^{\tilde{w}^2})-\tilde{w}^2]^{\frac{1}{2}}}{\tilde{w}}})\right)\frac{z}{b}\frac{1}{b+z} \\
\end{split}
\end{equation}
We now make use of the following expression that makes sense in distribution theory.
\begin{equation}
\underset{\epsilon\rightarrow 0}{\mathrm{lim}}\, \frac{1}{x+z\pm i\epsilon}=\mathcal{P}(\frac{1}{x+z})\mp i\pi \delta(-z)
\end{equation}
We define $\Delta_0(z)$ the jump of $\Psi(z)$ across the branch cut $ z \in ]-\infty,-\tilde{w}^2[$.
\begin{equation}
\label{jump}
\begin{split}
\Delta_0(z)&=\underset{\epsilon\rightarrow 0}{\mathrm{lim}}\, [\Psi(z-i\epsilon)-\Psi(z+i\epsilon)]\\
 &=\frac{4}{3}[\tilde{w}^2-W_0(-ze^{\tilde{w}^2})]^{3/2}-4[\tilde{w}^2-W_0(-ze^{\tilde{w}^2})]^{1/2}+2\tilde{w}\, \ln\left(\frac{\tilde{w}+[\tilde{w}^2-W_{0}(-ze^{\tilde{w}^2})]^{1/2}}{|\tilde{w}-[\tilde{w}^2-W_{0}(-ze^{\tilde{w}^2})]^{1/2}|}\right)  
\end{split}
\end{equation}
where we used the complex expression of $\arctan$ in terms of logarithm. 
Besides, we have the derivative of the jump
\begin{equation}
\Delta'_0(z)=-2\frac{[\tilde{w}^2-W_0(-ze^{\tilde{w}^2})]^{1/2}}{z}
\end{equation}
Note that we have introduced an absolute value in the logarithm of \eqref{jump}, since, as we will see,
the argument can change sign on the interval that we will consider. This does not affect the value
of the derivative.\\
~\\
From this, we define the analytic continuation of $\Psi(z)$ to a multivalued function on $z \in [-\tilde{w}^2,e^{-1-\tilde{w}^2}]$.
\begin{equation}
\forall z\in [-\tilde{w}^2,e^{-1-\tilde{w}^2}], \quad\Psi_{\mathrm{continued},0}(z):=\Psi(z)+\Delta_0(z)
\end{equation}
The upper boundary $e^{-1-\tilde{w}^2}$ comes from the definition of the real branch $W_0$ of the Lambert function.
Note that at the branching point $z=- \tilde w^2$, both $\Psi$ and $\Psi+\Delta_0$ are only once right-differentiable 
and since $\Delta_0(-\tilde{w}^2)=\Delta_0'(-\tilde{w}^2)=0$ their first derivatives coincide. Higher derivatives are
ill defined, i.e. $\Delta_0^{(2)}(-\tilde{w}^2)=+\infty$. \\

With some additional work, it is possible to find two other reals jumps that are the continuations of $\Delta_0(z)$. For this aim, we generalize \eqref{jump} by defining two complex numbers $z_1$ and $z_2$ and study the limit
\begin{equation}
\Delta(z_1,z_2)=\underset{\epsilon\rightarrow 0}{\mathrm{lim}}\, [\Psi(z_1-i\epsilon)-\Psi(z_2+i\epsilon)]
\end{equation}
There are two contributions in $\Delta(z_1,z_2)$, the first one comes from the principal values which cancels when $z_1=z_2$, the second one comes from the $\delta$ functions $\delta(-z_1)+\delta(-z_2)$ and yields a contribution $\frac{\Delta_0(z_1)+\Delta_0(z_2)}{2}$. As $z_1-i\epsilon$ and $z_2+i\epsilon$ are independent complex numbers, $\Delta_0(z_1)$ and $\Delta_0(z_2)$ are independent functions. In particular, they might not be defined on the same Riemann sheet, meaning that they can be analytically continued independently one from the other.\\
~\\
We can define a second real jump called $\Delta_{-1}(z)$ on $z \in ]0, e^{-1-\tilde{w}^2}]$ using the second branch of the Lambert function $W_{-1}$, i.e replacing $W_0$ by $W_{-1}$ in formula \eqref{jump}. The above remark about the independence of $z_1$ and $z_2$ leads us to define three possible continuations to $\Delta_0(z_1)+\Delta_0(z_2)$.
\begin{equation}
\Delta_0(z_1)+\Delta_0(z_2) \to
\begin{cases}
\Delta_{-1}(z_1)+\Delta_0(z_2)\\
\Delta_0(z_1)+\Delta_{-1}(z_2)\\
\Delta_{-1}(z_1)+\Delta_{-1}(z_2)
\end{cases}
\end{equation}
To cancel the principal values, we require to have $z_1=z_2=z$, which reduces the number of possible continuations
\begin{equation}
\Delta_0(z) \to
\begin{cases}
\frac{\Delta_{-1}(z)+\Delta_0(z)}{2}\\
\Delta_{-1}(z)
\end{cases}
\end{equation}

~\\
We can then define a second and a third continuations of $\Psi(z)$ on $z \in ]0, e^{-1-\tilde{w}^2}]$ as
\begin{equation}
\forall z\in  ]0, e^{-1-\tilde{w}^2}], \quad\Psi_{\mathrm{continued},-1}(z):=\Psi(z)+\Delta_{-1}(z) ,\quad \Psi_{\mathrm{continued},-1/2}(z):=\Psi(z)+\frac{\Delta_{-1}(z)+\Delta_{0}(z)}{2}
\end{equation}
By analogy with \cite{janas2016dynamical}, we call $\Psi_{\mathrm{continued},-1}(z)$ the symmetric second branch and $\Psi_{\mathrm{continued},-1/2}(z)$ the asymmetric second branch. Let us discuss the behavior at $z \to 0^+$. Since $\Psi(0)=0$ it is dominated by the
small $z>0$ behavior of $\Delta_{-1}(z)$. Recall the asymptotics of the second branch for small negative argument $
W_{-1}(-z)\sim_{z\rightarrow 0^+}\ln (z)$. Hence it is dominated by the first term in \eqref{jump} with
%which will give the exponents for the right tail of the distribution of $H$.
\begin{equation}
\Psi_{\mathrm{continued},-1}(z)\underset{z\rightarrow 0^+}{\sim} \frac{4}{3} [-\ln z]^{3/2} \qquad \Psi_{\mathrm{continued},-1/2}(z)\underset{z\rightarrow 0^+}{\sim} \frac{2}{3} [-\ln z]^{3/2} \label{asympt31} 
\end{equation}
We can also define the analytic partners of $G$ with the analytic partners of $\Psi$, we name them $G_0$, $G_{-1/2}$ and $G_{-1}$. We now wish to know if we can obtain all $H\in \mathbb{R}^+$ with these partners.
\section{6. Behavior of $G_0$, $G_{-1/2}$ and $G_{-1}$}

\subsection{6.1 $G_0$}

$G_0(z)$ is defined on $z \in [-\tilde{w}^2,e^{-1-\tilde{w}^2}]$ and can be written as $G_0(z)=z [\Psi'(z)+\Delta'_0(z)]^2+2\tilde{w}[\Psi'(z)+\Delta'_0(z)]$. Using $\Delta_0'(-\tilde{w}^2)=0$, we have the continuity relation $G(-\tilde{w}^2)=G_0(-\tilde{w}^2)$. Recalling the parametric relation $e^H=G_0(z)$ and 
observing numerically that $G_0(z)$ is monotonically increasing with $z$, 
as $z$ increases from $- \tilde w^2$ to $e^{-1-\tilde{w}^2}$, $H$ increases from
$H_c(\tilde w) = \ln G(- \tilde w^2)=\ln G_0(- \tilde w^2) \leq 0$ to a second critical value $H_{c2}(\tilde w) = \ln G_0(e^{-1-\tilde{w}^2})$. At this point
\begin{equation}
 \Delta'_0(e^{-1-\tilde{w}^2}) = - 2 \sqrt{1+ \tilde w^2} e^{1+\tilde w^2} \underset{\tilde{w}\rightarrow 0}{\sim}-2e
\end{equation}
In the stationary limit $\tilde{w}=0$ we can write more explicitly
\begin{equation}
\begin{split}
H_{c2}(0)&=\ln [(\Psi_0'(e^{-1})+\Delta'(e^{-1}))^2 e^{-1}]\\
&=2\ln[2e-\Psi_0'(e^{-1})]-1\\
\end{split}
\end{equation}
as given in the text, 
where $ \qquad \Psi'_0(e^{-1})=\underset{\tilde{w}\rightarrow 0}{\lim}\Psi'(e^{-1-\tilde{w}^2})=\frac{1}{\pi} \int_{0}^{+\infty} dy [1+\frac{1}{y}] \frac{\sqrt{y}}{e^{-1}+ye^y}$. We do not have a closed form for this integral, but numerically, we find $\Psi_0'(e^{-1})\simeq1.27213$, yielding the numerical estimate for $H_{c2}(0)$
\begin{equation}
H_{c2}(0)\simeq 1.85316
\end{equation}
In terms of the units of reference \cite{janas2016dynamical}, this would yield a critical height $\tilde{H}_{c2}=-2H_{c2}\simeq -3.7$ as predicted for the phase transition : we likely found an explicit exact expression for the critical height.
\subsection{6.2 $G_{-1/2}$}
$G_{-1/2}(z)$ is defined on $z \in ]0,e^{-1-\tilde{w}^2}]$ and can be written as $
G_{-1/2}(z)=z [\Psi'(z)+\frac{\Delta_{-1}'(z)+\Delta_{0}'(z)}{2}]^2+2\tilde{w}[\Psi'(z)+\frac{\Delta_{-1}'(z)+\Delta_{0}'(z)}{2}]
$. 
Using the regularity of the branching point of the Lambert function, we have the continuity relation $G_{-1/2}(e^{-1-\tilde{w}^2})=G_0(e^{-1-\tilde{w}^2})=e^{H_{c2}(\tilde w)}$. On the other side of the interval, we have using \eqref{asympt31} 
\begin{equation}
G_{-1/2}(z)\sim_{z\to 0^+} -\frac{ \ln z}{ z}
\end{equation}\\
We see numerically that $G_{-1/2}$ is monotonically decreasing, so as one decreases $z$ from $e^{-1-\tilde{w}^2}$ to 0, $G_{-1/2}(z)$ increases from $\exp (H_{c2}(\tilde w))$ to $+\infty$. Hence, for any given $H\in [H_{c2},+\infty[$ there is a unique solution $z(H)$ using $G_{-1/2}(z)$. 

\subsection{6.3 $G_{-1}$}
$G_{-1}(z)$ is defined on $z \in ]0,e^{-1-\tilde{w}^2}]$ and can be written as $
G_{-1}(z)=z [\Psi'(z)+\Delta'_{-1}(z)]^2+2\tilde{w}[\Psi'(z)+\Delta'_{-1}(z)]
$. 
Using the regularity of the branching point of the Lambert function, we have the continuity relation $G_{-1}(e^{-1-\tilde{w}^2})=G_0(e^{-1-\tilde{w}^2})=e^{H_{c2}(\tilde w)}$. On the other side of the interval, we have using \eqref{asympt31} 
\begin{equation}
G_{-1}(z)\sim_{z\to 0^+} -4\frac{ \ln z}{ z}
\end{equation}\\
We see numerically that $G_{-1}$ is monotonically decreasing, so as one decreases $z$ from $e^{-1-\tilde{w}^2}$ to 0, $G_{-1}(z)$ increases from $\exp (H_{c2}(\tilde w))$ to $+\infty$. Hence, for any given $H\in [H_{c2},+\infty[$ there is a unique solution $z(H)$ using $G_{-1}(z)$. 
\section{7. Analyticity of $\Phi(H)$ and summary }

\subsection{7.1 Analyticity at the point $H_c$} 

We first discuss analyticity at $H_c(\tilde w)$, which corresponds to the point $z=-\tilde w^2$. Let us examine, in the vicinity
and on both sides of $H_{c}(\tilde w)$, the pair of parametric
equations consisting of (i) Eqs. \eqref{supp1.13}-\eqref{defG2} (ii) $\Phi'(H) = - z \Psi'(z)$, $H<H_{c}$, 
$\Phi'(H) = - z \Psi_{\mathrm{continued},0}'(z)$ for $H>H_{c}$.  As discussed above since
$\Delta'_0(z=-\tilde w^2)=0$, $\Phi'(H)$ is continuous at $H=H_c$, with value $\Phi'(H_c(\tilde w))=\tilde w^2 \Psi'(-\tilde w^2)$. We now show continuity of the second derivative. Taking a derivative of (i) allows to express
$dH/dz$ as a function of the triplet $z, \Psi'(z), \Psi''(z)$. Taking a derivative of (ii), and using
this relation, one obtains $\Phi''(H)$ as a function of $z, \Psi'(z), \Psi''(z)$. As $z \to - \tilde w^2$, 
$\Psi''(z)$ diverges, and the expression for $\Phi''(H)$ has a finite limit depending only on
$z,\Psi'(z)$ and one finds

\be
\Phi''(H_c)  = \frac{\tilde{w}^2 \Psi '\left(-\tilde{w}^2\right)
   \left(\tilde{w} \Psi
   '\left(-\tilde{w}^2\right)-2\right)}{2 \tilde{w} \Psi
   '\left(-\tilde{w}^2\right)-2} = \frac{\Phi '\left(H_c\right) \left(2 \tilde{w}-\Phi
   '\left(H_c\right)\right)}{2 \left(\tilde{w}-\Phi
   '\left(H_c\right)\right)}
\ee 
since the expression is the same upon replacing $\Psi \to \Psi_{\mathrm{continued},0}$ the second derivative is
continuous $\Phi''(H_c)$ at $H_c$, and can be expressed from the first one. 
In principle this can be pushed to higher derivatives to show continuity of all by expanding the implicit relation \eqref{parametric2} up to any order.\\

Let us now examine the stationary case $\tilde w=0^+$, for which $H_c=0$. 
Let us recall the result \eqref{psitaylor}
\bea \label{psitaylor2}
\Psi(z) = \psi(u= \sqrt{z}) \quad , \quad \psi(u) = \sum_{n \geq 1} \frac{u^n}{n!} \psi^{(n)}(0) \quad , \quad 
\psi^{(n)}(0) = (-1)^{n-1}  \frac{2^{n-1}}{\sqrt{\pi}} \Gamma(\frac{n}{2}) (\frac{n}{2})^{\frac{n-3}{2}}
\eea 

We can now insert this expansion in the pairs of parametric equations
\bea
&& e^H = z \Psi'(z)^2 \quad , \quad \Phi(H)=\Psi(z) - 2 z \Psi'(z) \quad , \quad H<H_c=0 \\
&& e^H = z (\Psi'(z)- \frac{2}{z} [- W_0(-z)]^{1/2})^2 \quad , \quad \Phi(H)=\Psi(z) - 2 z \Psi'(z) 
+ \frac{4}{3} [- W_0(-z)]^{3/2}  \quad , \quad H>H_c=0
\eea 
Elimination of $z$ leads to an expansion of $\Phi(H)$ in powers of $H$ around $H=0$ on both sides. 
One can check order by order that inserting the values for the odd derivatives $\psi^{(2 q-1)}(0)$ 
given in \eqref{psitaylor2} yield identical Taylor series on both sides. This shows that 
$\Phi(H)$ is analytic at $H=0$. Inserting the values for the even derivatives $\psi^{(2 q)}(0)$ 
given in \eqref{psitaylor2} then allows to recover the results of \eqref{derPhiH0} for $\tilde w=0^+$,
showing that the calculation at $w=0$ matches the one at $w=0^+$. 

%that imposing identical Taylor equation on both sides (analyticity) fixed the values
%of $\psi^{(3)}(0)=2$, $\psi^{(5)}(0)=30$, and more generally of all odd derivatives. Then we obtain that
%$\Phi(H)$ is analytic at $H=0$. To recover the values of the derivatives obtained in 
%\eqref{derPhiH0} for $\tilde w=0$ requires $\psi''(0)=-2/\sqrt{\pi}$, and 
%$\psi^{(4)}(0)=-8 \sqrt{\frac{2}{\pi }}$. We have shown that these are indeed the values predicted 
%{\red REMAINS TO BE DONE !} 

\subsection{7.2 Non-analyticity of $\Phi(H)$ at $H=H_{c2}(\tilde{w})$}
Starting from the implicit representation for $\Phi(H)$
\begin{equation}
\Phi ( H)- 2\Phi '( H)=\Psi(e^{- H} \Phi '( H)(\Phi'(H)-2\tilde{w}))+2\tilde{w}\ln \left|1-\frac{\Phi'(H)}{2\tilde{w}} \right|
\end{equation}
we observe that the regularity of $\Phi(H)$ highly depends on the regularity of $\Psi(z)$. To have continuations of $\Phi(H)$ that are analytic, we at least require $\Psi(z)$ to have the same regularity as its continuations at the branching points.\\

At $H=H_{c2}(\tilde{w})$, we have $z=e^{-1-\tilde{w}^2}$, and the corresponding values for $\Delta_0(z)$, $\Delta_{-1}(z)$ and their derivatives are 
\begin{equation}
\begin{split}
\Delta_0(e^{-1-\tilde{w}^2})&=\Delta_{-1}(e^{-1-\tilde{w}^2})=\frac{4}{3}[\tilde{w}^2+1]^{3/2}-4[\tilde{w}^2+1]^{1/2}+2\tilde{w}\, \ln\left(\frac{\tilde{w}+[\tilde{w}^2+1]^{1/2}}{|\tilde{w}-[\tilde{w}^2+1]^{1/2}|}\right)  \\
 \Delta'_0(e^{-1-\tilde{w}^2}) &= \Delta'_{-1}(e^{-1-\tilde{w}^2}) = - 2 \sqrt{1+ \tilde w^2} e^{1+\tilde w^2} \\
\end{split}
\end{equation}
The second derivatives are ill defined, so we define $\epsilon$ close to $1$ such that $z=- \epsilon \; e^{-1-\tilde{w}^2}$, then we have for the second derivatives
\begin{equation}
\begin{split}
\Delta''_0(z)&=\frac{W_0\left(-ze^{\tilde{w}^2}\right)}{z^2 \sqrt{\tilde{w}^2-W_0\left(-ze^{\tilde{w}^2}\right)} \left(W_0\left(-ze^{\tilde{w}^2}\right)+1\right)}+\frac{2 \sqrt{\tilde{w}^2-W_0\left(-z e^{\tilde{w}^2}\right)}}{z^2}\\
&=\frac{W_0\left(-\epsilon e^{-1}\right)}{\epsilon^2 e^{-2-2\tilde{w}^2} \sqrt{\tilde{w}^2-W_0\left(-\epsilon e^{-1}\right)} \left(W_0\left(-\epsilon e^{-1}\right)+1\right)}+\frac{2 \sqrt{\tilde{w}^2-W_0\left(-\epsilon e^{-1}\right)}}{\epsilon^2 e^{-2-2\tilde{w}^2}}
\end{split}
\end{equation}
As $W_0(-e^{-1})=W_{-1}(-e^{-1})=-1$, we see that only the denominator of the first term diverges. Close to $y=-e^{-1}$, we have \cite{corless1996lambertw} $W_0(y)=-1+\sqrt{2(ey+1)}$ and $W_{-1}(y)=-1-\sqrt{2(ey+1)}$, so 
\begin{equation}
\begin{split}
\Delta''_0(z)&\underset{\epsilon \to 1}{\simeq}\frac{-1}{e^{-2-2\tilde{w}^2} \sqrt{\tilde{w}^2+1} \sqrt{2(1-\epsilon)}}+\frac{2 \sqrt{\tilde{w}^2+1}}{ e^{-2-2\tilde{w}^2}}\to -\infty\\\Delta''_{-1}(z)&\underset{\epsilon \to 1}{\simeq}\frac{1}{e^{-2-2\tilde{w}^2} \sqrt{\tilde{w}^2+1} \sqrt{2(1-\epsilon)}}+\frac{2 \sqrt{\tilde{w}^2+1}}{ e^{-2-2\tilde{w}^2}}\to +\infty
\end{split}
\end{equation}
As a consequence $\Delta''_{-1}$ and $\Delta''_{0}$ are ill defined at $z=e^{-1-\tilde{w}^2}$ but $\frac{1}{2}(\Delta''_{0}+\Delta''_{-1})$ is not, in fact
\begin{equation}
\frac{1}{2}(\Delta''_{0}+\Delta''_{-1})(e^{-1-\tilde{w}^2})=2 \sqrt{\tilde{w}^2+1} e^{2+2\tilde{w}^2}
\end{equation}
Therefore $\Psi_{\mathrm{continued},0}(z)$ and $\Psi_{\mathrm{continued},_1}(z)$ are only once differentiable while $\Psi_{\mathrm{continued},-1/2}(z)$ is twice differentiable at $z=e^{-1-\tilde{w}^2}$. This already shows that the regularity of $\Phi(H)$ at $H_{c2}(\tilde{w})$ depends on the continuation chosen for $\Psi(z$).\\

Recalling the system of parametric equations \eqref{parametric}, as we have at $z=e^{-1-\tilde{w}^2}$
\begin{equation}
\begin{cases}
\Psi_{\mathrm{continued},0}(z)=\Psi_{\mathrm{continued},-1}(z)=\Psi_{\mathrm{continued},-1/2}(z)\\
\Psi'_{\mathrm{continued},0}(z)=\Psi'_{\mathrm{continued},-1}(z)=\Psi'_{\mathrm{continued},-1/2}(z)
\end{cases}
\end{equation}
we are ensured that $\Phi(H)$ and $\Phi'(H)$ are continuous at $H=H_{c2}(\tilde{w})$ whatever branch we choose.\\

We can also obtain the parametric representation for $\Phi''(H)$ 
\begin{equation} \label{eq1} 
\Phi''(H)=-\frac{(\Psi'(z)+z\Psi''(z))(z\Psi'(z)^2+2\tilde{w}\Psi'(z))}{2\tilde{w}\Psi''(z)+\Psi'(z)^2+2z \Psi'(z) \Psi''(z)}
\end{equation}
If $\Psi''(z)$ is infinite, then the expression gets simplified
\begin{equation} \label{eq2} 
\Phi''(H)=-\frac{z(z\Psi'(z)^2+2\tilde{w}\Psi'(z))}{2\tilde{w}+2z \Psi'(z) }
\end{equation}
where $\Psi(z)$ stands for any branch of $\Psi(z)$, meaning $\Psi_{\mathrm{continued},(0,-\frac{1}{2},-1)}$.
We see that as $\Delta_0''(z)$ and $\Delta_{-1}''(z)$ are infinite, and $\Psi'_{\mathrm{continued},0}(z)=\Psi'_{\mathrm{continued},-1}(z)$, $\Phi''(H)$ will be continuous if we choose $\Delta_{-1}$ to be the continuation of $\Delta_0$ but $\Phi''(H)$ will have a discontinuity if we choose $\frac{1}{2}(\Delta''_{0}+\Delta''_{-1})$ as seen in Fig. \ref{phi'(H)}. We conjecture that this jump in the second derivative of $\Phi(H)$ at $H=H_{c2}(\tilde{w})$ is the phase transition observed in \cite{janas2016dynamical}.\\

As in section 7.1, Taylor expanding the implicit equation for $\Phi(H)$ given in \eqref{parametric2}, it is possible to see that $\Phi(H)$ is analytic at the branching point $H_{c2}(\tilde{w})$ if we do the replacement $\Psi_{\mathrm{continued},0}\to \Psi_{\mathrm{continued},-1}$. \\

We can now give more specific values for the stationary case $\tilde w=0$. Setting $\tilde w=0$ in \eqref{eq1}, \eqref{eq2}
and inserting $z=e^{-1}$, we find for all branches around $H_{c2}(\tilde{w})$
\bea
&& \Phi'(H_{c2}(0)) = -e^{-1} (\Psi'(e^{-1}) + \Delta_{0,-1}'(e^{-1})) = 2-e^{-1}\Psi'(e^{-1})=1.53201 \\
&& \Phi''(H_{c2}(0)) = \frac{1}{2} \Phi'(H_{c2}(0)) =0.76601  \qquad \; {\rm \it {( analytic)}} \\
&& \Phi''(H_{c2}(0)) = -\frac{(\Psi'(z)+z\Psi''(z))(z\Psi'(z))}{\Psi'(z)+2z  \Psi''(z)}=0.14997 \quad \; {\rm \it {(non\;  analytic)}}
\eea
and we recall $H_{c2}(0)= 2 \ln(2 e- \Psi'(e^{-1}))-1 = 1.85316$. The jump is then $\Phi''_{\rm analytic}-\Phi''_{\rm non\; analytic}=0.61603$.

\subsection{7.3 Summary}

To summarize, for a given $H\in [-\infty,\infty]$, the optimum
$z$ is determined from the equation
\begin{equation}
 e^H= G(z)\, 
\label{supp1.13}
\end{equation}
where the function $G(z)$ is given by

\begin{flalign} \label{defG2} 
&G(z)= z\Psi'(z)^2+2\tilde{w}\Psi'(z)\quad {\rm 
for}\quad 
z\in [-\tilde{w}^2,+\infty]\quad {\rm and}\quad H\leq H_c(\tilde{w})&\\
&G_0(z)= z [\Psi'(z)+\Delta'_0(z)]^2+2\tilde{w}[\Psi'(z)+\Delta'_0(z)] \quad {\rm for}\quad
z\in [\tilde{w}^2,e^{-1-\tilde{w}^2}]\quad {\rm and}\quad H_c(\tilde{w})\leq H \leq H_{c2}(\tilde{w}) \, . \nn&
\end{flalign} 
For $H>H_{c2}(\tilde{w})$, there exist two solutions $G_{-1/2}(z)$ and $G_{-1}(z)$ given by 

\begin{flalign} \label{defG3} 
&G_{-1}(z)= z [\Psi'(z)+\Delta'_{-1}(z)]^2+2\tilde{w}[\Psi'(z)+\Delta'_{-1}(z)]\quad {\rm for}\quad
z\in ]0,e^{-1-\tilde{w}^2}] \,  &\\
&G_{-1/2}(z)=z [\Psi'(z)+\frac{\Delta_{-1}'(z)+\Delta_{0}'(z)}{2}]^2+2\tilde{w}[\Psi'(z)+\frac{\Delta_{-1}'(z)+\Delta_{0}'(z)}{2}] \quad {\rm for}\quad
z\in ]0,e^{-1-\tilde{w}^2}] \, . \nn&
\end{flalign}
~\\
The function $\ln G(z)$ vs $z$ is plotted in Fig. (\ref{fig.wz}), with
the four elements $\ln G(z)$ (shown by solid blue line), $\ln G_0(z)$
(shown by the dashed red line) $\ln G_{-1/2}(z)$ (shown by the dot-dashed brown line) and $\ln G_{-1}(z)$ (shown by the dot-dashed green line). Note that the branching $G_0(z) \to G_{-1/2}(z)$ is continuous but not differentiable.

\begin{figure}[!h]
\begin{center}
\includegraphics[width = 1\linewidth]{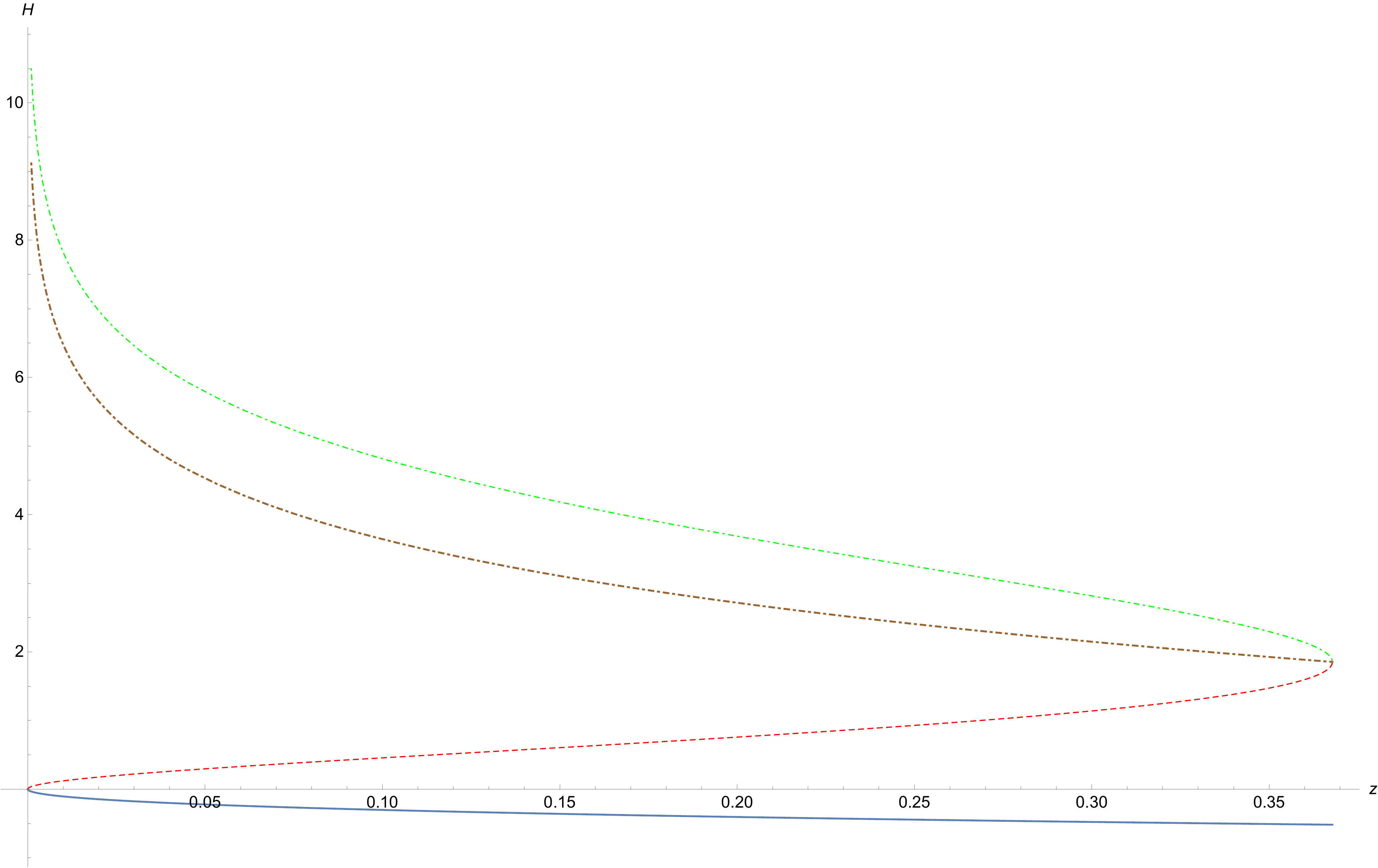}
\caption{The function $\ln [G(z)]$ vs $z$ plotted in the range $z\in 
[0,e^{-1}]$ for $\tilde{w}=0$. We have plotted the four functions $G,G_0,G_{-1/2},G_{-1}$ defined in 
\eqref{defG2} \eqref{defG3}. When $\tilde{w}>0$, the branching between $G$ and $G_0$ appears at $z=-\tilde{w}^2$ and $H_c(\tilde{w})<0$ and the second branching at $z=e^{-1-\tilde{w}^2}$ and $H_{c2}(\tilde{w})<H_{c2}(0)$. Having a non zero $\tilde{w}$ is equivalent to shifting the figure, but does not change its behavior.}
\label{fig.wz}
\end{center}
\end{figure}
To obtain now the continuations of the rate function $\Phi(H)$ for all $H$ we
use the first equation of \eqref{parametric2} replacing
$\Psi$ by $\Psi + \Delta_{-1}$ and $\Psi + \Delta_0$, $\Psi + \frac{1}{2}(\Delta_0+\Delta_{-1})$ respectively.
Using the
above definitions of $\Delta_0$ and $\Delta_{-1}$, we find that the rate function $\Phi(H)$ is determined from the parametric equations

\begin{flalign} \label{PhiParam} 
& \Phi(H)= \Psi(z)-2z \Psi'(z)+2\tilde{w}\ln \left|1+\frac{z\Psi'(z)}{2\tilde{w}} \right| \quad {\rm 
for}\quad 
z\in [-\tilde{w}^2,+\infty] &\\
&\Phi(H)=\Psi(z)-2z \Psi'(z)+\frac{4}{3}[\tilde{w}^2-W_0(-ze^{\tilde{w}^2})]^{\frac{3}{2}}+2\tilde{w}\ln \left|\frac{(1+\frac{z(\Psi'(z)+\Delta_0'(z))}{2\tilde{w}})(1-\frac{z\Delta'_0(z)}{2\tilde{w}})}{1+\frac{z\Delta'_0(z)}{2\tilde{w}}}\right|  \; {\rm for}\;
z\in [\tilde{w}^2,e^{-1-\tilde{w}^2}] \, . \nn &
\end{flalign}
On the interval $z\in ]0,e^{-1-\tilde{w}^2}]$, $\Phi(H)$ finally has two extensions, the first one being analytic and the second one being non-analytic
\begin{flalign} \label{PhiParam2} 
&\Phi(H)= \Psi(z)-2z \Psi'(z)+\frac{4}{3}[\tilde{w}^2-W_{-1}(-ze^{\tilde{w}^2})]^{\frac{3}{2}}+2\tilde{w}\ln \left|\frac{(1+\frac{z(\Psi'(z)+\Delta_{-1}'(z))}{2\tilde{w}})(1-\frac{z\Delta'_{-1}(z)}{2\tilde{w}})}{1+\frac{z\Delta'_{-1}(z)}{2\tilde{w}}}\right| \quad {\rm \it {(analytic)}}\nn &
\\
&\Phi(H)= \Psi(z)-2z \Psi'(z)+\frac{2}{3}[\tilde{w}^2-W_{0}(-ze^{\tilde{w}^2})]^{\frac{3}{2}}+\frac{2}{3}[\tilde{w}^2-W_{-1}(-ze^{\tilde{w}^2})]^{\frac{3}{2}}\nn &\\
&+ 2\tilde{w}\ln \left|(1+\frac{z(2\Psi'(z)+\Delta_{0}'(z)+\Delta_{-1}'(z))}{4\tilde{w}})\left(\frac{(1-\frac{z\Delta'_{-1}(z)}{2\tilde{w}})(1-\frac{z\Delta'_{0}(z)}{2\tilde{w}})}{(1+\frac{z\Delta'_{-1}(z)}{2\tilde{w}})(1+\frac{z\Delta'_{0}(z)}{2\tilde{w}})}\right)^{\frac{1}{2}}\right|  . \; {\rm \it {(non\;  analytic)}}\nn &
\end{flalign}
where $z$ should be replaced by the corresponding solution $z(H)$ from \eqref{supp1.13}-\eqref{defG2}. 
Note that the arguments of the logarithms are actually positive in each interval considered
hence the absolute value could be removed. In the limit $\tilde{w}=0$, this system can be simplified by setting all $\tilde w$'s in \eqref{PhiParam} to 0. The logarithmic factors smoothly vanish, as confirmed by numerics, and that way, we obtain the solution $\Phi(H)$ for $\tilde{w}=0$ which is the stationary case.\\

We represent in Fig. \ref{phi'(H)} the function $\Phi'(H)$ vs $H$ at $\tilde{w}=0$ for the exact solutions and all extensions discussed above. One easily identifies the non-analyticity at the point $H=H_{c2}(0)$ where $\Phi'(H)$ is continuous but not differentiable.\\

\begin{figure}[h!]
\begin{center}
\includegraphics[width = 0.9\linewidth]{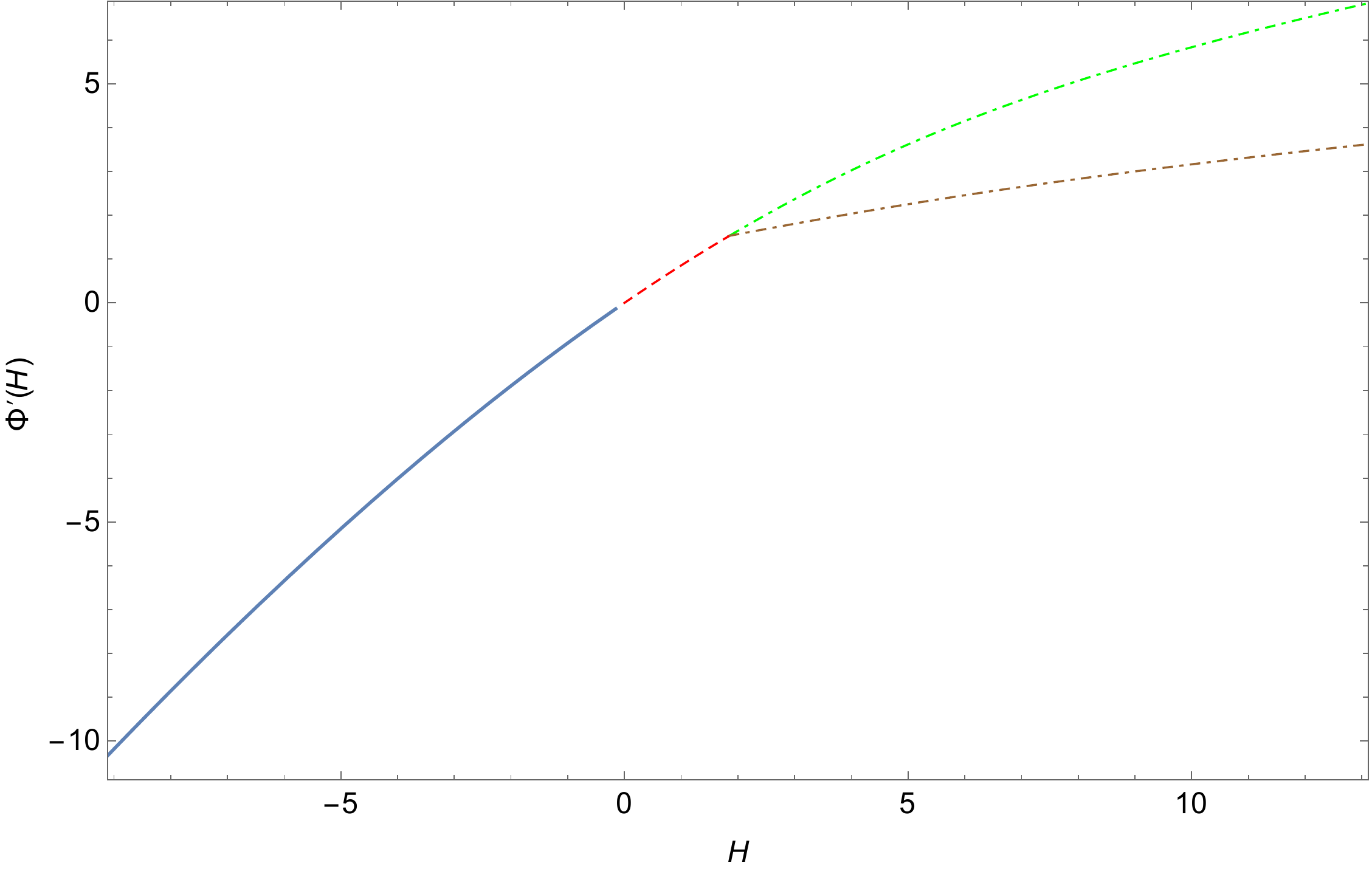}
\caption{The function $\Phi'(H)$ vs $H$ for $\tilde{w}=0$. The blue line corresponds to the exact solution for $H<0$, the dashed red line corresponds to a first analytic continuation for $0<H<H_{c2}$, the dot-dashed green line corresponds to a second symmetric analytic continuation for $H>H_{c2}$ and the dot-dashed brown line corresponds to a second asymmetric non-analytic continuation for $H>H_{c2}$, where $H_{c2}\approx 1.85316$. Note that at the point $H=H_{c2}$ the asymmetric continuation of $\Phi'(H)$ is continuous but not differentiable.}
\label{phi'(H)}
\end{center}
\end{figure}
%\bigskip
 
It is also possible to obtain a variational representation of $\Phi(H)$ in the stationary case, at $\tilde{w}=0$, for all branches as follows 
\bea
\label{PhiResult}
&&\Phi(H) = 
\begin{cases}
 \displaystyle \max_{\substack{z \in [0,+\infty[}} [\Psi(z)  -2\sqrt{ze^H} ], \; H \leq H_c(0)=0\\
\\
 \displaystyle \max_{z \in [0,e^{-1}]} [\Psi(z) +\frac{4}{3}[-W_0(-z)]^{\frac{3}{2}}-4[-W_0(-z)]^{\frac{1}{2}} +2 \sqrt{ ze^H} ], \; H_c(0)\leq H \leq H_{c2}(0)\\
 \\
 \displaystyle \min_{z \in ]0,e^{-1}]} [\Psi(z)+\frac{4}{3}[-W_{-1}(-z)]^{\frac{3}{2}}-4[-W_{-1}(-z)]^{\frac{1}{2}}  +2\sqrt{  ze^H}], \; H \geq H_{c2}(0) \qquad {\rm \it {( analytic)}}\\
  \\
 \displaystyle \min_{z \in ]0,e^{-1}]} [\Psi(z) +\frac{2}{3}[-W_{0}(-z)]^{\frac{3}{2}}+\frac{2}{3}[-W_{-1}(-z)]^{\frac{3}{2}}-2[-W_{0}(-z)]^{\frac{1}{2}}-2[-W_{-1}(-z)]^{\frac{1}{2}} +2\sqrt{  ze^H}], \; H \geq H_{c2}(0) \\
 \qquad \qquad {\rm \it {(non\;  analytic)}}
\end{cases}
%\hspace{-1cm}\text{with} \quad \Delta_{0,-1}(z)=\frac{4}{3}[\tilde{w}^2-W_{0,-1}(-ze^{\tilde{w}^2})]^{3/2}\\
%\hspace{-3.2cm}-4[\tilde{w}^2-W_{0,-1}(-ze^{\tilde{w}^2})]^{1/2}\\
%+2\tilde{w}\, \ln(\frac{\tilde{w}+[\tilde{w}^2-W_{0,-1}(-ze^{\tilde{w}^2})]^{1/2}}{\tilde{w}-[\tilde{w}^2-W_{0,-1}(-ze^{\tilde{w}^2})]^{1/2}})  
\eea
where we have inserted the explicit expressions of $\Delta_{0,-1}(z)$ from \eqref{Delta0}, \eqref{Delta1} setting
$\tilde w=0$. In addition, $H_{c}(0)= 0$ and $H_{c2}(0)=2\ln(2e-\Psi_0'(e^{-1}))-1\simeq 1.85316$, where $\Psi_0$ is the function $\Psi$ in the limit $\tilde w \rightarrow 0$.

One can understand the change of sign in front of $2\sqrt{z e^H}$ as follows : we first decrease $z$ from $+\infty$ to $0$ and then increase it to $e^{-1}$. In the complex $z$-plane, turning around $0$ induces a branch change in the square root function $\sqrt{z}\to -\sqrt{z}$. The change from a maximum to a minimum can be seen from a change of convexity in the argument of the variational problem.\\

Note that there exist some version of this variational representation for $\tilde w>0$ but we have not
attempted to write it explicitly. It is complicated by the fact that there is a new distinct
field $H^*(\tilde w)$, with $H_c(\tilde w) < H^*(\tilde w) < H_{c2}(\tilde w)$, which is the unique field 
at which $\Phi'(H^*(\tilde w))=\tilde w$. The expression for $\Phi'(H)$ changes form 
at this value, i.e. one has for all $H$ that $\Phi'(H)=\tilde w - \epsilon_H \sqrt{\tilde w^2 + z e^H}$  
with $\epsilon_H={\rm sgn}(H^*(\tilde w)-H)$. The field $H^*(\tilde w)$ is thus also the
unique field such that $z e^{H^*}=- \tilde w^2$, while for all other fields one has 
$\tilde w^2 + z e^H \geq 0$, and the square root changes sign there, 
very much as in Eq. \eqref{PhiResult}.

\subsection{7.4 Comparison with the data of Janas, Kamenev and Meerson \cite{janas2016dynamical}}
We compare in this section our exact expression for the rate function with the numerical estimates obtained by Janas, Kamenev and Meerson in \cite{janas2016dynamical}. The authors kindly provided us their numerical data enabling us to overlap our results with theirs in Fig. \ref{meersonkamenev}.\\

In our system of units, see \cite{footnote1}, the comparison is possible for a range $H \in [0,4]$ which comprises all continuations of $\Phi(H)$.	The data were provided for both symmetric and asymmetric WNT solutions, allowing us to test our hypothesis whether our analytic and non-analytic branches match these solutions.\\

The interpretation of Fig. \ref{meersonkamenev} is that our analytic branch matches point to point the symmetric WNT solution and that our non-analytic branch also matches point to point the asymmetric WNT solution for the interval considered $H \in [0,4]$. Further numerics would be required to allow a comparison outside $H \in [0,4]$ but according to the overlap of our exact result with numerical estimates, we are confident in saying that the branching point $H_{c2}$ is the critical field where a phase transition was observed in \cite{janas2016dynamical}.

\begin{figure}[h!]
\begin{center}
\includegraphics[width = 1\linewidth]{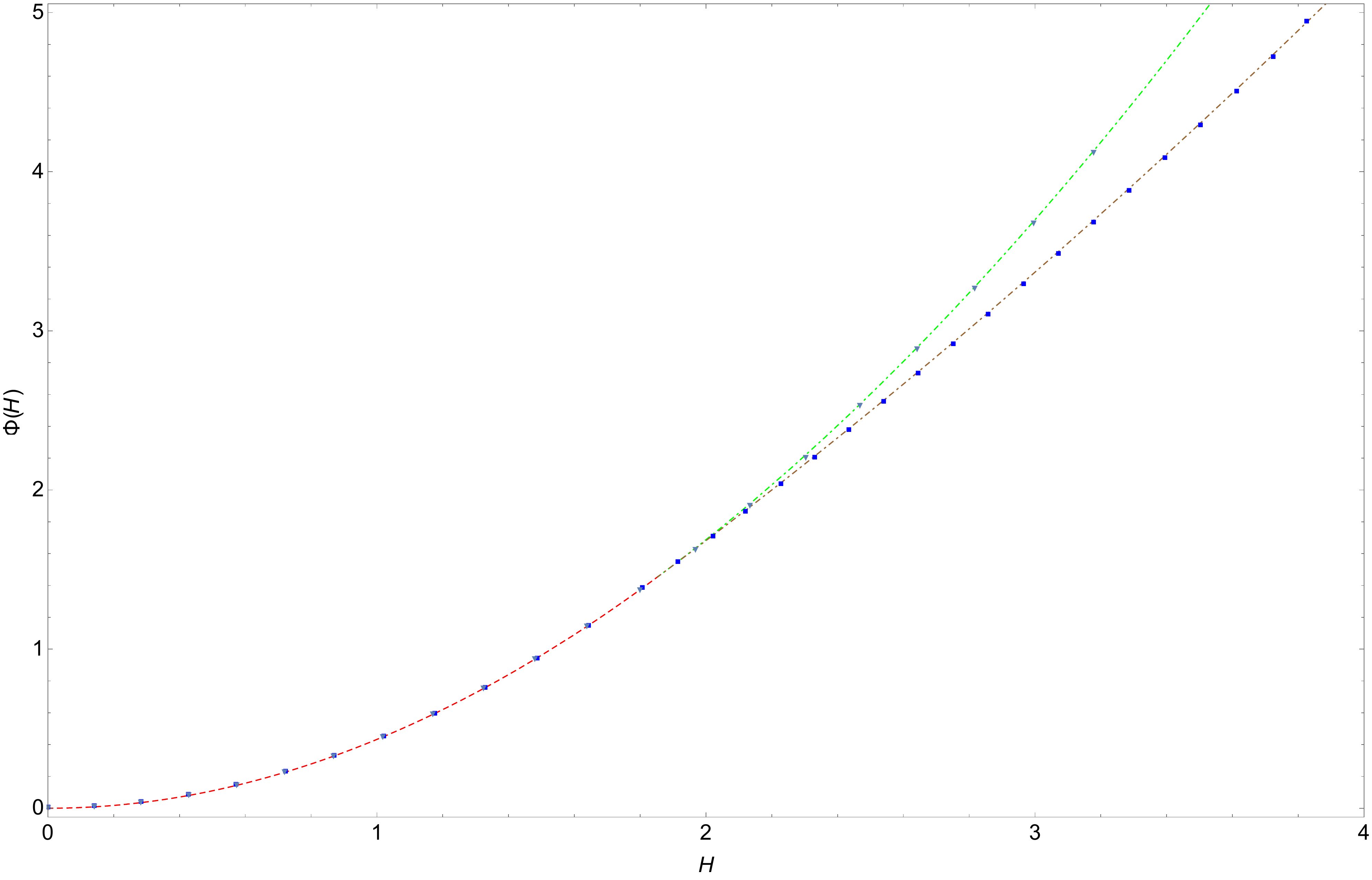}
\caption{The exact rate function $\Phi(H)$ summarized in \eqref{PhiResult} is compared with the numerical estimates from \cite{janas2016dynamical}. The dashed red line corresponds to a first analytic continuation of $\Phi(H)$ for $0<H<H_{c2}$, the dot-dashed green line corresponds to a second analytic continuation for $H>H_{c2}$ and the dot-dashed brown line corresponds to a second non-analytic continuation for $H>H_{c2}$, where $H_{c2}\approx 1.85316$. The blue squares represent the value of action obtained from the asymmetric WNT solution and the grey triangles represent the value of the action obtained from the symmetric WNT solution. The numerical estimates of \cite{janas2016dynamical} and our exact results match point to point for both branches.}
\label{meersonkamenev}
\end{center}
\end{figure}

\section{8. Asymptotic behavior of $\Phi(H)$} 

\subsection{8.1 Left tail $H\rightarrow -\infty$}

We are looking for a power law growth for the large deviation function at large negative $H$ of the form
\begin{equation}
\Phi(H)\underset{H\rightarrow -\infty}{\simeq} c_-[-H]^{\beta_1},
\end{equation}
with $\beta_1$ and $c_-$ positive reals. Using the fact that $\Psi$ has a logarithmic asymptotic for large positive arguments, $\Psi(z)\underset{z\rightarrow +\infty}{\sim} \frac{4}{15 \pi} [\ln z]^{5/2}$ and that its derivative behaves as $\Psi'(z)\underset{z\rightarrow +\infty}{\sim} \frac{2}{3 \pi} \frac{ [\ln z]^{3/2}}{z}$ leads, using \eqref{parametric} and \eqref{parametric2}, to the parametric system of equations
\begin{equation}
\begin{cases}
 e^H\sim \frac{4}{9\pi^2}\frac{ [\ln z]^{3}}{ z}\\
c_-[-H]^{\beta_1} \sim \frac{4}{15 \pi} [\ln z]^{5/2}
\end{cases}
\end{equation}
Combining these equations and identifying the coefficients we obtain the exponent and coefficient of the left tail.
\begin{equation}
c_-=\frac{4}{15 \pi}  \qquad
\beta_1=\frac{5}{2}
\end{equation}
This tail is valid for all $\tilde{w}$, in particular it is valid for the stationary and droplet IC's. Using the asymptotics \eqref{asymptPsi} for $\Psi(z)$ and the equations \eqref{parametric} we obtain the
subleading corrections of the left tail valid for fixed finite $\tilde w$ as
\be
\Phi(H)\underset{H\rightarrow -\infty}{\simeq} \frac{4}{15 \pi}[-H]^{\frac{5}{2}}+\frac{4}{3\pi}[-H]^{\frac{3}{2}}\ln(-H) +\frac{4+6\ln(\frac{4}{9\pi^2})}{9\pi}[-H]^{\frac{3}{2}}+\mathcal{O}([-H]^{\frac{1}{2}}\ln(-H)^2)
\ee

\subsection{8.2 Right tail $H\rightarrow +\infty$ for the symmetric second branch}

We are looking for a power law growth for the large deviation function at large positive $H$ of the form
\begin{equation}
\Phi(H)\underset{H\rightarrow +\infty}{\sim} c_+ H^{\beta_2}
\end{equation}
with $\beta_2$ and $c_+$ positive reals. Using the fact that $\Psi_{\rm continued,-1}$ has a logarithmic asymptotic for small positive argument, $\Psi_{\rm continued,-1}(z)\sim_{z \to 0^+} \frac{4}{3}[-\ln z]^{3/2}$ (see above) and that its derivative behaves as $\Psi'_{\rm continued, -1}(z) \sim_{z \to 0^+} -2 \frac{[-\ln z]^{1/2}}{z}$ leads to the parametric system of equation
(using \eqref{PhiParam}, or equivalently again \eqref{parametric}, \eqref{parametric2} with the replacement
$\Psi \to \Psi + \Delta_{-1}$) \begin{equation}
\begin{cases}
 e^H\sim -4\frac{ \ln z}{ z}\\
c_+ [H]^{\beta_2} \sim \frac{4}{3} [-\ln z]^{3/2}
\end{cases}
\end{equation}

By identification, we obtain the exponent and coefficient of the right tail :
\begin{equation}
c_+=\frac{4}{3} \qquad
\beta_2=\frac{3}{2}
\end{equation}
This tail is also valid for all $\tilde{w}$, in particular it is valid for the symmetric branch of the stationary IC and droplet. Using the asymptotics for $W_{-1}$ given in section 1, we obtain the subdominant corrections
to the right tail of the symmetric branch valid at fixed finite $\tilde w$ as
\be
\Phi(H)\underset{H\rightarrow +\infty}{\simeq} \frac{4}{3}H^{\frac{3}{2}}
 - (4 \ln 2) H^{\frac{1}{2}}+\mathcal{O}(\ln H)
\ee

\subsection{8.3 Right tail $H\rightarrow +\infty$ for the asymmetric second branch}

We are looking for a power law growth for the large deviation function at large positive $H$ of the form
\begin{equation}
\Phi(H)\underset{H\rightarrow +\infty}{\sim} c_+ H^{\beta_3}
\end{equation}
with $\beta_3$ and $c_+$ positive reals. Using the fact that $\Psi_{\rm continued,-1/2}$ has a logarithmic asymptotic for small positive argument, $\Psi_{\rm continued,-1/2}(z)\sim_{z \to 0^+} \frac{2}{3}[-\ln z]^{3/2}$ (see above) and that its derivative behaves as $\Psi'_{\rm continued, -1}(z) \sim_{z \to 0^+} - \frac{[-\ln z]^{1/2}}{z}$ leads to the parametric system of equation
(using \eqref{PhiParam}, or equivalently again \eqref{parametric}, \eqref{parametric2} with the replacement
$\Psi \to \Psi + \frac{\Delta_{-1}+\Delta_0}{2}$) \begin{equation}
\begin{cases}
 e^H\sim -\frac{ \ln z}{ z}\\
c_+ [H]^{\beta_3} \sim \frac{2}{3} [-\ln z]^{3/2}
\end{cases}
\end{equation}

By identification, we obtain the exponent and coefficient of the right tail :
\begin{equation}
c_+=\frac{2}{3} \qquad
\beta_3=\frac{3}{2}
\end{equation}
This tail is also valid for all finite $\tilde{w}$, in particular it is valid for the asymmetric branch of the stationary IC. Using the asymptotics for $W_{-1}$ given in section 1, we obtain the subdominant corrections
to the right tail of the asymmetric branch as
\be
\Phi(H)\underset{H\rightarrow +\infty}{\simeq} \frac{2}{3}H^{\frac{3}{2}}
+\mathcal{O}(\frac{\ln H}{H^{\frac{1}{2}}})
\ee 
While the leading term is valid for any fixed $\tilde w$, we have indicated the subdominant one in the case
$\tilde w=0$. 

\section{9. Useful check : the droplet limit}

We present in this section useful checks that allow us to recover the droplet limit at each step of our reasoning.
\subsection{9.1 The deformed Airy function}
We start by re-introducing the deformed Airy function with the proper scaling of our problem.
\begin{equation}
\label{suppdrop1}
\begin{split}
&\mathrm{Ai}_{\Gamma}^{\Gamma}(\tilde{a}t^{-1/3},t^{-1/3},w,w)=\frac{1}{2\pi}\int\limits_{-\infty+i\epsilon}^{+\infty+i\epsilon}\mathrm{exp}(\frac{iz^3}{3}+i\tilde{a}t^{-1/3}z)\frac{\Gamma(it^{-1/3}z+w)}{\Gamma(-it^{-1/3}z+w)} dz \\
\end{split}
\end{equation}
For large $w$ the ratio of Gamma functions converges towards a power law
\begin{equation}
\frac{\Gamma(it^{-1/3}z+w)}{\Gamma(-it^{-1/3}z+w)}\underset{{w\to +\infty}}{\simeq}  \exp \left(it^{-1/3}z \ln(w^2)\right)
\end{equation}
Inserting this asymptotics into the integrand of \eqref{suppdrop1}, we recognize the Airy function with argument $(\tilde{a}+\ln(w^2))t^{1/3}$.
\begin{equation}
\label{suppdrop2}
\mathrm{Ai}_{\Gamma}^{\Gamma}(\tilde{a}t^{-1/3},t^{-1/3},w,w)\underset{{w\to +\infty}}{\simeq}  \mathrm{Ai}\left((\tilde{a}+\ln(w^2))t^{-1/3}\right)
\end{equation}
The limit \eqref{suppdrop2} also points out a misprint in Ref. \cite{SasamotoStationary2}, Eq. (2.15), where
the shift $\ln(w^2)$ is missing in the asymptotics of the deformed Airy function.

\subsection{9.2 The deformed Airy kernel and exact Fredholm representation}

As the kernel of the Fredholm determinant related to the droplet IC is the Airy kernel, the convergence of the deformed Airy function to the Airy function also gives the convergence of the kernels. Therefore, up to the shift $\ln(w^2)$ that we can
incorporate in the definition of $H$, we are able to obtain the droplet IC in the limit of large $w$.\\

Starting from the exact Fredholm representation at the point $x=0$ of the generating function of $e^{\tilde{H}}$
\begin{eqnarray}
&&\bigg\langle \exp \left( - e^{\tilde H - s t^{1/3}}  \right) \bigg\rangle = Q_t(s) \quad , \quad Q_t(s) := {\rm Det}[ I -  \bar K_{t,s}] 
\end{eqnarray}
in terms of the kernel $\bar K_{t,s}(v,v')=K_{\rm Ai, \Gamma}(v,v')\sigma_{t,s}(v')$ (see Section 0.) with
\be \begin{split} 
 K_{\rm Ai, \Gamma}(v,v') := \int_{0}^{+\infty} dr \, \mathrm{Ai}_\Gamma^\Gamma(r+v,t^{-\frac{1}{3}},w,w) \mathrm{Ai}_\Gamma^\Gamma(r+v',t^{-\frac{1}{3}},w,w) 
\end{split} \ee
Using \eqref{suppdrop2}, the asymptotics of the kernel for large $w$ is 
\begin{equation}
 K_{\rm Ai, \Gamma}(v,v') \underset{{w\to +\infty}}{\simeq}  K_{\rm Ai}(v+\ln(w^2)t^{-1/3},v'+\ln(w^2)t^{-1/3}) 
\end{equation}
where $K_{\rm Ai}$ is the Airy kernel entering in the Fredholm determinant giving the generating function of the droplet IC. Noting that $ \sigma_{t,s}(v')=\sigma_{t,s+\ln(w^2)t^{-1/3}}(v'+\ln(w^2)t^{-1/3})$, it yields 
\begin{equation}
\bar K_{t,s}(v-\ln(w^2)t^{-1/3},v'-\ln(w^2)t^{-1/3})\underset{{w\to +\infty}}{\simeq}  K_{\rm Ai}(v,v') \sigma_{t,s+\ln(w^2)t^{-1/3}}(v')
\end{equation}
Coming back to \cite{le2016exact}, and defining $ Q^{\rm drop}_t(s) := {\rm Det}[ I -  \bar K^{\rm drop}_{t,s}]$ the Fredholm determinant associated to the droplet IC with $\bar K^{\rm drop}_{t,s}(v,v')=K_{\rm Ai}(v,v')\sigma_{t,s}(v')$, we obtain
\begin{equation}
Q_t(s) \underset{{w\to +\infty}}{\simeq} Q^{\rm drop}_t(s+\ln(w^2)t^{-1/3})
\label{kernelasymptotics}
\end{equation}
The moment generating function $\bigg\langle \exp \left( - e^{\tilde H - s t^{1/3}}  \right) \bigg\rangle$ tell us that shifting $s$ is equivalent to shifting $\tilde H$, which is itself equivalent to shifting $H$. 

Furthermore, by a saddle point analysis, from the PDF of $\chi$ or from \eqref{saddlechi}, one sees that for large $w$, $\chi$ is almost surely a deterministic variable $\chi=-\ln(2w)$. Combining this information with \eqref{kernelasymptotics}, we have
\begin{equation}
\bigg\langle \exp \left( - \frac{w}{2}	e^{ H - s t^{1/3}}  \right) \bigg\rangle \underset{{w\to +\infty}}{\simeq} Q^{\rm drop}_t(s)
\end{equation}
Defining $H_{\rm droplet}=H+\ln(\frac{w}{2})+\ln{\sqrt{4\pi t}}$, we fully recover the result of \cite{le2016exact}, i.e the droplet IC. This is perfectly consistent with the exact property that the solution $h(x,t)$ of the KPZ equation
with the initial condition \eqref{ic} converges to the droplet solution in the following sense
\bea
\frac{w}{2} e^{h(x,t)} \underset{t \to 0}{\to} e^{B(x)} \frac{w}{2} e^{- w|x|} \simeq_{w \to + \infty} \delta(x) 
\eea 
and from the difference of definitions of $H$ here and $H_{drop}$ in \cite{le2016exact} by
a term $\frac{1}{2} \ln (4 \pi t)$. Note that all the above considerations are valid for arbitrary time $t>0$.

\subsection{9.3 Convergence of the large deviation function $\Psi(z)$ to its droplet limit}
As claimed in the text, in the limit $\tilde{w}=+\infty$, it is also possible to find the short time estimate of the Fredholm determinant of the droplet IC by noticing the following limit, from \eqref{resQ},
\bea
\lim_{\tilde w \to +\infty} \Psi(\tilde w^2 z) = 
 \frac{1}{\pi} \int_{0}^{+\infty} dy \sqrt{y} \ln\left(1 + ze^{-y} \right) = - \frac{1}{\sqrt{4 \pi}} {\rm Li}_{5/2}(-z)=\Psi_{\rm drop}(z)
\eea 

\subsection{9.4 The analytic partner of the large deviation function $\Psi(z)$}

The analytic partner of $\Psi$ was obtained by adding the jump \eqref{jump} following the change of Riemann sheet to the function $\Psi$.
\begin{equation}
\Delta_0(z)=\frac{4}{3}[\tilde{w}^2-W_0(-ze^{\tilde{w}^2})]^{3/2}-4[\tilde{w}^2-W_0(-ze^{\tilde{w}^2})]^{1/2}+2\tilde{w}\, \ln\left(\frac{\tilde{w}+[\tilde{w}^2-W_{0}(-ze^{\tilde{w}^2})]^{1/2}}{|\tilde{w}-[\tilde{w}^2-W_{0}(-ze^{\tilde{w}^2})]^{1/2}|}\right)  
\end{equation}

For negative $z$, in the limit of large $\tilde{w}$ using the logarithmic asymptotics of $W_0$ for large positive argument, we find that $\Delta_0(z) = \frac{4}{3} [-\ln(-z)]^{3/2}$, which is the analytic continuation used for the droplet IC in \cite{le2016exact}.\\

\section{10. Short time expansion of the stochastic heat equation}
 Here we sketch the calculation of the cumulants of $Z=e^H$ at short time,
which provides a useful test of our method, we provide more details at the end of the section. The KPZ equation \eqref{eq:KPZ} in our units
\eqref{units} is equivalent to the stochastic heat equation (SHE)
\bea
\partial_{t} Z(x,t)= \partial_x^2 Z(x,t) + \sqrt{2} \xi(x,t) Z(x,t) 
\eea 
with $\langle \xi(x,t) \xi(x',t') \rangle= \delta(x-x')\delta(t-t')$, 
and the initial condition $Z(x,t=0)=e^{B_w(x)}$ where $B_w(x):=B(x)-w|x|$, and $B(x)$ a
two-sided unit Brownian motion.

Here we want to calculate the cumulants of $Z(0,t)$. We will thus rescale time
and space as $t \to t \tau$ and $x \to \sqrt{t} x$, and use scaling properties of the white noise
and the Brownian to obtain
\bea
\partial_\tau Z(x,\tau) = \partial_x^2 Z(x,\tau) + t^{1/4} \eta(x,\tau) Z(x,\tau) + \delta(\tau) e^{t^{1/4} B_{\tilde w}(x)}  \label{SHE2}
\eea 
where $\langle \eta(x,\tau) \eta(x',\tau) \rangle= 2 \delta(x-x')\delta(\tau-\tau')$, 
and $B_{\tilde w}(x)=B(x)-\tilde w|x|$ is another Brownian. We have incorporated the
initial condition into the equation, with $Z(x,\tau<0)=0$. Now $t$ appears explicitly as a small
parameter and we want to calculate the cumulants of $Z(0,\tau=1)$. We now use schematic notations. Eq. \eqref{SHE2} is solved as
\bea \label{solu1} 
Z = G \cdot ( t^{1/4} \eta Z + \delta e^{t^{1/4} B}  ) 
\eea 
where $\cdot$ means convolution, $(G \cdot f)(x,\tau)= \int_{x'\tau'} G_{x-x',\tau-\tau'} f(x',\tau')$, where
$\int_y \equiv \int dy$,
while multiplication is simple multiplication i.e. $(\eta Z)(x,\tau)= 
\eta(x,\tau) Z(x,\tau)$. Here $G_{x,\tau}:=(4 \pi \tau)^{-1/2} \exp(-x^2/(4 \tau)) \theta(\tau)$ is the
free propagator, $\delta$ denotes the delta function in time $\delta(\tau)$, and we
dropped the index $\tilde w$ in $B_{\tilde w}$. Eq \eqref{solu1} is solved perturbatively as
\bea
&& Z_0= G \cdot \delta e^{t^{1/4} B} \quad , \quad  Z_1 = G \cdot ( t^{1/4} \eta Z_0 ) =  t^{1/4} G \cdot  \eta G \cdot \delta e^{t^{1/4} B} \\
&& Z_2 = G \cdot ( t^{1/4} \eta Z_1 ) = \sqrt{t} G \cdot  \eta G \cdot  \eta G \cdot \delta e^{t^{1/4} B}
\eea 
The first moment is thus - using the average of the geometric Brownian motion - (brackets now denote averages over both $\eta$ and $B$)
\bea
\langle Z_{0,\tau=1} \rangle = \langle G \cdot \delta e^{t^{1/4} B}  \rangle = \int_{y} G_{y,1} e^{ t^{1/2} (1-\tilde w) |y|} 
= 1 + \mathcal{O}(t^{1/2})
\eea 

Anticipating that the leading behavior of the second cumulant is $\mathcal{O}(t^{1/2})$,
we need only in the second moment
\bea
Z_{0,\tau=1}^2 \simeq (Z_0 + Z_1)^2 = (G \cdot \delta e^{t^{1/4} B})^2 + 2 t^{1/4} (G \cdot  \eta G \cdot \delta e^{t^{1/4} B})
(G \cdot \delta e^{t^{1/4} B}) + \sqrt{t} (G \cdot  \eta G \cdot \delta e^{t^{1/4} B})^2 
\eea 
which leads to the second cumulant as 
\bea
&& \langle Z^2 \rangle^c = \langle Z_{x,\tau=1}^2 \rangle^c \simeq t^{1/2} \left( \langle (G \cdot \delta B)^2 \rangle
+  \langle (G \cdot  \eta G \cdot \delta 1)^2 \rangle \right) \nn \\
&& \underset{\tilde w=0}{=} 2 \sqrt{t}  \left( \int_{y>0,y'>0} G_{y,1} G_{y',1} \min(y,y') + \int_0^1 d\tau G_{y,1-\tau}^2 \right)
%&& =  \sqrt{t}  ( \frac{2-\sqrt{2}}{\sqrt{\pi }} + \sqrt{\frac{2}{\pi }} ) 
=   \frac{2}{\sqrt{\pi }} \sqrt{t} \label{Z2new} 
\eea 
where we have used that $\int dy' G_{y',(1-\tau)}= 1$, i.e. $G \cdot \delta 1=1$ and that $B$ is a double-sided Brownian motion, meaning that $B(y)$ and $B(y')$ are independent if $y$ and $y'$ do not share the same sign.
We give here only the result for $\tilde w=0$, but we have checked the result for
all $\tilde w$ against the method of the previous sections. The two integral contributions are represented by the two-point connected diagrams in Fig. \ref{z2c}.
\begin{figure}[h!]
\begin{center}
\includegraphics[scale=.4]{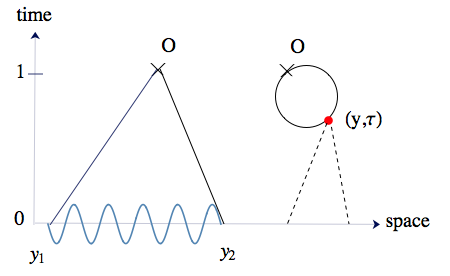}
\caption{Plot of the two-point connected diagrams contributing in the second connected moment $ \langle Z^2 \rangle^c$. The solid black lines are the free propagators $G$, the blue curved line is the double-sided Brownian propagator $\langle BB \rangle$, the dashed black lines represent the relation $G.\delta 1=1$ the red dot is the white noise propagator $\langle \eta \eta \rangle$ and the cross $X$ is the final point of coordinate $(x=0,\tau=1)$. Note that the Brownian propagator lies on the axis of zero-time as it comes from the initial condition.}
\vspace{-0.5cm}
\label{z2c}
\end{center}
\end{figure}

We now turn to the third cumulant $\langle Z^3 \rangle^c$ for which there are 
a priori 4 terms which are $\mathcal{O}(t)$ and non-vanishing
\bea
\langle Z^3 \rangle^c = \langle  Z_0^3 \rangle^c + 3 \langle  Z_0 Z_1^2 \rangle^c + 3 \langle Z_2 Z_0^2 \rangle^c
+ 3 \langle Z_2 Z_1^2 \rangle^c
\eea  
Here the subscript $c$ means connected w.r.t. to 3 points, hence the term $3 \langle Z_2 Z_0^2 \rangle^c =  3 t \langle (G \cdot \eta G \cdot \eta) (G \cdot \delta B)^2 \rangle^c = 0$ vanishes. 
There are thus three non-zero terms, represented by three connected diagrams, see Fig. \ref{z3c}.
\begin{figure}[h!]
\begin{center}
\includegraphics[scale=.4]{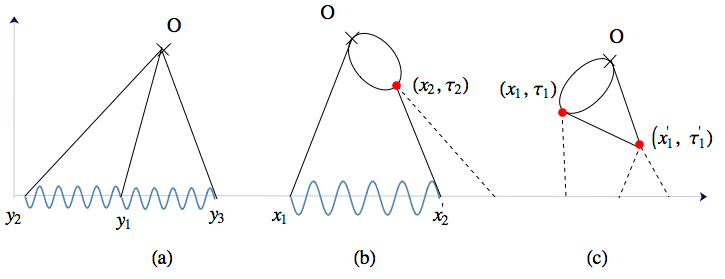}
\caption{Plot of the three-point connected diagrams contributing in the second connected moment $ \langle Z^2 \rangle^c$. (a) represents $I_1$, (b) $I_2$ and (c) $I_3$. The solid black lines are the free propagators $G$, the blue curved line is the double-sided Brownian propagator $\langle BB \rangle$, the dashed black lines represent the relation $G.\delta 1=1$ the red dot is the white noise propagator $\langle \eta \eta \rangle$ and the cross $X$ is the final point of coordinate $(x=0,\tau=1)$.}
\label{z3c}
\vspace{-0.5cm}
\end{center}
\end{figure}

\bea
&& I_1=\langle Z_0^3 \rangle^c 
=  \frac{3}{2} t \int_{y_1,y_2,y_3} G_{y_1,1} G_{y_2,1} G_{y_3,1} \langle B(y_1)^2 B(y_2) B(y_3)\rangle^c \nn \\
&& \underset{\tilde w=0}{=} 3 t \int_{y_1,y_2,y_3} G_{y_1,1} G_{y_2,1} G_{y_3,1}  \langle B(y_1) B(y_2) \rangle 
\langle B(y_1) B(y_3) \rangle = 3 t (\frac{1}{6}+\frac{1}{\pi }-\frac{2 \sqrt{2}}{\pi
   }+\frac{\sqrt{3}}{\pi })
\eea 
where $c$ means connected w.r.t. the three points $y_1,y_2,y_3$, and 
\bea
&& I_2 = 3 \langle Z_0 Z_1^2 \rangle^c %=  %\frac{3}{2} t \overline{ (G \cdot \delta B)^2 (G \cdot \eta) (G \cdot \eta) }^c+
 %6  t \overline{ (G \cdot \delta B) (G \cdot \eta G \cdot \delta B) (G \cdot \eta) }^c 
= 6  t \langle (G \cdot \delta B) (G \cdot \eta G \cdot \delta B) (G \cdot \eta) \rangle^c \nn \\
%&& = 6 t \overline{ G_{x_1,1} B_{x_1} G_{x_2,\tau_2} \eta_{x_2,\tau_2} G_{x'_2-x_2,1-\tau_2} B_{x'_2}
%G_{x_3,\tau_3} \eta_{x_3,\tau_3} }^c \\
&& \underset{\tilde w=0}{=} 12 t \int_{x_1,x_2,x_2',0<\tau_2<1}  G_{x_1,1} G^2_{x_2,\tau_2} G_{x'_2-x_2,1-\tau_2} \langle B(x_1) B(x'_2) \rangle 
= 12 t  ( -\frac{1}{24}+\frac{1}{4 \pi }+\frac{1}{\sqrt{2} \pi
   }-\frac{\sqrt{3}}{4 \pi } ) 
\eea 
using again that $G \cdot \delta 1=1$ and that $\langle (G \cdot \delta B)^2 (G \cdot \eta) (G \cdot \eta) \rangle^c=0$.
\bea
&& I_3 = 3 \langle Z_2 Z_1^2 \rangle^c %=  3 t \overline{ (G \cdot \eta G \cdot \eta) (G \cdot \eta)^2}^c \\
%&& 
= 3 t  \int_{x_1,x_1',x_2,x_3,0<\tau_1,\tau'_1,\tau_2,\tau_3<1} \langle(G_{0x_1\tau_1}\eta_{x_1\tau_1} G_{x'_1-x_1,\tau'_1-\tau_1}\eta_{x'_1\tau'_1})
(G_{0x_2\tau_2}\eta_{x_2\tau_2}) (G_{0x_3\tau_3}\eta_{x_3\tau_3}) \rangle^c \nn \\
&& = 24 t  \int_{x_1,x_1',0<\tau_1,\tau'_1<1}
G_{x_1\tau_1}^2 G_{x'_1-x_1,\tau'_1-\tau_1} G_{x'_1\tau'_1} = 2 t 
\eea 
where we have reversed time integrations, and used again that $G \cdot \delta 1=1$. Here the $c$ means
\bea
\langle\eta_{x_1\tau_1} \eta_{x'_1\tau'_1} \eta_{x_2\tau_2} \eta_{x_3\tau_3} \rangle^c
= 4 \delta_{x_1 \tau_1,x_2 \tau_2} \delta_{x'_1 \tau'_1,x_3 \tau_3}  + 
4 \delta_{x_1 \tau_1,x_3 \tau_3} \delta_{x'_1 \tau'_1,x_2 \tau_2}
\eea
i.e. disconnected (w.r.t. $1=1',2,3$) subgraph when performing Wick's theorem
are set to 0. This diagram is the only one appearing in the flat IC calculation. In total we 
find, after substantial cancellations, the third cumulant as
\bea
\langle Z^3 \rangle^c = I_1+I_2+I_4 = (2 + \frac{6}{\pi}) t \label{Z3new} 
\eea

Having obtained here $\langle Z^q \rangle^c$ here for $q=1,2,3$ by an independent method, we can check the consistency
with our saddle point method. The cumulants of $Z$ are encoded in the function $R(z)$, defined in \eqref{defR},
as
$\langle Z^q \rangle^c= (-1)^{q-1} t^{(q-1)/2} R^{(q)}(0)$. From \eqref{LegendreR} 
and \eqref{parametric} we see that $R'(X) = e^H = G(z)$ and that 
$- z \Psi'(z) = \Phi'(H) = - X e^H= - X R'(X)$ hence we $R(X)$ must satisfy
\bea
R'(\frac{z}{z \Psi'(z) + 2 \tilde w}) = G(z) = z \Psi'(z)^2 + 2 \tilde w \Psi'(z) 
\eea 
Expanding around $z=0$ we obtain the derivatives $R^{(q)}(0)$ as polynomials of the
derivatives $\Psi^{(q')}(0)$, $q' \leq q$, for which we
have an explicit expression \eqref{beautiful}. We then arrive at
\bea
&& \langle Z \rangle = e^{\tilde{w}^2} \text{Erfc}\left(\tilde{w}\right) + \mathcal{O}(t^{1/2}) \\
&& \langle Z^2 \rangle^c   
= -\frac{e^{2 \tilde{w}^2}
   \left(\text{Erfc}\left(\tilde{w}\right)^2+\left(4
   \tilde{w}^2-1\right) \text{Erfc}\left(\sqrt{2}
   \tilde{w}\right)\right)+2 \sqrt{\frac{2}{\pi }}
   \tilde{w}}{2 \tilde{w}^2} t^{1/2} + \mathcal{O}(t) \label{Z2predict} 
\eea 
as well as the first two terms at small $\tilde w$ (up to terms $\mathcal{O}\left(\tilde{w}^2\right)$ and higher orders in $t$)
\bea
&& \langle Z \rangle \simeq 1-\frac{2 \tilde{w}}{\sqrt{\pi}} \quad , \quad  \langle Z^2 \rangle^c  \simeq \left( \frac{2}{\sqrt{\pi  }}-\frac{2 (1+\pi ) \tilde{w}}{\pi} \right) t^{1/2} \\
&&  \langle Z^3 \rangle^c \simeq \left(  \left(2+\frac{6}{\pi}\right)-\frac{4 \left(1+\left(3+2 \sqrt{2}\right)
   \pi \right) \tilde{w}}{\pi^{3/2}} \right) t , \quad 
   \langle Z^4 \rangle^c  \simeq \left( \frac{8 \left(3+\left(3+\sqrt{2}\right) \pi \right)}{\pi^{3/2}}-\frac{12 \left(1+2 \pi  \left(3+4
   \sqrt{2}+2 \pi \right)\right) \tilde{w}}{\pi^2} \right) t^{3/2} \nn
\eea
which agree with the above results \eqref{Z2new}, \eqref{Z3new}. We have also checked
explicitly \eqref{Z2predict} for arbitrary $\tilde w$, but will not give details here. \\

Let us conclude by noting the general relation between the generating function $R(X)$ of the cumulants of
$Z$ and the generating function $\phi(p)$ of the cumulants of
$H$. Using relations given in the the previous sections we obtain
\bea
e^{\phi'(p)} = R'(X) \quad , \quad p = - X R'(X)
\eea 
Hence expanding $e^{\phi'(- X R'(X))} = R'(X)$ in powers of $X$ around $X=0$ we easily
obtain the $\phi^{(q)}(0)$ as rational fractions of the $R^{(q')}(0)$ for $q' \leq q$. Using 
$\langle Z^q \rangle^c= (-1)^{q-1} t^{(q-1)/2} R^{(q)}(0)$ and
$\langle H^q \rangle^c= t^{(q-1)/2} \phi^{(q)}(0)$, we then obtain the desired relations.
Note that these relations are general (independent of the form of $\Psi(z)$) : the present
method is equivalent, but more powerful, than the one given in Appendix B of \cite{flatshorttime}

\end{widetext}

\end{document}